\documentclass[11pt]{amsart}
\usepackage{graphicx}
\usepackage[curve,matrix,arrow,color]{xy}
\usepackage[mathscr]{eucal}

\usepackage[all,dvips,color]{xypic}
\usepackage[usenames,dvipsnames]{color}
\xyoption{line}

\usepackage{epsf}
\usepackage{epsfig}
\usepackage{amssymb}
\usepackage{amsthm}
\usepackage{amsmath}
\usepackage{epstopdf}

\usepackage{pstricks,pst-fill,pst-text,pst-node,pstricks-add}
\usepackage{psfrag}

\usepackage[OT2,T1]{fontenc}
\newcommand{\mifody}{%
  \renewcommand\rmdefault{wncyr}%
  \renewcommand\sfdefault{wncyss}%
  \renewcommand\encodingdefault{OT2}%
  \normalfont
  \selectfont}

\advance\textheight\topskip
\numberwithin{equation}{section}
\makeatletter
\@addtoreset{equation}{section}
\@addtoreset{subsubsection}{section}

\def\@secnumfont{\bfseries}
\def\subsubsection{\@startsection{subsubsection}{3}%
  \z@{.5\linespacing\@plus.7\linespacing}{-.5em}%
  {\normalfont\bfseries}}
\def\paragraph{\@startsection{paragraph}{4}%
  \z@\z@{-\fontdimen2\font}%
  \normalfont\bfseries}
\def\subparagraph{\@startsection{subparagraph}{5}%
  \z@\z@{-\fontdimen2\font}%
  \normalfont\bfseries}

\makeatother

\newcommand{\rme}{{\rm e}}

\swapnumbers

\newcommand{\stf}{t}

 \newcommand{\arc}{ \ \psset{xunit=2mm,yunit=2mm}\begin{pspicture}(0,0)(0.75,1)
\psellipticarc[linecolor=black,linewidth=1.0pt]{-}(0,1.0)(0.5,1.42){180}{360}
\end{pspicture}}
\newcommand{\thl}{\psset{xunit=2mm,yunit=2mm}\begin{pspicture}(0,0)(0.75,1)
\psline[linecolor=black,linewidth=1.0pt](0,-0.5)(0,1)
\end{pspicture}}
\newcommand{\fug}{\delta}

\newcommand{\im}{\mathrm{im}}

\newcommand{\one}{\boldsymbol{1}}
\newcommand{\tensor}{\otimes}

\newcommand{\fus}{\times_{f}}
\newcommand{\q}{\mathfrak{q}}
\newcommand{\ffrac}[2]{\mbox{\footnotesize$\displaystyle\frac{#1}{#2}$}}

\newcommand{\UresSL}[1]{\overline{U}_{\q} s\ell(#1)}

\newcommand{\TL}[1]{TL_{#1}}
\newcommand{\TLq}[1]{TL_{\q,#1}}

\newcommand{\Res}{\mathrm{Res}}
\newcommand{\Ind}{\mathrm{Ind}}
\newcommand{\inv}{\mathrm{inv}}
\newcommand{\modd}{\,\mathrm{mod}\,}
\newcommand{\modl}{|}
\newcommand{\reldim}[1]{\{{#1}\}}
\newcommand{\lev}{\ell}

\newcommand{\dd}{\mathsf{d}}

\newcommand{\sg}{\mathrm{sg}}
\newcommand{\cent}{\mathfrak{Z}}

\newcommand{\LQG}{U_{\q} s\ell(2)}

\newcommand{\half}{%
  \mathchoice{\ffrac{1}{2}}{\frac{1}{2}}{\frac{1}{2}}{\frac{1}{2}}}

 \newcommand{\leftact}{\triangleright}
  \newcommand{\rightact}{\triangleleft}

\newcommand{\FFmodTL}[2]{\mathsf{K}_{#1;#2}}

\newcommand{\Hilb}{\mathcal{H}}
\newcommand{\chVv}{\Hilb_{N}}

\newcommand{\veven}[1]{|v^{\text{even}}\rangle}
\newcommand{\vodd}[1]{|v^{\text{odd}}\rangle}

\newcommand{\vacr}{|\text{vac}\rangle}

\newcommand{\primb}{\varphi}
\newcommand{\primt}{\psi}
\newcommand{\primr}{\rho}

\newcommand{\priml}{\xi}

\newcommand{\oN}{\mathbb{N}}
\newcommand{\oC}{\mathbb{C}}

\newcommand{\step}{\mathrm{step}}
\newcommand{\Endo}{\mathrm{End}}

\newcommand{\Vir}{\mathcal{V}}

\newcommand{\gl}{\mathsf{gl}}

\newcommand{\bb}{\beta}

\newcommand{\Hom}{\mathrm{Hom}}

\newcommand{\Endq}{\mathrm{End}_{\rule{0pt}{6.5pt}%
{\LQG}}}

\newcommand{\K}{\mathsf{K}}
\newcommand{\F}{\mathsf{F}}
\newcommand{\f}{\mathsf{f}}
\newcommand{\E}{\mathsf{E}}
\newcommand{\h}{\mathsf{h}}
\newcommand{\e}{\mathsf{e}}

\newcommand{\stprp}{v}

\newcommand{\bmodP}{\mathbb{P}}

\newcommand{\catUq}{\mathscr{C}}
\newcommand{\catTL}{\mathscr{D}}
\newcommand{\fhom}{\mathscr{H}}
\newcommand{\funt}{\mathscr{T}}

\newcommand{\myar}{\ar@{-->}@[|(1.0)]}

\newcommand{\XX}{\mathcal{X}}
\newcommand{\PP}{\mathcal{P}}
\newcommand{\repX}{\XX}

\newcommand{\modM}{\mathcal{M}}
\newcommand{\modN}{\mathcal{W}}
\newcommand{\modNbar}{\mathcal{W}^*}
\newcommand{\modWeyl}{\modN}
\newcommand{\modWeylj}[1]{\modWeyl_{#1}}
\newcommand{\modWeyls}{\modNbar}

\newcommand{\IrrTL}[1]{\mathsf{X}_{#1}}
\newcommand{\tIrrTL}[1]{\tilde{\mathsf{X}}_{#1}}
\newcommand{\PrTL}[1]{\mathsf{P}_{#1}}
\newcommand{\StTL}[1]{\mathsf{S}_{#1}}
\newcommand{\StTLn}[2]{\mathsf{S}_{#1}[#2]}
\newcommand{\tStTL}[1]{\tilde{\mathsf{S}}_{#1}}

\newcommand{\VX}{\mathscr{X}}
\newcommand{\VK}{\mathscr{K}}
\newcommand{\VP}{\mathscr{P}}
\newcommand{\cVP}{\mathscr{\check P}}
\newcommand{\Verma}{\mathscr{V}}

\newcommand{\Hmod}{\mathscr{H}}

\newcommand{\Braket}[1]{\left<#1  \right>}

\renewcommand{\geq}{\,{\geqslant}\,}
\renewcommand{\leq}{\,{\leqslant}\,}

\newtheorem{Thm}[subsection]{Theorem}

\newtheorem{Prop}[subsection]{Proposition}
\newtheorem{prop}[subsubsection]{Proposition}

\newtheorem{conj}[subsubsection]{Conjecture}

\theoremstyle{definition}

\begin{document}

\title[Lattice fusion rules and logarithmic OPEs]{Lattice fusion rules and logarithmic operator product expansions}

\author{A.M.~Gainutdinov}
\address{Institut de Physique Th\'eorique, CEA Saclay,
 91191 Gif Sur Yvette, France}
\email{azat.gaynutdinov@cea.fr}

\author{R.~Vasseur}
\address{Institut de Physique Th\'eorique, CEA Saclay, 91191 Gif Sur Yvette, France\newline\mbox{}\;\;\,
LPTENS, 24 rue Lhomond, 75231 Paris, France}
\email{romain.vasseur@cea.fr}
  
\begin{abstract}
The interest in Logarithmic Conformal Field Theories (LCFTs) has been growing over the last few years
thanks to recent developments coming from various approaches. A particularly fruitful point of view
consists in considering lattice models as regularizations for such quantum field theories.
The indecomposability then encountered in the representation theory of the corresponding \textit{finite-dimensional} associative
algebras exactly mimics the Virasoro indecomposable modules expected to arise in the 
continuum limit.  
In this paper, we study in detail the so-called Temperley-Lieb (TL) {\it fusion functor} introduced in physics by Read and Saleur [Nucl. Phys.  B {\bf 777}, 316 (2007)].
Using quantum group results, we provide rigorous calculations of the fusion of various TL modules at roots of unity cases. 
Our results are illustrated by many explicit examples relevant for physics. We discuss how indecomposability
arises in the ``lattice'' fusion and compare the mechanisms involved with similar observations in the corresponding
field theory. We also discuss the physical meaning of our lattice fusion rules in terms
of indecomposable operator-product expansions of quantum fields.
\end{abstract}

\maketitle

\section{Introduction}

The study of logarithmic conformal field theories (LCFTs) through their lattice realizations has proven
to be useful in the past few years. The main features of such quantum field theories are indeed already 
present when one considers {\it non-local} observables in usual statistical systems at their critical 
point -- such as geometrical properties of self-avoiding walks or percolation clusters. The description of such observables
has led to the extension of the minimal models into non-rational theories, 
including observables described not only by irreducible Virasoro representations, but also 
by larger indecomposable, yet reducible, representations. Indecomposability is the main feature of a 
logarithmic CFT. In general, it leads to non-diagonalizability of the dilatation operator $L_0$ and to 
logarithmic singularities in correlation functions.

Applications of LCFTs range from disordered systems in condensed matter physics to string theory~\cite{Cardylog,GurarieLudwig1,Zirnbauer,SQHE,RBIM,FLN,PSU122}. Two-dimensional 
geometrical problems such as percolation or self-avoiding walks provide probably the most natural examples of LCFTs with central 
charge $c=0$. 
More interestingly maybe, LCFTs are expected to appear
in disordered critical points~\cite{Cardylog,GurarieLudwig1}. In particular, critical points of non-interacting 
fermions in 2+1 dimensions are believed to be described by logarithmic $c=0$ theories. This is the case for example
in the transition between plateaux in the Integer Quantum Hall Effect (see {\it e.g.}~\cite{Zirnbauer})
or in its cousin, related to the percolation problem, the Spin Quantum Hall Effect~\cite{SQHE}. 
Another example of interesting disordered critical point is provided by the random-bond Ising model
at the Nishimori point, whose underlying CFT is still unknown (see {\it e.g.}~\cite{RBIM} and references therein). 
In the modern language of random non-interacting fermionic systems, this problem is closely related to the
so-called class~D of symmetry, with broken time reversal and no spin-rotation symmetry~\cite{SymClass}.
On a very different matter, 2D super-symmetric sigma models (on compact Kahler manifolds) beyond the topological 
sector and 4D gauge theories~\cite{FLN} provide other interesting physical applications of LCFTs.
LCFTs probably also play an important role in the AdS/CFT correspondence, as they describe massless limit
of non-linear sigma models with non-compact target spaces (see {\it e.g.}~\cite{PSU122}). Besides concrete physical applications, LCFTs attracted recently considerable interest in the mathematics community~\cite{[HLZ],[AM1],[TW]} as well.

A fundamental question about LCFTs is to determine the operator product expansions (OPEs) of their quantum fields. Thanks to the conformal symmetry, the OPEs are essentially determined by the fusion rules for the corresponding modules over the Virasoro algebra.
Mainly, two different direct lines of attack have been pursued to tackle the problem of computing fusion rules. 
The first one, which is probably the most straightforward, uses  the  the so-called
Nahm--Gaberdiel--Kausch (NGK) algorithm~\cite{Nahm,KauschGaberdiel} based on  `geometrical' comultiplication for Virasoro generators~\cite{MS89}.
 This approach was applied to many (chiral) LCFTs like percolation with success~\cite{Flohr,MathieuRidout,MathieuRidout1}. 
The usual output of the repeated fusion consists in 
\textit{staggered} modules~\cite{Rohsiepe}
which are indecomposable Virasoro modules with a diamond-shape subquotient structure and non-diagonalizable action of the $L_0$ generator.
The systematic study of these staggered modules was achieved recently in~\cite{KytolaRidout}. 
The second approach is to consider lattice regularizations of LCFTs 
using 2D statistical systems~\cite{PRZ,PRPolymers,JSBlob} or 1+1D quantum spin chains~\cite{ReadSaleur07-1,ReadSaleur07-2} at their critical point. 
This is strongly motivated by the known regularization of the stress-energy tensor in terms of  Temperley-Lieb (TL) algebra generators~\cite{KooSaleur}.
This ``lattice'' approach also relies on the similarity between the representation
theory of the associative algebras
on the lattice side and that of
the Virasoro algebra in the continuum. As it turns out, the structure of the projective TL modules mimics exactly
that of some staggered modules over Virasoro. To be more precise, one can construct an inductive system of categories of TL modules with terms labeled by the number of sites in the spin-chain, the inductive limit then belongs to the category of Virasoro modules. 
The study of such ``lattice'' algebras then allows to understand the complicated
structure of the continuum limit. This was partially  done in~\cite{ReadSaleur07-1,ReadSaleur07-2} 
and independently, with less emphasize on algebraic aspects, in the work~\cite{PRZ}. 
Using quantum-group considerations, the authors of~\cite{ReadSaleur07-1,ReadSaleur07-2}
were able to obtain several fusion rules for the projective TL modules. Meanwhile,
 many consistent fusion rules were conjectured using both lattice considerations~\cite{RP1,RP2} and  their combination with the NGK
algorithm~\cite{JR}. Both approaches yield a consistent picture of chiral LCFTs. However, to the best of our knowledge 
there is lack of a clear, well-defined framework to compute fusion from lattice models in a systematic and effective way, without relying on the somewhat cumbersome NGK algorithm.

In this paper, we consider 1+1D quantum spin chains with Heisenberg nearest-neighbor interaction given by TL generators. Such systems admit a large symmetry (or TL centralizing) algebra given by a representation of the quantum group $\LQG$~\cite{PasquierSaleur}. This provides the well-known quantum version of the classical Schur--Weyl duality between the symmetric group action and its centralizer $SU(2)$.
 We revisit and generalize the fusion procedure suggested in~\cite{ReadSaleur07-1,ReadSaleur07-2}
using recent results on quantum groups~\cite{[BFGT],[BGT]} at roots of unity cases. We obtain rigorously the fusion rules for standard, irreducible, and projective modules
over the Temperley-Lieb algebra at any root of unity. These results are also supported by very simple lattice calculations on a few sites. Using these exact lattice results, we conjecture general fusion rules
for Virasoro modules and interpret them physically in terms of the fields that appear in the continuum limit of our lattice models.
Interestingly, these fusion calculations parallel a well known approach that consists in considering LCFTs as limits of ordinary, non-logarithmic, CFTs~\cite{Gurarie2, GurarieLudwig2, KoganNichols, VJS} (see also~\cite{Cardylog} for a similar discussion in the language of replicas). Within this approach, logarithms and indecomposability arise to compensate for divergent terms in OPEs as one goes to logarithmic (or critical) points.  We note that the existence and uniqueness of logarithmic OPEs (including the associativity condition) were actually stated, as a theorem, in the fundamental serie of papers~\cite{[HLZ]}. We believe our lattice description of fusion provides a very natural lattice regularization
of such indecomposable (logarithmic) OPEs.

It is worthwhile to mention that there is another, though more abstract and less straightforward, approach to compute  fusion rules for modules over Virasoro (and its extended cousins like triplet $\mathcal{W}$-algebras and affine Lie algebras)  based on the so-called Kazhdan--Lusztig (KL) correspondence. The paradigm here is coming from a belief that to any vertex-operator algebra, one can associate a quantum group with an ``equivalent'' representation theory including fusion rules data. A first example of such an equivalence was established between affine $\widehat{s\ell}(2)$ at a negative integer level and a quantum group at a root of unity case~\cite{KL}. While in rational CFTs there are also  strong connections between fusion rules and braiding structures for quantum-group and  Virasoro modules~\cite{GSRb}, similar correspondences were also observed in the context of affine   $\widehat{s\ell}(2)$ modules with positive integer level~\cite{Fink}. In the logarithmic cases, 
 examples of KL correspondences were also established in the triplet $\mathcal{W}$-algebra context~\cite{[FGST],[FGST2],[FGST4]} and in the Virasoro context~\cite{[BFGT]}. Here, the construction of corresponding quantum groups is based on the screening charges in a free-field realization and the quantum groups are identified with full symmetry algebras (centralizers) of the chiral algebras.
This approach is  probably less straightforward than the other two mentioned above (see nevertheless recent successes~\cite{ST,Semikh-fus} clarifying this direction) but it shares strong similarities with the lattice approach here and in~\cite{ReadSaleur07-1,ReadSaleur07-2}  where the main computational tool relies on the centralizer concept as well, \textit{i.e.}, is provided by the quantum version of the Schur--Weyl duality. We thus believe that our paper is a further step towards the understanding the KL correspondences.

The plan of our paper is as follows. We begin with a general preliminary Sec.~\ref{SecGen1}, 
in which we introduce the fundamental object, called {\it fusion functor}, that we shall study throughout this paper.
We will expose our general strategy in this first section by working out a few simple examples to illustrate our methods.
Sec.~\ref{sec:bimod} contains technical details about the representation theory of the TL algebra
and its centralizer algebra, the quantum group $\LQG$. These considerations will be used in Sec.~\ref{Sec::fusion}
to obtain rigorously general  fusion rules for the TL algebra. We  then illustrate those results
through very simple lattice calculations, chosen for their relevance to physics. 
The reader not interested in mathematical details may skip Sec.~\ref{sec:bimod} and jump directly
to the fusion results of Sec.~\ref{Sec::fusion}.
Finally, we discuss the implications of our findings in terms of indecomposable OPEs in Sec.~\ref{Sec::ScalingLimit}.
 We will mainly focus on
 $(p,p+1)$ and $(1,p)$ theories for physical purposes, although
our results could apply to any $(p,p')$ theory, by choosing appropriate (non-primitive) root of unity value for the $\q$ parameter in the quantum group. Because we restrict our study to the case of the Temperley-Lieb algebra, our results will apply only
to Virasoro modules with conformal weights lying in the first row of the infinite Kac table (for an analog continuum construction,
see~\cite{MathieuRidout,MathieuRidout1}).

\subsection{Notations}
For convenience, we collect here some notations that we shall use throughout this paper:
\begin{itemize}
\item[--] $\TLq{N}$ Temperley-Lieb algebra on $N$ sites with loop fugacity $\q+\q^{-1}$,
\item[--] $\Vir(p',p)$  Virasoro algebra with central charge $c_{p',p}$,
\item[--] $\IrrTL{j}$ irreducible module over $\TLq{N}$,
\item[--] $\StTL{j}$ standard module over $\TLq{N}$,
\item[--] $\PrTL{j}$ projective module over $\TLq{N}$,
\item[--] $\Verma_{h}$ Verma module  over the Virasoro algebra  with conformal weight $h$,
\item[--] $\VX_j = \VX_{1,1+2j}$ irreducible module over the Virasoro algebra with conformal weight $h_{1,1+2j}$,
\item[--] $\VK_j = \VK_{1,1+2j}$ Kac module over the Virasoro algebra with conformal weight $h_{1,1+2j}$,
\item[--] $\VP_j = \VP_{1,1+2j}$ staggered Virasoro module with conformal weight $h_{1,1+2j}$.
\end{itemize}


\section{XXZ spin chain, Temperley--Lieb algebra and fusion functor}\label{SecGen1}
Let us begin with a preliminary section in order to explain our general strategy
on a few simple examples, general results and more technical details will be
given in the rest of the paper. We introduce in Sec.~\ref{sec:XXZ-def} notations that we will use in  the paper and  we recall some well-known results about the XXZ spin chain 
as a representation of the Temperley-Lieb (TL) algebra.
In Sec.~\ref{sec:TL-fusion} and~\ref{paragraphOPElattice}, we introduce the concept
of fusion of two TL modules and illustrate it with a few explicit calculations. 
To conclude this first section, we describe in Sec.~\ref{sec:scal-lim-intro}
the corresponding construction in the continuum limit and discuss how
our lattice construction in the logarithmic case somewhat parallels 
the well-known ``catastrophes'' encountered when considering
operator product expansions.

\subsection{XXZ spin chain and loop models}\label{sec:XXZ-def}

Our strategy will be to study finite quantum 1D lattice models whose continuum limits
are described by LCFTs. The idea of doing so probably goes back
to~\cite{ReadSaleur07-2}, who studied the structure of XXZ spin chains
and supersymmetric models~\cite{ReadSaleur01} on the lattice. To fix
 ideas, we shall focus here on a $\LQG$-invariant open XXZ spin
chain~\cite{PasquierSaleur} of length $N = 2 L$, which is essentially the
XXZ spin-$\half$ chain with additional boundary terms
\begin{equation}\label{XXZ_H}
\displaystyle H = \frac{1}{2} \sum_{i=1}^{N-1} \left( \sigma^x_i
\sigma^x_{i+1} + \sigma^y_i \sigma^y_{i+1} + \ffrac{\q + \q^{-1}}{2}
\sigma^z_i \sigma^z_{i+1} \right) + \ffrac{\q - \q^{-1}}{4} \left(
\sigma^z_1 - \sigma^z_{N} \right),
\end{equation}
with Hilbert space $\mathcal{H}_{N}=(\mathbb{C}^2)^{\otimes N}$.
This Hamiltonian can be rewritten (up to a constant term) as $ H = - \sum_{i=1}^{N-1} e_i$
with the densities
\begin{equation}\label{TL-XXZ}
\displaystyle e_i = \ffrac{\q + \q^{-1}}{4} - \frac{1}{2} \left(
\sigma^x_i \sigma^x_{i+1} + \sigma^y_i \sigma^y_{i+1} + \ffrac{\q +
\q^{-1}}{2} \sigma^z_i \sigma^z_{i+1} \right) - \ffrac{\q - \q^{-1}}{4}
\left( \sigma^z_i - \sigma^z_{i+1} \right).
\end{equation}
Straightforward calculations show that this operator satisfy the relations of the so-called Temperley-Lieb (TL) algebra $\TLq{N}$
\begin{subequations} \label{TLdef}
\begin{eqnarray}
\left[ e_i , e_j \right] &=&0 \ (\left|i-j \right| \geq 2 )\\
e_i ^2 &=& \fug e_i\\
e_i e_{i \pm 1} e_i &=& e_i
\end{eqnarray}
\end{subequations}
with
\begin{equation}
\displaystyle{ \fug=\q+\q^{-1}}.
\end{equation}
This algebra is best understood diagrammatically~\cite{M0}. Introducing the notation
$$
\begin{pspicture}(0.,0.)(1.0,1.0)
	\psline[](0.,0.)(0.,0.6)
	\psline[](0.5,0.)(0.5,0.6)
	\rput{0}(-0.7,0.25){$e_i=$}
\end{pspicture}\hdots \ \ \
\begin{pspicture}(0.,0.)(1.0,1.0)
	\psellipticarc{-}(0.25,0.)(0.25,-0.25){180}{0}
	\psellipticarc{-}(0.25,0.6)(0.25,0.25){180}{0}
	\rput{0}(0.,-0.3){$^i$}
	\rput{0}(0.5,-0.3){$^{i+1}$}
\end{pspicture} \hdots \ \ \
\begin{pspicture}(0.,0.)(1.0,1.0)
	\psline[](0.,0.)(0.,0.6)
	\psline[](0.5,0.)(0.5,0.6)
\end{pspicture},
$$
the equations~\eqref{TLdef} can now be interpreted geometrically, the
composition law corresponding to stacking the diagrams of the $e_i$'s
where it is assumed that every closed loop carries a weight $\fug$ -- the fugacity of a loop.
Within this geometrical setup, the algebra $\TLq{N}$ itself 
can be thought of as an algebra of diagrams.
 It is well known~\cite{M0,M1} that when  $\q$ is generic,
{\it i.e.} not a root of unity, the TL representation theory is  semi-simple. For a (half-)integer $0\leq j\leq N/2$, we define a standard module
$\StTL{j}[N]$, which is irreducible for $\q$ generic, as the span of link diagrams with $2j$ through-lines (also called ``strings'') 
which are not allowed to be contracted by the Temperley-Lieb generators. 
The action of the generators on these modules is again interpreted as stacking
the various diagrams. 
The dimension of the standard modules does not depend on $\q$ and can be easily computed
within this geometrical setup
\begin{equation}\label{eqDimStdTL}
\dim(\StTL{j}[N])\equiv \dd_j = \binom{N}{N/2+j} -
\binom{N}{N/2+j+1},\qquad
\text{and we set}\; \dd_j=0\; \text{for}\; 2j>N.
\end{equation}
Note that $j$ must be half integer when $N$ is odd.
For $N=4$ for instance, there are four standard modules 
$
\StTL{2}[4]=
\{ \
\psset{xunit=2mm,yunit=2mm}
\begin{pspicture}(0,0)(3,1)
\psline[linewidth=1.0pt](0,-0.5)(0,1)
\psline[linewidth=1.0pt](1,-0.5)(1,1)
\psline[linewidth=1.0pt](2,-0.5)(2,1)
\psline[linewidth=1.0pt](3,-0.5)(3,1)
\end{pspicture} \
\}
$,
$
\StTL{1}[4]=
\{ \ 
\psset{xunit=2mm,yunit=2mm}
\begin{pspicture}(0,0)(3,1)
\psline[linewidth=1.0pt](0,-0.5)(0,1)
\psellipticarc[linecolor=black,linewidth=1.0pt]{-}(1.5,1.0)(0.5,1.42){180}{360}
\psline[linewidth=1.0pt](3,-0.5)(3,1)
\end{pspicture}
\ ,\, \psset{xunit=2mm,yunit=2mm}
\begin{pspicture}(0,0)(3,1)
 \psellipticarc[linecolor=black,linewidth=1.0pt]{-}(0.5,1.0)(0.5,1.42){180}{360}
\psline[linewidth=1.0pt](2,-0.5)(2,1)
\psline[linewidth=1.0pt](3,-0.5)(3,1)
\end{pspicture}
\ , \, \psset{xunit=2mm,yunit=2mm}
\begin{pspicture}(0,0)(3,1)
\psline[linewidth=1.0pt](0,-0.5)(0,1)
\psline[linewidth=1.0pt](1,-0.5)(1,1)
\psellipticarc[linecolor=black,linewidth=1.0pt]{-}(2.5,1.0)(0.5,1.42){180}{360}
\end{pspicture} \
\}
$, and
$
\StTL{0}[4]=
\{ \
\psset{xunit=2mm,yunit=2mm}
\begin{pspicture}(0,0)(3,1)
 \psellipticarc[linecolor=black,linewidth=1.0pt]{-}(1.5,1.0)(1.5,1.42){180}{360}
 \psellipticarc[linecolor=black,linewidth=1.0pt]{-}(1.5,1.0)(0.5,0.71){180}{360}
\end{pspicture}
,\, \psset{xunit=2mm,yunit=2mm}
\begin{pspicture}(0,0)(3,1)
 \psellipticarc[linecolor=black,linewidth=1.0pt]{-}(0.5,1.0)(0.5,1.42){180}{360}
 \psellipticarc[linecolor=black,linewidth=1.0pt]{-}(2.5,1.0)(0.5,1.42){180}{360}
\end{pspicture} \
\}
$. To be fully explicit, let us give a few examples of the action of the TL generators on $\StTL{1}[4]$:
$e_2 \ \psset{xunit=2mm,yunit=2mm}
\begin{pspicture}(0,0)(3,1)
\psline[linewidth=1.0pt](0,-0.5)(0,1)
\psellipticarc[linecolor=black,linewidth=1.0pt]{-}(1.5,1.0)(0.5,1.42){180}{360}
\psline[linewidth=1.0pt](3,-0.5)(3,1)
\end{pspicture} = \fug \ \psset{xunit=2mm,yunit=2mm}
\begin{pspicture}(0,0)(3,1)
\psline[linewidth=1.0pt](0,-0.5)(0,1)
\psellipticarc[linecolor=black,linewidth=1.0pt]{-}(1.5,1.0)(0.5,1.42){180}{360}
\psline[linewidth=1.0pt](3,-0.5)(3,1)
\end{pspicture}\ $, 
$e_2 \ \psset{xunit=2mm,yunit=2mm}
\begin{pspicture}(0,0)(3,1)
 \psellipticarc[linecolor=black,linewidth=1.0pt]{-}(0.5,1.0)(0.5,1.42){180}{360}
\psline[linewidth=1.0pt](2,-0.5)(2,1)
\psline[linewidth=1.0pt](3,-0.5)(3,1)
\end{pspicture} = \ \psset{xunit=2mm,yunit=2mm}
\begin{pspicture}(0,0)(3,1)
\psline[linewidth=1.0pt](0,-0.5)(0,1)
\psellipticarc[linecolor=black,linewidth=1.0pt]{-}(1.5,1.0)(0.5,1.42){180}{360}
\psline[linewidth=1.0pt](3,-0.5)(3,1)
\end{pspicture} \ $, and $e_3 \ \psset{xunit=2mm,yunit=2mm}
\begin{pspicture}(0,0)(3,1)
 \psellipticarc[linecolor=black,linewidth=1.0pt]{-}(0.5,1.0)(0.5,1.42){180}{360}
\psline[linewidth=1.0pt](2,-0.5)(2,1)
\psline[linewidth=1.0pt](3,-0.5)(3,1)
\end{pspicture} = 0$.

 When $\q$ is generic, the Hilbert space of the Hamiltonian densities~\eqref{TL-XXZ} nicely decomposes
onto the standard modules of $\TLq{N}$
\begin{equation}\label{Hilb-decomp-gen}
\displaystyle \Hilb_{N} |_{\rule{0pt}{7.5pt}%
\TLq{N}} \cong \bigoplus_{j=(N\modd2)/2}^{N/2} (2j+1) \StTL{j}[N],
\end{equation}
where the degeneracies $2j+1$ correspond to the dimension of the
so-called Weyl modules (which are also generically irreducible) over the
centralizer of $\TLq{N}$, which is a finite-dimensional
homomorphic image of the quantum group $\LQG$ called the $\q$-Schur algebra. 
Recall that for  an associative algebra $A$ and its representation space
  $\Hilb$, \textit{the centralizer} of $A$ is the algebra
  $\cent_{A}$ of all commuting operators $[\cent_{A},A]=0$,
  {\it i.e.}, the centralizer is defined as the algebra of intertwiners $\cent_{A}=\Endo_{A}(\Hilb)$.
 The centralizer of a fully reducible representation, like the one in~\eqref{Hilb-decomp-gen}, is obviously  a semi-simple algebra because its action on multiplicities is fully reducible as well. We can thus consider the space $\Hilb_N$ as a semi-simple bi-module over the  pair of commuting algebras denoted  by the exterior\footnote{The exterior product of two algebras is the usual tensor product over  complex numbers. We use the symbol~$\boxtimes$ in order not to overload the paper with the usual notation $\otimes$.} tensor product $\TLq{N}\boxtimes\LQG$
\begin{equation}\label{Hilb-decomp-gen-bimod}
\displaystyle \Hilb_{N} |_{\rule{0pt}{7.5pt}%
\TLq{N}\boxtimes\LQG} \cong \bigoplus_{j=(N\modd2)/2}^{N/2} \StTL{j}[N]\boxtimes\modWeylj{j},
\end{equation}
where the first algebra which is $\TLq{N}$ acts on the left tensorands $\StTL{j}[N]$, while the second algebra which is $\LQG$ acts on the right components which are Weyl modules of dimension $2j+1$ denoted by $\modWeylj{j}$, and these $\LQG$-modules do not depend on $N$.
We finally note that the decomposition~\eqref{Hilb-decomp-gen-bimod} is usually referred to as the quantum version of the classical Schur--Weyl duality between the symmetric group action and its centralizer $U s\ell(2)$.

\medskip

 Things become more intricate when $\q$ is a root of unity, which corresponds
to most of the physically relevant cases. We shall denote $\q=\mathrm{e}^{i\pi /p}$ in this case,
and we will use the following denomination, borrowed from the Potts model terminology, 
for the several physically relevant cases:
 dense polymers ($p=2$), percolation ($p=3$), Ising model ($p=4$), {\it etc}.
In these cases, the algebra $\TLq{N}$ is non-semisimple and the decomposition~\eqref{Hilb-decomp-gen} is no longer true. A full analysis of the structure of the XXZ spin-chain at all (primitive) roots of unity will be given below in Sec.~\ref{sec:bimod}.

\subsection{Temperley-Lieb fusion rules}\label{sec:TL-fusion}

We now describe a procedure allowing to compute fusion rules on the lattice~\cite{ReadSaleur07-1,ReadSaleur07-2}.
Our approach relies on the so-called induction functor which associates with any pair of  modules over the algebras $\TLq{N_1}$ and $\TLq{N_2}$ a module over the bigger algebra $\TLq{N_1+N_2}$. 
 
Let us start by giving a formal definition.
Let $M_1$ and $M_2$ be two modules over $\TLq{N_1}$ and $\TLq{N_2}$ respectively, with the same fugacity $\fug$. Then, the tensor product $M_1\tensor M_2$ is a module over the product $\TLq{N_1}\tensor\TLq{N_2}$ of the two algebras. We note that this product of algebras is naturally a subalgebra in $\TLq{N_1+N_2}$.
\textit{The fusion functor}~$\fus$ on two modules $M_1$ and $M_2$  is then defined as the  module induced from this subalgebra, {\it i.e.} 
\begin{equation}\label{fusfunc-def}
M_1\fus M_2 = \TLq{N_1+N_2}\tensor_{\TLq{N_1}\tensor\TLq{N_2}} M_1\tensor M_2,
\end{equation}
where the balanced product $\tensor_A$ (of right and left modules)  over an algebra $A$ is defined as a quotient of the usual tensor product by the relations $v_1\rightact a\tensor v_2 = v_1\tensor a\leftact v_2$ for all $a\in A$, where the left and right actions of $A$ are denoted by $\leftact$ and $\rightact$, respectively. In other words, we simply allow any element from $A$  to pass through the tensor-product symbol from right to left and {\it vice versa}. In our context, the algebra $A$ is $\TLq{N_1}\tensor\TLq{N_2}$ and we consider $\TLq{N_1+N_2}$ as a bimodule over itself, with the left and right actions given by the multiplication, and in particular it is a right module over the subalgebra $A$.  The space $M_1\fus M_2$ in~\eqref{fusfunc-def} is then a left module over $\TLq{N_1+N_2}$. 

In all that follows, we will consider families of modules that can be defined for any $N$, so we will note $M_1[N_1]$ and $M_2[N_2]$, the modules $M_1$ and $M_2$ being for example the standard modules $\StTL{j}$. 
 We will show in the following (in Sec.~\ref{sec:TL-fusion-gen}) that the fusion module $M_1[N_1]\fus M_2[N_2]$ depends only on $N=N_1+N_2$, up to an isomorphism. For any pair of left modules $M_1$ and $M_2$ over $\TLq{N_1}$ and $\TLq{N_2}$ and for any choice of $N_1$ and $N_2$ such that $N_1+N_2=N$, we shall call \textit{fusion rules} the decomposition of the induced module (into indecomposable direct summands) defined by the fusion functor.

As we shall see shortly, it turns out that the fusion rules are stable when $N$ is growing,
and thus persist in the ``thermodynamic limit'' $N \rightarrow \infty$.
The fusion functor acting on a pair of standard TL modules is of most physical interest, 
as it will turn out to have a deep connection with the operator product expansions of primary fields
in the corresponding conformal field theory. A large part of our paper is devoted 
to the computation of various fusions rules for the Temperley-Lieb algebra at $\q$ root of unity.

Actually, several examples of fusion rules can be easily inferred from very simple calculations.
 We will denote the generators of the algebra $\TLq{N_1}\tensor\TLq{N_2}$ by $e_j$, with $1\leq j\leq N_1-1$ and $N_1+1\leq j\leq N_1+N_2-1$, in accordance with the natural embedding of this product into $\TLq{N_1+N_2}$. 

Let us begin with two almost trivial examples to illustrate the calculation of fusion rules.

\subsubsection{Example I}
The fusion $\StTLn{0}{2}\fus\StTLn{1}{2}$ results in a three-dimensional $\TLq{4}$-module with the basis
\begin{equation}\label{fus:StTL-1}
\StTLn{0}{2}\fus\StTLn{1}{2} = \langle\,\arc\tensor\thl\thl, e_2 \arc\tensor\thl\thl, e_3e_2\arc\tensor\thl\thl\,\rangle
\end{equation}
It is easy to convince oneself that these states are the only ones allowed, because of the relation $e_1e_2\arc\tensor\thl\thl=\fug^{-1}e_1e_2e_1\arc\tensor\thl\thl=\fug^{-1}e_1\arc\tensor\thl\thl=\arc\tensor\thl\thl$, which follows from the relation on $2$ sites $\arc = \fug^{-1} e_1\arc$. So for example, we have $e_1e_3e_2\arc\tensor\thl\thl = 0$.
The module defined in~\eqref{fus:StTL-1} is (isomorphic to) $\StTL{1}[4]$, we thus denote $\StTLn{0}{2}\fus\StTLn{1}{2} = \StTL{1}[4]$. 
A similar calculation shows that $\StTLn{0}{2}$ ``acts'' by $\fus$ on $\StTLn{0}{2}$ also as the identity $\StTLn{0}{2}\fus\StTLn{0}{2} = \StTL{0}[4]$. 

\subsubsection{Example II} Next, we consider the fusion
$\StTLn{0}{4}\fus\StTLn{1}{2}$ in the case $\q=\mathrm{e}^{i \pi/3}$ ($p=3$, percolation), 
where $\StTLn{0}{4}$ has the basis $\{a=\psset{xunit=2mm,yunit=2mm}
\begin{pspicture}(0,0)(3,1)
 \psellipticarc[linecolor=black,linewidth=1.0pt]{-}(0.5,1.0)(0.5,1.42){180}{360}
 \psellipticarc[linecolor=black,linewidth=1.0pt]{-}(2.5,1.0)(0.5,1.42){180}{360}
\end{pspicture},b=\psset{xunit=2mm,yunit=2mm}
\begin{pspicture}(0,0)(3,1)
 \psellipticarc[linecolor=black,linewidth=1.0pt]{-}(0.5,1.0)(0.5,1.42){180}{360}
 \psellipticarc[linecolor=black,linewidth=1.0pt]{-}(2.5,1.0)(0.5,1.42){180}{360}
\end{pspicture}-\begin{pspicture}(0,0)(3,1)
 \psellipticarc[linecolor=black,linewidth=1.0pt]{-}(1.5,1.0)(1.5,1.42){180}{360}
 \psellipticarc[linecolor=black,linewidth=1.0pt]{-}(1.5,1.0)(0.5,0.71){180}{360}
\end{pspicture} \}$, which results in a nine-dimensional $\TLq{6}$-module with the basis
\begin{multline}\label{fus:StTL-2}
\StTLn{0}{4}\fus\StTLn{1}{2} = \langle\, a\tensor\thl\thl,\, b\tensor\thl\thl,\, e_4 a\tensor\thl\thl,\, e_4 b\tensor\thl\thl,\, e_3e_4 b\tensor\thl\thl,\\
 e_5e_4 b\tensor\thl\thl,\, e_5e_4 a\tensor\thl\thl,\, e_5e_3e_4 b\tensor\thl\thl,\, e_4e_5e_3e_4 b\tensor\thl\thl\,\rangle.
\end{multline}
Note that we have the following relations 
\begin{equation*}
e_3e_4 a\tensor\thl\thl = a\tensor\thl\thl,\quad e_2e_3e_4 b\tensor\thl\thl = e_2e_3e_4 (a-e_2 a)\tensor\thl\thl = (a-b)\tensor\thl\thl - e_4(a-b)\tensor\thl\thl,
\end{equation*}
among other simple consequences following from them. It is easy to check that the module defined in~\eqref{fus:StTL-2} is $\StTL{1}[6]$. This shows that $\StTLn{0}{4}$ acts again as the identity for the fusion product, even for $p=3$ where it is an indecomposable module with the following subquotient structure $\IrrTL{0}\to\IrrTL{2}=\langle a\rangle\to \langle b\rangle$ where $\IrrTL{j}$ denotes the irreducible module over $\TLq{N}$ which is in the top of $\StTLn{j}{N}$ by definition, and the arrow represents the action of the TL algebra. The general subquotient structure of the standard modules and other details about the non-generic representation theory of the TL algebra will be given in Sec.~\ref{sec:TL-repnongeneric}.

\medskip
When $\q$ is not a root of unity, it is quite easy to convince oneself that the fusion rules for the TL standard modules 
follow a simple $s\ell(2)$ spin addition rule\footnote{A direct argument is given by considering a filtration of the induced module $\StTLn{j_1}{N_1=2j_1}\fus\StTLn{j_2}{N_2=2j_2}$ by the subspaces indexed by the number $j$ of through-lines which obviously takes integer values from $|j_1-j_2|$ up to $j_1+j_2$. This is demonstrated in the next example. Then, using a semi-simplicity argument we deduce the direct sum decomposition. For other values of $N_1$ and $N_2$, the decomposition can be proven in a similar way.}
\begin{equation}\label{eqFusionGen}
\StTLn{j_1}{N_1}\fus\StTLn{j_2}{N_2} = \bigoplus_{j=|j_1-j_2|}^{j_1+j_2} \StTLn{j}{N_1+N_2},
\end{equation}
for $2j_1 \leq N_1$ and $2j_2 \leq N_2$. This relation to $s\ell(2)$ is not a coincidence
and will actually be crucial to derive general formulas in non-generic situations. 
For now, we just note that~\eqref{eqFusionGen} is just a particular case of the general
relations that we will derive in Sec.~\ref{Sec::fusion}.

\subsection{A simple example of indecomposable fusion}\label{paragraphOPElattice}

We now turn to a more interesting example, but also slightly more involved, where the fusion yields
indecomposable modules that consist of a gluing of two standard modules. These indecomposable modules
are called {\it projective} modules and will be described in details in Sec.~\ref{sec:TL-repnongeneric}.
We will discuss how the Hamiltonian $H=-\sum_{i} e_i$ becomes non-diagonalizable in this case,
a property that arises in the resolution of
a ``catastrophe'' when taking the limit from a generic value of $\q$ or $\fug$ to a critical value.

Let us consider the fusion $\StTLn{1}{2}\fus\StTLn{1}{2}$,
 where $\StTLn{1}{2}$ has the basis $\{ \ \psset{xunit=2mm,yunit=2mm}
\begin{pspicture}(0,0)(1.5,1)
\psline[linecolor=black,linewidth=1.0pt](0,-0.5)(0,1)
\psline[linecolor=black,linewidth=1.0pt](1,-0.5)(1,1)
\end{pspicture}
\}$ with $e_1 \ \psset{xunit=2mm,yunit=2mm}
\begin{pspicture}(0,0)(1.5,1)
\psline[linecolor=black,linewidth=1.0pt](0,-0.5)(0,1)
\psline[linecolor=black,linewidth=1.0pt](1,-0.5)(1,1)
\end{pspicture}=0$.
The induction results in a six-dimensional $\TLq{4}$-module with the basis
\begin{equation}\label{fus:StTL-3}
\StTLn{1}{2}\fus\StTLn{1}{2} = \langle\, l,\, e_2 l,\, e_1e_2 l,\,  e_3e_2 l,\, e_1e_3e_2 l,\, e_2e_1e_3e_2 l\,\rangle,
\end{equation}
with $l=\psset{xunit=2mm,yunit=2mm}
\begin{pspicture}(0,0)(1.5,1)
\psline[linecolor=black,linewidth=1.0pt](0,-0.5)(0,1)
\psline[linecolor=black,linewidth=1.0pt](1,-0.5)(1,1)
\end{pspicture} \otimes
\psset{xunit=2mm,yunit=2mm}
\begin{pspicture}(0,0)(1.5,1)
\psline[linecolor=black,linewidth=1.0pt](0,-0.5)(0,1)
\psline[linecolor=black,linewidth=1.0pt](1,-0.5)(1,1)
\end{pspicture}$.
This module is decomposed for $\q$ generic as
\begin{equation}\label{fus:StTL-3-gen}
\StTLn{1}{2}\fus\StTLn{1}{2} = \StTL{0}[4] \oplus \StTL{1}[4] \oplus \StTL{2}[4],
\end{equation}
where the two-dimensional invariant subspace $\StTL{0}[4]$ is spanned by $e_1e_3e_2 l$ and $e_2e_1e_3e_2 l$ which may be identified with the link states $\psset{xunit=2mm,yunit=2mm}
\begin{pspicture}(0,0)(3,1)
 \psellipticarc[linecolor=black,linewidth=1.0pt]{-}(0.5,1.0)(0.5,1.42){180}{360}
 \psellipticarc[linecolor=black,linewidth=1.0pt]{-}(2.5,1.0)(0.5,1.42){180}{360}
\end{pspicture}$ and $\psset{xunit=2mm,yunit=2mm}
\begin{pspicture}(0,0)(3,1)
 \psellipticarc[linecolor=black,linewidth=1.0pt]{-}(1.5,1.0)(1.5,1.42){180}{360}
 \psellipticarc[linecolor=black,linewidth=1.0pt]{-}(1.5,1.0)(0.5,0.71){180}{360}
\end{pspicture}$, respectively.
Meanwhile, the invariant one-dimensional subspace $\StTL{2}[4]$ is spanned, after solving a simple system of linear equations, by
\begin{equation}\label{fus:StTL-3-inv}
\inv(\fug) = l + \ffrac{1}{\fug^2-2}\Bigl( e_1 e_2l + e_3 e_2 l - \fug e_2 l + \ffrac{1}{\fug^2-1}(e_2e_1e_3e_2 l - \fug e_1e_3e_2 l )\Bigr),
\end{equation}
with $e_j\inv(\fug)=0$, for $j=1,2,3$.
Therefore, three remaining linearly independent states contribute to the three-dimensional irreducible direct summand isomorphic to  $\StTL{1}[4]$ because the algebra is semisimple for generic $\q$.

Note that the fusion states can be identified with link states in the following way:
$
l=
\psset{xunit=2mm,yunit=2mm}
\begin{pspicture}(0,0)(3,1)
\psline[linecolor=red,linewidth=1.0pt](0,-0.5)(0,1)
\psline[linecolor=red,linewidth=1.0pt](1,-0.5)(1,1)
\psline[linecolor=blue,linewidth=1.0pt](2,-0.5)(2,1)
\psline[linecolor=blue,linewidth=1.0pt](3,-0.5)(3,1)
\end{pspicture}
$\,, 
$
e_2 l = 
\psset{xunit=2mm,yunit=2mm}
\begin{pspicture}(0,0)(3,1)
\psline[linecolor=red,linewidth=1.0pt](0,-0.5)(0,1)
\psellipticarc[linecolor=black,linewidth=1.0pt]{-}(1.5,1.0)(0.5,1.42){180}{360}
\psline[linecolor=blue,linewidth=1.0pt](3,-0.5)(3,1)
\end{pspicture}
$\,, 
$ 
e_1e_2 l = \psset{xunit=2mm,yunit=2mm}
\begin{pspicture}(0,0)(3,1)
 \psellipticarc[linecolor=black,linewidth=1.0pt]{-}(0.5,1.0)(0.5,1.42){180}{360}
\psline[linecolor=red,linewidth=1.0pt](2,-0.5)(2,1)
\psline[linecolor=blue,linewidth=1.0pt](3,-0.5)(3,1)
\end{pspicture}
$\,, 
$
 e_3e_2 l = \psset{xunit=2mm,yunit=2mm}
\begin{pspicture}(0,0)(3,1)
\psline[linecolor=red,linewidth=1.0pt](0,-0.5)(0,1)
\psline[linecolor=blue,linewidth=1.0pt](1,-0.5)(1,1)
\psellipticarc[linecolor=black,linewidth=1.0pt]{-}(2.5,1.0)(0.5,1.42){180}{360}
\end{pspicture}
$\,, 
$
 e_1e_3e_2 l = \psset{xunit=2mm,yunit=2mm}
\begin{pspicture}(0,0)(3,1)
 \psellipticarc[linecolor=black,linewidth=1.0pt]{-}(0.5,1.0)(0.5,1.42){180}{360}
 \psellipticarc[linecolor=black,linewidth=1.0pt]{-}(2.5,1.0)(0.5,1.42){180}{360}
\end{pspicture}
$\,, and 
$e_2e_1e_3e_2 l = \psset{xunit=2mm,yunit=2mm}
\begin{pspicture}(0,0)(3,1)
 \psellipticarc[linecolor=black,linewidth=1.0pt]{-}(1.5,1.0)(1.5,1.42){180}{360}
 \psellipticarc[linecolor=black,linewidth=1.0pt]{-}(1.5,1.0)(0.5,0.71){180}{360}
\end{pspicture}
$\,.
Note that we use colors to keep track of the original algebras and their  modules, red through-lines correspond to the left $\StTLn{1}{2}$ in $\StTLn{1}{2}\fus\StTLn{1}{2}$ while the blue ones correspond to the right $\StTLn{1}{2}$. We only keep these colors for convenience but these are not really necessary as we can always split the through lines into two halves (the number of through lines is even in this case) and assign the red color to the leftmost ones while the other lines should be blue.
One can then compute the action of the Temperley-Lieb generators on these states using the usual TL rules with a slight modification: when a Temperley-Lieb
generator acts on two through lines with two different colors, it comes with a weight 1 instead of 0. In other words, one can fuse a red trough-line with a blue one with weight 1. For example, one has $e_2 \ \psset{xunit=2mm,yunit=2mm}
\begin{pspicture}(0,0)(3,1)
\psline[linecolor=red,linewidth=1.0pt](0,-0.5)(0,1)
\psline[linecolor=red,linewidth=1.0pt](1,-0.5)(1,1)
\psline[linecolor=blue,linewidth=1.0pt](2,-0.5)(2,1)
\psline[linecolor=blue,linewidth=1.0pt](3,-0.5)(3,1)
\end{pspicture} =  \psset{xunit=2mm,yunit=2mm}
\begin{pspicture}(0,0)(3,1)
\psline[linecolor=red,linewidth=1.0pt](0,-0.5)(0,1)
\psellipticarc[linecolor=black,linewidth=1.0pt]{-}(1.5,1.0)(0.5,1.42){180}{360}
\psline[linecolor=blue,linewidth=1.0pt](3,-0.5)(3,1)
\end{pspicture}$ while $e_1 \  \psset{xunit=2mm,yunit=2mm}
\begin{pspicture}(0,0)(3,1)
\psline[linecolor=red,linewidth=1.0pt](0,-0.5)(0,1)
\psline[linecolor=red,linewidth=1.0pt](1,-0.5)(1,1)
\psline[linecolor=blue,linewidth=1.0pt](2,-0.5)(2,1)
\psline[linecolor=blue,linewidth=1.0pt](3,-0.5)(3,1)
\end{pspicture} = 0$.
With these rules in hand, the calculations become easy and can be done geometrically, we shall use these notations as they are less cumbersome.

We see that the submodules $\StTL{0}[4]$ and $\StTL{1}[4]$ (or their basis elements) have a well-defined limit $\fug\to1$ ($p=3$, percolation) while the invariant $\inv(\fug)$ spanning $\StTL{2}[4]$ is not defined at the limit -- the state in~\eqref{fus:StTL-3-inv} has a diverging term at $\fug\to1$ or $\fug\to\sqrt{2}$ -- or in other words the system of linear equations 
\begin{equation}\label{fus:StTL-3-system}
e_j\bigl(\psset{xunit=2mm,yunit=2mm}
\begin{pspicture}(0,0)(3,1)
\psline[linecolor=red,linewidth=1.0pt](0,-0.5)(0,1)
\psline[linecolor=red,linewidth=1.0pt](1,-0.5)(1,1)
\psline[linecolor=blue,linewidth=1.0pt](2,-0.5)(2,1)
\psline[linecolor=blue,linewidth=1.0pt](3,-0.5)(3,1)
\end{pspicture}
 +a \
\psset{xunit=2mm,yunit=2mm}
\begin{pspicture}(0,0)(3,1)
\psline[linecolor=red,linewidth=1.0pt](0,-0.5)(0,1)
\psellipticarc[linecolor=black,linewidth=1.0pt]{-}(1.5,1.0)(0.5,1.42){180}{360}
\psline[linecolor=blue,linewidth=1.0pt](3,-0.5)(3,1)
\end{pspicture}
 + b \
\psset{xunit=2mm,yunit=2mm}
\begin{pspicture}(0,0)(3,1)
\psline[linecolor=red,linewidth=1.0pt](0,-0.5)(0,1)
\psline[linecolor=blue,linewidth=1.0pt](1,-0.5)(1,1)
\psellipticarc[linecolor=black,linewidth=1.0pt]{-}(2.5,1.0)(0.5,1.42){180}{360}
\end{pspicture}
 + c \
\psset{xunit=2mm,yunit=2mm}
\begin{pspicture}(0,0)(3,1)
 \psellipticarc[linecolor=black,linewidth=1.0pt]{-}(0.5,1.0)(0.5,1.42){180}{360}
\psline[linecolor=red,linewidth=1.0pt](2,-0.5)(2,1)
\psline[linecolor=blue,linewidth=1.0pt](3,-0.5)(3,1)
\end{pspicture}
 +d \
\psset{xunit=2mm,yunit=2mm}
\begin{pspicture}(0,0)(3,1)
 \psellipticarc[linecolor=black,linewidth=1.0pt]{-}(0.5,1.0)(0.5,1.42){180}{360}
 \psellipticarc[linecolor=black,linewidth=1.0pt]{-}(2.5,1.0)(0.5,1.42){180}{360}
\end{pspicture}
 + e \ 
\psset{xunit=2mm,yunit=2mm}
\begin{pspicture}(0,0)(3,1)
 \psellipticarc[linecolor=black,linewidth=1.0pt]{-}(1.5,1.0)(1.5,1.42){180}{360}
 \psellipticarc[linecolor=black,linewidth=1.0pt]{-}(1.5,1.0)(0.5,0.71){180}{360}
\end{pspicture}
\bigr)=0, \quad \text{with}\; j=1,2,3,
\end{equation}
 has no solutions at $p=3,4$, with the normalization chosen such that the coefficient in front of \,$\psset{xunit=2mm,yunit=2mm}
\begin{pspicture}(0,0)(3,1)
\psline[linecolor=red,linewidth=1.0pt](0,-0.5)(0,1)
\psline[linecolor=red,linewidth=1.0pt](1,-0.5)(1,1)
\psline[linecolor=blue,linewidth=1.0pt](2,-0.5)(2,1)
\psline[linecolor=blue,linewidth=1.0pt](3,-0.5)(3,1)
\end{pspicture}$ equals $1$. On the other hand, we observe that a new invariant appears when $p=3$ in the submodule $\StTL{0}[4] \subset \StTLn{1}{2}\fus\StTLn{1}{2}$, it generates a proper one-dimensional submodule isomorphic to  $\StTL{2}[4]$ spanned by the difference $\psset{xunit=2mm,yunit=2mm}
\begin{pspicture}(0,0)(3,1)
 \psellipticarc[linecolor=black,linewidth=1.0pt]{-}(1.5,1.0)(1.5,1.42){180}{360}
 \psellipticarc[linecolor=black,linewidth=1.0pt]{-}(1.5,1.0)(0.5,0.71){180}{360}
\end{pspicture}-\psset{xunit=2mm,yunit=2mm}
\begin{pspicture}(0,0)(3,1)
 \psellipticarc[linecolor=black,linewidth=1.0pt]{-}(0.5,1.0)(0.5,1.42){180}{360}
 \psellipticarc[linecolor=black,linewidth=1.0pt]{-}(2.5,1.0)(0.5,1.42){180}{360}
\end{pspicture}$. Note that in principle, $\StTL{0}[4]$  might not be a direct summand anymore in the decomposition of $\StTLn{1}{2}\fus\StTLn{1}{2}$ when $p=3$, actually, we will see that this module is glued into a bigger indecomposable module.
 
The divergence observed above indicates that we should prepare our basis before taking the limit  in order to eliminate the ``$\fug~\to~1$ catastrophe''  in~\eqref{fus:StTL-3-inv}. Indeed, the divergence in $\fug\to1$ can be resolved by taking an appropriate combination of $\inv(\fug)$ with a state that has the same eigenvalue (which is $0$ here) with respect to the Hamiltonian $H=-(e_1+e_2+e_3)$ in the limit $\fug=1$. We thus need first to diagonalize $H$ on the submodule $\StTL{0}[4]$ which contains such a state in the limit -- the one spanning the one-dimensional submodule $\StTL{2}[4]$ in $\StTL{0}[4]$ at $\fug=1$. A simple calculation shows that the two eigenvectors are $v_{\pm}=\psset{xunit=2mm,yunit=2mm}
\begin{pspicture}(0,0)(3,1)
 \psellipticarc[linecolor=black,linewidth=1.0pt]{-}(1.5,1.0)(1.5,1.42){180}{360}
 \psellipticarc[linecolor=black,linewidth=1.0pt]{-}(1.5,1.0)(0.5,0.71){180}{360}
\end{pspicture}+a_{\pm}\psset{xunit=2mm,yunit=2mm}
\begin{pspicture}(0,0)(3,1)
 \psellipticarc[linecolor=black,linewidth=1.0pt]{-}(0.5,1.0)(0.5,1.42){180}{360}
 \psellipticarc[linecolor=black,linewidth=1.0pt]{-}(2.5,1.0)(0.5,1.42){180}{360}
\end{pspicture}$ with the eigenvalues $h_{\pm}(\fug) = -\frac{3\fug\pm\sqrt{8+\fug^2}}{2}$ and $a_{\pm}=-h_{\pm}(\fug)-\fug$. The state $v_+$ corresponds to the vacuum (groundstate) with eigenvalue $h_+\to-3$ while the state $v_-$ gives the next-to-vacuum excitation (the ``stress energy tensor'' in the CFT language) with (yet non-shifted) eigenvalue $h_-\to0$ in the limit. Therefore, we should consider the state $v_-$ and combine it with the ill-defined $\inv(\fug)$. Moreover, we note that the limit $v_-\xrightarrow{\fug\to1}\psset{xunit=2mm,yunit=2mm}
\begin{pspicture}(0,0)(3,1)
 \psellipticarc[linecolor=black,linewidth=1.0pt]{-}(1.5,1.0)(1.5,1.42){180}{360}
 \psellipticarc[linecolor=black,linewidth=1.0pt]{-}(1.5,1.0)(0.5,0.71){180}{360}
\end{pspicture}-\psset{xunit=2mm,yunit=2mm}
\begin{pspicture}(0,0)(3,1)
 \psellipticarc[linecolor=black,linewidth=1.0pt]{-}(0.5,1.0)(0.5,1.42){180}{360}
 \psellipticarc[linecolor=black,linewidth=1.0pt]{-}(2.5,1.0)(0.5,1.42){180}{360}
\end{pspicture}$ coincides with the (limit of the) ``excitation'' $(e_2e_1e_3e_2 l - \fug e_1e_3e_2 l )$ in front of the diverging part of $\inv(\fug)$ in~\eqref{fus:StTL-3-inv}. This observation suggests us to resolve the catastrophe by taking the combination
\begin{equation}
t(\fug) = \inv(\fug) - \ffrac{1}{(\fug^2-2)(\fug^2-1)}\bigl(e_2e_1e_3e_2 l + a_- e_1e_3e_2 l\bigr),
\end{equation}
where $a_-=\fug/2-\sqrt{2+(\fug/2)^2}$. Now, the state $t(\fug)$ is well-defined and the divergence is resolved in the state
\begin{align}
t(\fug) & = \psset{xunit=2mm,yunit=2mm}
\begin{pspicture}(0,0)(3,1)
\psline[linecolor=red,linewidth=1.0pt](0,-0.5)(0,1)
\psline[linecolor=red,linewidth=1.0pt](1,-0.5)(1,1)
\psline[linecolor=blue,linewidth=1.0pt](2,-0.5)(2,1)
\psline[linecolor=blue,linewidth=1.0pt](3,-0.5)(3,1)
\end{pspicture}
 + 
 \ffrac{1}{\fug^2-2}\Bigl( \psset{xunit=2mm,yunit=2mm}
\begin{pspicture}(0,0)(3,1)
\psline[linecolor=red,linewidth=1.0pt](0,-0.5)(0,1)
\psline[linecolor=blue,linewidth=1.0pt](1,-0.5)(1,1)
\psellipticarc[linecolor=black,linewidth=1.0pt]{-}(2.5,1.0)(0.5,1.42){180}{360}
\end{pspicture} + \ \begin{pspicture}(0,0)(3,1)
 \psellipticarc[linecolor=black,linewidth=1.0pt]{-}(0.5,1.0)(0.5,1.42){180}{360}
\psline[linecolor=red,linewidth=1.0pt](2,-0.5)(2,1)
\psline[linecolor=blue,linewidth=1.0pt](3,-0.5)(3,1)
\end{pspicture} -\fug \ \psset{xunit=2mm,yunit=2mm}
\begin{pspicture}(0,0)(3,1)
\psline[linecolor=red,linewidth=1.0pt](0,-0.5)(0,1)
\psellipticarc[linecolor=black,linewidth=1.0pt]{-}(1.5,1.0)(0.5,1.42){180}{360}
\psline[linecolor=blue,linewidth=1.0pt](3,-0.5)(3,1)
\end{pspicture}
   + 
\ffrac{h_-}{\fug^2-1} \psset{xunit=2mm,yunit=2mm}
\begin{pspicture}(0,0)(3,1)
 \psellipticarc[linecolor=black,linewidth=1.0pt]{-}(0.5,1.0)(0.5,1.42){180}{360}
 \psellipticarc[linecolor=black,linewidth=1.0pt]{-}(2.5,1.0)(0.5,1.42){180}{360}
\end{pspicture} \Bigr) \notag \\
 & = \psset{xunit=2mm,yunit=2mm}
\begin{pspicture}(0,0)(3,1)
\psline[linecolor=red,linewidth=1.0pt](0,-0.5)(0,1)
\psline[linecolor=red,linewidth=1.0pt](1,-0.5)(1,1)
\psline[linecolor=blue,linewidth=1.0pt](2,-0.5)(2,1)
\psline[linecolor=blue,linewidth=1.0pt](3,-0.5)(3,1)
\end{pspicture}
 + \ffrac{1}{\fug^2-2}\Bigl( \psset{xunit=2mm,yunit=2mm}
\begin{pspicture}(0,0)(3,1)
\psline[linecolor=red,linewidth=1.0pt](0,-0.5)(0,1)
\psline[linecolor=blue,linewidth=1.0pt](1,-0.5)(1,1)
\psellipticarc[linecolor=black,linewidth=1.0pt]{-}(2.5,1.0)(0.5,1.42){180}{360}
\end{pspicture} + \ \begin{pspicture}(0,0)(3,1)
 \psellipticarc[linecolor=black,linewidth=1.0pt]{-}(0.5,1.0)(0.5,1.42){180}{360}
\psline[linecolor=red,linewidth=1.0pt](2,-0.5)(2,1)
\psline[linecolor=blue,linewidth=1.0pt](3,-0.5)(3,1)
\end{pspicture} -\fug \ \psset{xunit=2mm,yunit=2mm}
\begin{pspicture}(0,0)(3,1)
\psline[linecolor=red,linewidth=1.0pt](0,-0.5)(0,1)
\psellipticarc[linecolor=black,linewidth=1.0pt]{-}(1.5,1.0)(0.5,1.42){180}{360}
\psline[linecolor=blue,linewidth=1.0pt](3,-0.5)(3,1)
\end{pspicture}   - \ffrac{4}{3 \fug + \sqrt{8 + \fug^2}}\psset{xunit=2mm,yunit=2mm}\begin{pspicture}(0,0)(3,1)
 \psellipticarc[linecolor=black,linewidth=1.0pt]{-}(0.5,1.0)(0.5,1.42){180}{360}
 \psellipticarc[linecolor=black,linewidth=1.0pt]{-}(2.5,1.0)(0.5,1.42){180}{360}
\end{pspicture} \Bigr)
\end{align}
with the limit 
\begin{equation}\label{fus:StTL-3-log}
t\equiv t(1) = \psset{xunit=2mm,yunit=2mm}
\begin{pspicture}(0,0)(3,1)
\psline[linecolor=red,linewidth=1.0pt](0,-0.5)(0,1)
\psline[linecolor=red,linewidth=1.0pt](1,-0.5)(1,1)
\psline[linecolor=blue,linewidth=1.0pt](2,-0.5)(2,1)
\psline[linecolor=blue,linewidth=1.0pt](3,-0.5)(3,1)
\end{pspicture}
 - \
\begin{pspicture}(0,0)(3,1)
 \psellipticarc[linecolor=black,linewidth=1.0pt]{-}(0.5,1.0)(0.5,1.42){180}{360}
\psline[linecolor=red,linewidth=1.0pt](2,-0.5)(2,1)
\psline[linecolor=blue,linewidth=1.0pt](3,-0.5)(3,1)
\end{pspicture} - \psset{xunit=2mm,yunit=2mm}
\begin{pspicture}(0,0)(3,1)
\psline[linecolor=red,linewidth=1.0pt](0,-0.5)(0,1)
\psline[linecolor=blue,linewidth=1.0pt](1,-0.5)(1,1)
\psellipticarc[linecolor=black,linewidth=1.0pt]{-}(2.5,1.0)(0.5,1.42){180}{360}
\end{pspicture} + \psset{xunit=2mm,yunit=2mm}
\begin{pspicture}(0,0)(3,1)
\psline[linecolor=red,linewidth=1.0pt](0,-0.5)(0,1)
\psellipticarc[linecolor=black,linewidth=1.0pt]{-}(1.5,1.0)(0.5,1.42){180}{360}
\psline[linecolor=blue,linewidth=1.0pt](3,-0.5)(3,1)
\end{pspicture} + \ffrac{2}{3} \ \psset{xunit=2mm,yunit=2mm}
\begin{pspicture}(0,0)(3,1)
 \psellipticarc[linecolor=black,linewidth=1.0pt]{-}(0.5,1.0)(0.5,1.42){180}{360}
 \psellipticarc[linecolor=black,linewidth=1.0pt]{-}(2.5,1.0)(0.5,1.42){180}{360}
\end{pspicture}.
\end{equation}

Borrowing the terminology of LCFT, we say that the state $t$ from~\eqref{fus:StTL-3-log} 
is the ``logarithmic partner'' of the ``stress-energy tensor'' $T=\psset{xunit=2mm,yunit=2mm}\begin{pspicture}(0,0)(3,1)
 \psellipticarc[linecolor=black,linewidth=1.0pt]{-}(0.5,1.0)(0.5,1.42){180}{360}
 \psellipticarc[linecolor=black,linewidth=1.0pt]{-}(2.5,1.0)(0.5,1.42){180}{360}
\end{pspicture}$ \ - \ $\psset{xunit=2mm,yunit=2mm}
\begin{pspicture}(0,0)(3,1)
 \psellipticarc[linecolor=black,linewidth=1.0pt]{-}(1.5,1.0)(1.5,1.42){180}{360}
 \psellipticarc[linecolor=black,linewidth=1.0pt]{-}(1.5,1.0)(0.5,0.71){180}{360}
\end{pspicture}$. Indeed, we find a Jordan cell between these two states\footnote{In this equation, one should of course appropriately shift the Hamiltonian by the Dirac sea contribution in order to have a non-trivial diagonal action corresponding to the  equation $Ht=2t + T$ familiar in LCFT.}
\begin{equation}
H t = \ffrac{2}{3}T.
\end{equation}
We will also say that $T$ is the ``descendant'' of the vacuum state $\vacr = \psset{xunit=2mm,yunit=2mm}
\begin{pspicture}(0,0)(3,1)
 \psellipticarc[linecolor=black,linewidth=1.0pt]{-}(1.5,1.0)(1.5,1.42){180}{360}
 \psellipticarc[linecolor=black,linewidth=1.0pt]{-}(1.5,1.0)(0.5,0.71){180}{360}
\end{pspicture} \ +2 \ \psset{xunit=2mm,yunit=2mm}
\begin{pspicture}(0,0)(3,1)
 \psellipticarc[linecolor=black,linewidth=1.0pt]{-}(0.5,1.0)(0.5,1.42){180}{360}
 \psellipticarc[linecolor=black,linewidth=1.0pt]{-}(2.5,1.0)(0.5,1.42){180}{360}
\end{pspicture}$ as the standard module $\StTL{0}$ has the following indecomposable structure at $\fug=1$:
$\StTL{0} = \vacr \to T$ where we recall that the arrow corresponds to the action of the TL algebra.

We see that the standard modules $\StTL{0}[4]$ and $\StTL{2}[4]$ arising
in the generic fusion rules are ``glued'' together at $\fug=1$
into a bigger indecomposable module with the TL action given by the diagram $t\to\vacr\to T$. The subquotient structure of this module reads  $\IrrTL{2}\to\IrrTL{0}\to\IrrTL{2}$, where each subquotient is one-dimensional and we recall that  $\IrrTL{j}$ denotes the irreducible  top of $\StTLn{j}{N}$. This module is the first example we encounter of the so-called
{\it projective modules} over the TL algebra. We shall denote it $\PrTL{2}[4]$. Finally, the fusion rules at $\fug=1$ reads
\begin{equation}\label{fus:StTL-3-perc}
\StTLn{1}{2}\fus\StTLn{1}{2} =  \StTL{1}[4] \oplus \PrTL{2}[4],\qquad \text{for}\; p=3.
\end{equation}

The case for $p=4$ can be analyzed in a similar way, the only difference being that in that case $\StTL{0}[4]$ is irreducible in the limit and now the standard modules $\StTL{1}[4]$ and $\StTL{2}[4]$ arising
in the generic fusion  are  ``glued'' together at $\fug=\sqrt{2}$
into a bigger indecomposable which has the different subquotient structure $\IrrTL{2}\to\IrrTL{1}\to\IrrTL{2}$  (see more details in Sec.~\ref{subsec:example-one}).

Several remarks are in order:

\begin{itemize}

\item First of all, it is worth pointing out that the fusion~\eqref{fus:StTL-3-perc} has the same filtration by standard modules $\StTL{j}$ as in~\eqref{fus:StTL-3-gen}. This is actually true for any $N_1$ and $N_2$. The only crucial thing that happens when $\q$ is a root of unity is that the terms of the filtration are composed in different ways -- somehow to get a maximal number of bigger indecomposable projective modules in the fusion.

\item We have seen that the Hamiltonian $H = - \sum_i e_i$ at $\fug=1$ has a Jordan cell mapping the top of the 
projective module $\PrTL{2}[4]$ to its socle. This is actually a well-known fact in TL representation theory\footnote{Although to the
best of our knowledge, there is still no proof of this statement.}
that the particular TL combination  $\sum_i e_i$ becomes non-diagonalizable when evaluated onto 
indecomposable projective modules of this form. Note that the percolation problem (that corresponds to $\fug=1$) 
naturally yields Jordan cells and indecomposable modules in the fusion process, this contrasts with specific
geometrical formulations that may or may not show Jordan cells depending on the precise formulation
of the models~\cite{DJS,JordanLoop,VJS}.

\item The reader may have noticed the similarity between our approach to lattice fusion (based
on the work of Read and Saleur~\cite{ReadSaleur07-1,ReadSaleur07-2}) and the integrable boundary conditions used by Pearce-Rasmussen-Zuber
(see {\it e.g.}~\cite{PRZ, RP1, RP2}). We feel that it is necessary to clarify the link between these two approaches
as they actually correspond to the very same thing. It is indeed not hard to see that our previous computation of the fusion/induction,
where we refered to through-lines of two different colors, corresponds exactly to the so-called $(1,s)$ boundary conditions
used in~\cite{PRZ} (in the case we considered, to $(1,2)$ conditions on both sides of a strip). 
However, we believe that the algebraic methods we use provide a much more powerful 
tool as they allow us to obtain exact and rigorous results.
We will show in the following how one can derive many exact fusion rules for simple, standard and projective modules
over the TL algebra using quantum groups results only. 
Finally, we also claim that other boundary conditions $(r,s)$ could be described by appropriate fusion functors associated with boundary extensions of the TL algebra such as the so-called blob algebra~\cite{JSBlob}. This will be discussed elsewhere.

\end{itemize}

 \subsection{Scaling limit and Virasoro fusion rules}\label{sec:scal-lim-intro}

We finish this preliminary section by mentioning the analogous construction in the continuum limit (boundary) CFT. 
 It is not  clear how the continuum limit can be taken in a
mathematically rigorous way for any~$\q$, but roughly speaking, we take the
eigenvectors of $H$ in the spin-chain that have low-energy eigenvalues only, and we
expect that the inner products among these vectors can be made to tend
to some limits. Further, if we focus on long wavelength Fourier
components of the set of local generators $e_j$, we expect their limits to exist, and their commutation
relations to tend to those of the Virasoro generators $L_n$ (it was shown explicitly for free fermion systems, for the Ising chain in~\cite{KooSaleur} and  for the XX model in~\cite{GRS1}), in the sense of strong
convergence of operators in this basis of low-energy
eigenvectors. Then, the modules over the TL algebra restricted to
the low-energy states become, now in the scaling limit\footnote{The two notions -- continuum and scaling limits -- are essentially the same and below we will mostly use the more algebraic one which is the scaling limit, see also a similar discussion in~\cite{GRS1}.}, modules over the Virasoro algebra at appropriate central charge.

 \begin{figure}
\begin{center}
\includegraphics[width=10.0cm]{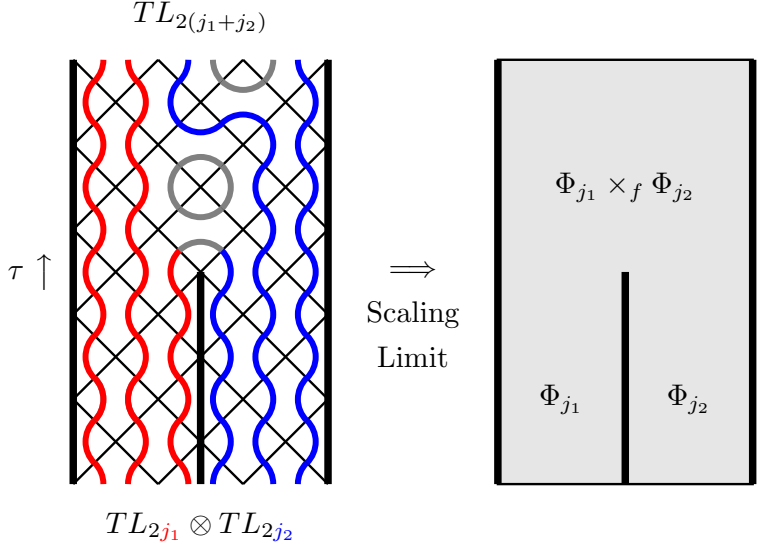}

\end{center}
  \caption{Physical interpretation of the fusion functor of two standard TL modules $\StTL{j_1}[N_1]$ and $\StTL{j_2}[N_2]$ (in the picture, 
  $N_1=2j_1$ and $N_2=2j_2$ so that both standard modules are one-dimensional). The induction procedure can be seen as an event in imaginary
  time $\tau$, consisting in ``joining'' the two standard modules by acting with an additional TL generator. In the scaling limit, we expect
  this construction to coincide with the usual fusion procedure or OPE of boundary fields, here $\Phi_{j_1}$ and $\Phi_{j_2}$, living in the corresponding Virasoro modules.}
  \label{FigFusion}
\end{figure}

It was argued~\cite{ReadSaleur07-1,ReadSaleur07-2}, in a rather heuristic way, that the lattice fusion operation 
we just described corresponds to the fusion of quantum 
fields (or corresponding Virasoro modules) in the continuum limit.
The idea is to consider the induction process as an event in (imaginary) time~$\tau$.
This event corresponds to joining two ``spin chains'', or rather their strips in the continuum limit,
each one carrying a representation of the TL algebra that will eventually become 
representations of the group of conformal transformations in the interior of the strips, after exponentiating the Virasoro algebra
in the continuum limit. Because of the additional TL generator that will join the two spin chains, or any pair of TL modules, one expects a single copy
of the conformal group
to emerge, which contains 
the tensor product of the conformal groups associated with the two initial strips.
Therefore, the induction process over the Temperley-Lieb algebra corresponds in 
the continuum limit to the induction 
over the group of conformal transformations in the corresponding regions. Time evolution is then implemented by acting
with the full transfer matrix containing the additional generator -- or equivalently in the strong anisotropy limit
by the Hamiltonian $H=-\sum_i e_i$. 
 The transfer matrix\footnote{
 We note that this transfer matrix does not generate a family of commuting conserved charges but rather can be thought of as the operator that constructs the two-dimensional Potts model partition function for example~\cite{M0}. Its strong anisotropy limit is indeed described by the quantum 1+1D Hamiltonian $H=-\sum_i e_i$.} $T = \prod_{i \ {\rm odd}} (1+e_i) \prod_{i \ {\rm even}} (1+e_i)$ is then conveniently represented in terms of plaquettes
(face operators)
$$
\begin{pspicture}(0,0)(10,2)
\psset{xunit=8mm,yunit=8mm}

 \rput[Bc](0.5,1){$1+ \ e_i \ $}
 \rput[Bc](1.5,1){$=$}
\psline[linecolor=black](2.25,1)(3,1.75)
\psline[linecolor=black](3,1.75)(3.75,1)
\psline[linecolor=black](2.25,1)(3,0.25)
\psline[linecolor=black](3,0.25)(3.75,1)

 \rput[Bc](4.5,1){$=$}

\psline[linecolor=black](5.25,1)(6,1.75)
\psline[linecolor=black](6,1.75)(6.75,1)
\psline[linecolor=black](5.25,1)(6,0.25)
\psline[linecolor=black](6,0.25)(6.75,1)

\psellipticarc[linecolor=black](5.25,1)(0.53033,0.53033){315}{45}
\psellipticarc[linecolor=black](6.75,1)(0.53033,0.53033){135}{225}

 \rput[Bc](7.5,1){$+  \ $}

\psline[linecolor=black](8.25,1)(9,1.75)
\psline[linecolor=black](9,1.75)(9.75,1)
\psline[linecolor=black](8.25,1)(9,0.25)
\psline[linecolor=black](9,0.25)(9.75,1)

\psellipticarc[linecolor=black](9,0.25)(0.53033,0.53033){45}{135}
\psellipticarc[linecolor=black](9,1.75)(0.53033,0.53033){225}{315}

\end{pspicture}
$$
As $\tau \rightarrow \infty$, we expect to obtain  out-states in the fusion of the two initial representations. 
From a geometrical point of view, the joining in time of the two spin-chains (strips) is conveniently represented
by a ``slit-strip''. A cartoon representation of this argument is given in Fig.~\ref{FigFusion}. We believe that a rigorous algebraic-geometry approach to fusion rules, like the one in~\cite{FT} where fusion is also based on induced modules, should be useful to clarify these heuristic arguments.

Note finally that the slit-strip can be mapped by a Schwarz--Christoffel transformation~\cite{GSchW} onto the upper half plane, where both sides and the slit of the strip are mapped onto the real line.
Then, the incoming and outcoming states correspond to fields localized at points on the boundary of the half plane.
One can then recover the usual interpretation of the fusion as OPE of the boundary fields. When $\q$ is generic, it is easy to check this
process indeed yields the fusion rules expected by the $\LQG$ symmetry. When $\q$ is a root of unity, we still have to identify 
the fields corresponding to the (limits of the) different modules over TL in the fusion (standard, projective, irreducible, {\it etc}). We shall come back to this question later. 
  
To understand  the connection with the Virasoro modules in more details, we describe shortly what happens to the lattice fusion rules when 
one considers the scaling limit $N_1 \to \infty$ and $N_2 \to \infty$. First of all, we will show in Sec.~\ref{Sec::fusion}
that the fusion rules are stable when the scaling limit is taken so one can formally obtain fusion rules when $N$ tends to $\infty$.
The keypoint is then to identify the TL modules considered in the fusion with Virasoro modules in the limit.
As an example, let us discuss how the TL standard modules become  so-called Kac modules over the Virasoro algebra when the
scaling limit is taken~\cite{PasquierSaleur}. For $\q=\mathrm{e}^{i \pi/p}$, we introduce the following formula
for the central charge
\begin{equation}\label{eqCentralCharge}
\displaystyle c_{p-1,p} = 1 - \frac{6}{p (p-1)}.
\end{equation}
The Kac formula at central charge $c_{p-1,p}$ reads
\begin{equation}
\displaystyle h_{r,s} = \frac{ \left(p r - (p-1)s \right)^2 - 1}{4 p (p-1)}.
\end{equation}
Using Bethe ansatz and keeping only low-lying excitations, it can be then shown that the generating function of the spectrum in the module $\StTL{j}[N]$ has the following limit~\cite{PasquierSaleur}
\begin{equation}\label{gen-func-lim}
\displaystyle \lim_{N\to\infty} \sum_{{\rm states} \ i} q^{\ffrac{N}{2\pi v_F} \left(E_i(N) - N e_{\infty} \right)} = q^{-c/24} \dfrac{q^{h_{1,1+2j}}-q^{h_{1,-1-2j}}}{\prod_{n=1}^{\infty} \left( 1 - q^n\right)},
\end{equation}
where $v_F=\frac{\pi \sin \gamma}{\gamma}$ (with $2\cos{\gamma}=\q+\q^{-1}=\fug$) is the Fermi velocity, the central charge $c=c_{p-1,p}$, $E_i(N)$ is the eigenvalue of the $i^{th}$ (counted from the vacuum) eigenstate of $H=-\sum_i e_i$, and $e_{\infty} = \lim_{N \rightarrow \infty} E_0(N)/N$, with $ E_0(N)$ the groundstate energy. The expression on the right-hand side of~\eqref{gen-func-lim} coincides 
with the Virasoro character ${\rm Tr}\, q^{L_0 - c/24}$ of the \textit{Kac module} with conformal weight $h_{1,1+2j}$ defined
as a quotient of the covering Verma module as $\VK_{1,1+2j} \equiv \Verma_{h_{1,1+2j}}/\Verma_{h_{1,-1-2j}}$.
We use here the standard notation $\Verma_{h}$ for the Virasoro Verma module generated from the highest-weight state of weight $h$~\cite{YellowBook}.
We already see at the level of generating functions and characters that we have a deep correspondence between the TL  and Virasoro algebras in the scaling limit,
where the (properly rescaled) Hamiltonian $H$ becomes the $L_0$ generator. As was mentioned above, it is even possible to construct
other Fourier modes by taking appropriate combinations of TL generators on the lattice that will tend (in a sense that can be made rigorous 
in some cases) to other Virasoro generators $L_n$ in the limit~\cite{KooSaleur,GRS1}.
In a similar fashion, one can argue that the  irreducible and projective TL-modules correspond 
respectively to irreducible and {\it staggered} modules~\cite{Rohsiepe, KytolaRidout} on the Virasoro side, see also Sec.~\ref{Sec::ScalingLimit}
 below for more details.

Having in mind that the standard TL modules give a sort of regularization of Kac modules over Virasoro, we can consider the TL fusion functor 
construction given above as a regularization of Operator Product Expansion (OPE) of the quantum fields in the corresponding (boundary) CFT. 
The precise physical meaning in terms of quantum fields of the TL fusion rules will be discussed in Sec.~\ref{Sec::ScalingLimit}.
We will also argue that our calculation of TL fusion by taking limits to degenerate points corresponds precisely
to a similar approach to LCFTs that consists in considering limits of generic (non logarithmic) CFTs~\cite{Gurarie2, Cardylog, KoganNichols, VJS}.
OPEs of primary fields associated with Kac modules in generic cases satisfy the fusion~\cite{BPZ} (see also a recent proof in~\cite{FZ}, Prop.~2.24)
\begin{equation}\label{eqFusionKacGeneric}
\displaystyle \VK_{1,1+2j_1} \fus \VK_{1,1+2j_2} = \bigoplus_{j=|j_1-j_2|}^{j_1+j_2} \VK_{1,1+2j}
\end{equation}
that corresponds exactly to the lattice formula~\eqref{eqFusionGen}. It may happen
that OPE coefficients (either the structure constants or the coefficients in front
of the descendants) become ill-defined when the central charge approaches a ``logarithmic'' value.
One possibility to solve this ``catastrophe'' is to look for other fields in the OPE
that will ``collide'' with the diverging term in order to cancel out the divergence.
This should be reminiscent of our lattice calculation of Sec.~\ref{paragraphOPElattice}.
The general pattern leading to logarithmic singularities in OPEs is actually quite simple. 
Let us assume that the OPE contains a term $\epsilon^{-1} A(\epsilon) z^{-\Delta(\epsilon)}$ appearing
as an amplitude in front of a field, where $A(\epsilon)$ is a regular function in $\epsilon$, with a finite non-zero limit as $\epsilon \to 0$.
This term is then apparently diverging when $\epsilon \to 0$. This ill-defined amplitude will typically be canceled by another OPE contribution 
$\epsilon^{-1} \tilde{A}(\epsilon) z^{-\tilde{\Delta}(\epsilon)}$ with $\Delta(0)=\tilde{\Delta}(0)$ and $A(0)=\tilde{A}(0)$.
Taking the limit will then yield logarithms
\begin{equation}
\dfrac{1}{\epsilon} \left( \dfrac{A(\epsilon)}{z^{\Delta(\epsilon)}} - \dfrac{\tilde{A}(\epsilon)}{z^{\tilde{\Delta}(\epsilon)}} \right) \longrightarrow A(0) \left(\tilde{\Delta}'(0)-\Delta'(0) \right)z^{-\Delta(0)}\log z,
\end{equation}
where $\Delta'(\epsilon)$ is the derivative of the function $\Delta(\epsilon)$.
Detailed examples related to our lattice results will be discussed in Sec.~\ref{Sec::ScalingLimit}.
 
 \medskip
 
 Finally, it is worth emphasizing that the general relation between the lattice, finite-dimensional,
Temperley-Lieb algebra on the one hand, and the infinite-dimensional Virasoro algebra on the other hand, remains at its infancy.
From a mathematical point of view, there is no proof of such relation, and so our results for Virasoro fusion rules 
deduced from the lattice are obviously conjectures, even though we are able to calculate rigorously our TL fusion rules. 
Nevertheless, as discussed in this section, there are strong evidences in the physics literature in favor of
this relation between TL and Virasoro, and we will see in the following that the general fusion rules that we obtain
are in perfect agreement with our few explicit calculations of indecomposable OPEs for Virasoro fields in Sec.~\ref{Sec::ScalingLimit}.
We thus believe that our paper provides yet another step
towards understanding this very intriguing connection.


\section{Bimodule structure of XXZ and representation theory of TL\\ at any root of unity}\label{sec:bimod}

We now come back to the lattice and show how general results can be obtained from algebraic considerations.
 Our aim in this section is to describe the bimodule structure of the XXZ spin-chain over the
pair of mutual centralizers for any root of unity, thus extending the results
of~\cite{ReadSaleur07-2} for $p=2,3$.
We shall then discuss how one can use these results to study TL representation theory at root of unity, 
the applications to the study of fusion rules for a wide family of modules over $\TLq{N}$ will be described in the next section.
The reader not interested in mathematical details may skip this section and jump directly
to the fusion results of the next section.

Two crucial points~\cite{M1} which we use in analyzing root of unity
 cases are 
\begin{enumerate}
\item the XXZ
representation~\eqref{TL-XXZ} of $\TLq{N}$ is faithful and
self-contragredient for any value of the fugacity $\fug$, 
\item  at any root of unity, the centralizer of $\TLq{N}$ is
given by the representation of the quantum group $\LQG$ with divided
powers -- the so-called Lusztig quantum group~\cite{Lusztig}, see its defining
relations in~\cite{[BFGT]}.
\end{enumerate}

Before going into details, we recall some basic facts about the
representation theory of $\LQG$ in the finite-dimensional case
and then use it in the decomposition of the XXZ spin-chain.
We finally describe the representation theory of the TL algebra for any root of unity.


\subsection{Quantum group representation theory} Let us first recall the parametrization $\q=\rme^{i\pi/p}$ in terms of $p\geq2$ integer. We
introduce the following standard notation for $\q$-numbers and $\q$-factorials
\begin{equation*}
[n]=\ffrac{\q^n-\q^{-n}}{\q-\q^{-1}},\qquad [n]!=[1][2]\dots[n].
\end{equation*}
We first recall some results about the Lusztig version of the quantum group $\LQG$, essentially following~\cite{[BFGT]}. We will not give here defining relations, they can be found in~\cite{[BFGT]}, but note only that $\LQG$, as an associative algebra, can be described as a semidirect product $\UresSL2 \rtimes U s\ell(2)$ of the so-called restricted or `small' quantum group  $\UresSL2$ and the universal enveloping  algebra of the usual $s\ell(2)$ Lie algebra. The former is a finite-dimensional algebra with
the  generators $\E$, $\F$, and
$\K^{\pm1}$ and with defining relations which are  the standard relations for the quantum $s\ell(2)$:
\begin{equation}\label{Uq-relations}
  \K\E\K^{-1}=\q^2\E,\quad
  \K\F\K^{-1}=\q^{-2}\F,\quad
  [\E,\F]=\ffrac{\K-\K^{-1}}{\q-\q^{-1}},
\end{equation}
along with the additional relations
\begin{equation}\label{root-rel}
  \E^{p}=\F^{p}=0,\quad \K^{2p}=\one.
\end{equation}
Note that the dimension of $\UresSL2$ is $2p^3$.
In $\LQG$, the $\UresSL2$ part forms a Lie ideal -- commutators of any element from $\LQG$ with an element from $\UresSL2$ belong to $\UresSL2$ -- and the $s\ell(2)$ acts by exterior derivatives on it such that the generators $\E$ and $\F$ form a doublet while the Cartan element $\K$ commutes with all the generators of the $U s\ell(2)$ part.
We denote in what follows the generators of $U s\ell(2)$ by small letters $\e$, $\f$, and $\h$, as opposed to the generators of $\UresSL2$ which are denoted by capital ones. The generators of $U s\ell(2)$ are obtained in the limit from generic value to a root of unity value of $\q$ as the divided powers $\f\sim \lim_{\q\to \rme^{i\pi/p}}\F^p/[p]!$ and $\e\sim \lim_{\q\to \rme^{i\pi/p}}\E^p/[p]!$. They satisfy the usual $s\ell(2)$-relations:
\begin{equation*}\label{sl2-rel}
  [\h,\e]=\e,\qquad[\h,\f]=-\f,\qquad[\e,\f]=2\h.
\end{equation*}

In order to study the decomposition of the XXZ spin-chain Hilbert space, the set of modules over $\LQG$ that we need to consider consists of
irreducible modules $\XX_{s,r}$, for any pair of integers $1\leq s\leq p$ and
$r\geq1$, their projective covers $\PP_{s,r}$ 
and Weyl modules $\modWeyl_{s,r}$, with $1\leq s\leq p-1$ and
$r\geq1$. We describe them below.

 The irreducible module $\XX_{s,r}$ can be constructed as a tensor product 
of $s$-dimensional irreducible $\UresSL2$-module and $r$-dimensional irreducible
$s\ell(2)$-module, {\it i.e.}, its dimension is $sr$ and both subalgebras commute on the irreducible module. This module has highest
weights $(-1)^{r-1}\q^{s-1}$ and $\frac{r-1}{2}$ with respect to $\K$ and $\h$
generators, respectively. The module $\XX_{s,r}$
is spanned by elements $\stprp_{n,m}$, $0\leq n\leq s{-}1$,
$0\leq m\leq r{-}1$, where $\stprp_{0,0}$ is the highest-weight
vector and the left action of the algebra on $\XX_{s,r}$ is
given~by
\begin{align}
  \K \stprp_{n,m} &=
  (-1)^{r-1} \q^{s - 1 - 2n} \stprp_{n,m},\qquad
  &\h\, \stprp_{n,m} &=  \half(r-1-2m)\stprp_{n,m},\label{basis-lusz-irrep-1}\\
  \E \stprp_{n,m} &=
  (-1)^{r-1} [n][s - n]\stprp_{n - 1,m},\qquad
  &\e\, \stprp_{n,m} &=  m(r-m)\stprp_{n,m-1},\label{basis-lusz-irrep-2}\\
  \F \stprp_{n,m} &= \stprp_{n + 1,m},\qquad
  &\f\, \stprp_{n,m} &=  \stprp_{n,m+1},\label{basis-lusz-irrep-3}
\end{align}
where we set $\stprp_{-1,m}=\stprp_{n,-1}
=\stprp_{s,m} =\stprp_{n,r}=0$.

Weyl modules over $\LQG$ are obtained as limits of irreducible modules from generic value  of $\q$ to a root of unity value. It is well known~\cite{ChPr} that as long as the dimension of these modules does not exceed $p$ or is divisible by $p$ they remain irreducible, otherwise, they are no longer irreducible. We will denote Weyl modules by $\modWeyl_{s,r}$ and  set in  what follows $\modWeyl_{s,0}\equiv\XX_{p-s,1}$.
The Weyl modules $\modWeyl_{s,r}$ with $1\leq s\leq p-1$ and $r\geq1$, which are generically irreducible modules of dimension $p(r+1)-s$, have a single proper submodule $\XX_{s,r}$ when $\q$ is a root of unity, {\it i.e.}, we have the diagram for the action of $\LQG$
\begin{equation}\label{diag:Weyl}
\modWeyl_{s,r} :\quad \XX_{p-s,r+1}\longrightarrow\XX_{s,r}\qquad \text{(Weyl module)}.
\end{equation}
Here and in diagrams below, arrows indicate the action of the generators -- states from the source of the arrow are mapped to states in its sink. In the diagram~\eqref{diag:Weyl}, the arrow corresponds to the action of the two generators $\F$ and $\E$ such that $\F$ ({\it resp.} $\E$) maps a state from $\XX_{p-s,r+1}$, with $s\ell(2)$-spin projection $j$ (with $-r/2\leq j\leq r/2$), to a corresponding state in $ \XX_{s,r}$, with $s\ell(2)$-spin projection $j-\half$ ({\it resp.}  $j+\half$). The subalgebra $U s\ell(2)$ acts semisimply as it should in a finite-dimensional module. Explicit formulas for the $\LQG$-action in a given basis can be found in~\cite{[BGT]}\footnote{We note that translation between notations in this paper and the ones in~\cite{[BGT]} is the following: for the Weyl modules $\modWeyl_{s,r}\leftrightarrow\mathsf{N}_{s,r}$, for contragredient Weyl modules $\modWeyls_{s,r}\leftrightarrow\text{\mifody\sf I}_{s,r}$, for irreducible modules  $\XX_{s,r}\leftrightarrow\mathsf{X}_{s,r}$, and for projectives $\PP_{s,r}\leftrightarrow\mathsf{P}_{s,r}$.}.   We also introduce a notation for the module $\modWeyls_{s,r}$ contragredient to the Weyl module $\modWeyl_{s,r}$, {\it i.e.} with the subquotient structure 
\begin{equation}\label{diag:cWeyl}
\mbox{}\qquad\qquad\quad\modWeyls_{s,r} :\quad\XX_{s,r}\longrightarrow \XX_{p-s,r+1}\qquad \text{(contragredient Weyl module)},
\end{equation}
where $\modWeyls_{s,r}$ is the vector space of linear functions $\modWeyl_{s,r}\to\oC$.

We next recall that a projective cover of an irreducible module  is a maximal (finite-dimensional, in our case) indecomposable  module that can be mapped onto the irreducible module, and onto any indecomposable having this irreducible at its top. It was shown in~\cite{[BFGT]} that 
the projective covers  of $\XX_{s,r}$
 which are denoted by $\PP_{s,r}$ have the following subquotient structure:
\begin{equation}\label{schem-proj}
   \xymatrix@C=5pt@R=15pt@M=2pt{%
    &\XX_{s,1}\ar[d]_{}
    \\
    &\XX_{p - s, 2}\ar[d]^{}
    \\
    &\XX_{s,1}
  }\qquad\qquad\qquad
  \xymatrix@=12pt{
    &&\XX_{s,r}
    \ar@/^/[dl]
    \ar@/_/[dr]
    &\\
    &\XX_{p - s, r - 1}\ar@/^/[dr]
    &
    &\XX_{p - s, r + 1}\ar@/_/[dl]
    \\
    &&{\XX_{s,r}}&
  }
\end{equation}
where $r\geq2$ and $1\leq s\leq p-1$. 
Once again, the arrows correspond to the action of the generators $\F$ and $\E$ only, and these generators act in a similar way as in the Weyl modules above. Explicit formulas for the $\LQG$-action in $\PP_{s,r}$ can again be found in~\cite{[BGT]}. We note also that whenever $s=p$ the projective covers $\PP_{p,r}=\XX_{p,r}$ are irreducible.

\subsection{XXZ decomposition over $\LQG$}

We recall then that $\LQG$ is a Hopf algebra and hence has comultiplication formulas $\Delta:\LQG\to\LQG\tensor\LQG$ for all its generators, they can be found  in~\cite{[BFGT]}. Tensor products of any two $\LQG$-modules are thus also modules over $\LQG$. They can in particular be decomposed onto direct summands. The set of irreducible modules and their
projective covers described above  is closed under their tensor product
decompositions. We give formulas borrowed from~\cite{[BGT]} in App.~\ref{appendix:tensor-prod-LQG}. 

As a module over $\LQG$, the XXZ spin-chain $\Hilb_N$ is the $N$-times folded tensor product $\XX_{2,1}^{\tensor N}$ of the fundamental representation of $\LQG$. Therefore, we can use repeatedly the formula~\eqref{fusion-XX} for tensor products of two irreducible modules  in order to decompose $\Hilb_N$ for all $2\leq N\leq p$. At $N=p$, for the first time a projective module appears in $\Hilb_N$. We then  use also the formula~\eqref{fusion-XP}  for tensor products of an irreducible with a projective module. From the decompositions~\eqref{fusion-XX}
and~\eqref{fusion-XP}, we see first of all that the only irreducible direct summands that appear in $\Hilb_N$ are $\XX_{s,1}$ with $N\modd2+1\leq s\leq p-1$ such that $(s+N)\modd 2=1$. All other direct summands in $\Hilb_N$ are projective covers $\PP_{s,r}$ with $r\geq1$ and $1\leq s\leq p$ such that $(p(r+1)-s+N)\modd 2=1$ and $p(r+1)-s\leq N+1$.

 The direct summands $\XX_{s,1}$ and $\PP_{s,r}$ appear in $\Hilb_N$ with some multiplicities. These multiplicities turn out to be the dimensions of irreducible modules over the centralizer of $\LQG$ which is the $\TLq{N}$ algebra. Actually, when $\q$ is a root of unity,  the standard modules $\StTL{j}$ over the TL algebra are no
longer irreducible and have a proper maximal submodule. Taking a quotient by this submodule we obtain an irreducible TL module which we denote as $\IrrTL{j}$. 
Each module $\XX_{s,1}$ or $\PP_{s,r}$ therefore appears in $\Hilb_N$ with a multiplicity which is nothing but the dimension of a corresponding irreducible TL module $\IrrTL{j}$.
To find out which $\IrrTL{j}$ module corresponds to $\XX_{s,1}$ or  $\PP_{s,r}$, we use
the correspondence between
the standard TL modules $\StTL{j}$ and 
the contragredient Weyl  $\LQG$-modules of dimension $2j+1$. This correspondence is well-known in the context of Hecke and $\q$-Schur algebras under the name \textit{$\q$-Schur--Weyl duality}~\cite{D93}, see also a good review in~\cite{MWood}, and it is also called \textit{Ringel} duality in the much wider context of quasi-hereditary algebras and their full tilting modules~\cite{DlRin}. We already mentioned this correspondence after~\eqref{Hilb-decomp-gen} in the semi-simple case, for which a Weyl module is of course isomorphic to its contragredient.
 
For any $\q$, the $\q$-Schur--Weyl duality can be obtained using the usual $\Hom$-functor   from
the category $\catUq$ of left finite-dimensional $\LQG$-modules\footnote{Rigorously, the $\q$-Schur--Weyl  duality is stated between $\TLq{N}$
and the corresponding $\q$-Schur algebra -- the finite-dimensional image of $\LQG$ under the XXZ representation  -- which is represented faithfully on the spin-chain.}
 to the category $\catTL$
of all left $\TLq{N}$-modules. Let $M$ be a $\LQG$-module, \textit{i.e.}, $M\in\mathrm{Ob}(\catUq)$\footnote{We recall that a category $C$ consists of its objects (or modules in our case) which form the set $\mathrm{Ob}(C)$ and composable (homo)morphisms between them, and a functor between two categories $C$ and $D$ gives a map from $\mathrm{Ob}(C)$ to $\mathrm{Ob}(D)$ and relates in a consistent way morphisms between objects in the source category with morphisms in the target one, see more precise statements {\it e.g.} in~\cite{Kassel}.}, then this functor is defined as
\begin{equation}\label{fhom-def}
\fhom: \catUq \to \catTL,\qquad M\mapsto\Hom_{\rule{0pt}{7.5pt}%
\LQG}(\chVv,M),
\end{equation}
where we consider the spin-chain vector space $\chVv$ as a left module over $\LQG$ and as a right
module over $\TLq{N}$ (using the anti-involution $\cdot^{\dagger}: (e_i)^{\dagger}=e_{N-i}$ on $\TLq{N}$ we define the right TL-action `$\rightact$'  by $v\rightact e_i = (e_i)^{\dagger}v$, for any $v\in\chVv$; note also that any anti-involution gives actually an isomorphism of the TL algebra because the defining relations are invariant under an inversion of the order of the products). The left action of $\TLq{N}$ on $\fhom(M)$ is
defined as $a\leftact\phi(v) = \phi(v\rightact a)$ for any $v\in\chVv$
and any $\phi$ from the \textrm{Hom} space. We denote by $\leftact$ and
$\rightact$ left and right actions of $\TLq{N}$, respectively. We recall
also that $\fhom$ maps any morphism $f:
M\to M'$ in the category $\catUq$ to a morphism in $\catTL$ defined by
the composition $f\circ\phi$ for any $\phi\in\Hom_{\rule{0pt}{3.5pt}%
\LQG}(\chVv,M)$. 

Then, the functor $\fhom$ maps the contragredient Weyl module with `spin' $j$ to the standard TL module $\StTL{j}$ with $2j$ through lines, and {\it vice versa}, a Weyl module corresponds to the contragredient standard TL module $\StTL{j}^*$.  Recall that the  (left) $A$-module $V^*$ contragredient to a left $A$-module $V$ is  the vector space of linear functions $V\to\oC$ with the left action of the algebra $A$ given by $a\leftact f(v) = f(a^{\dagger}v)$, where we used an anti-involution $\cdot^{\dagger}$ on $A$ for any $v\in V$ and $f\in V^*$.
The contragredient module is then described by a diagram where all arrows are inverse with respect to the diagram of the initial module. Such an inversion of arrows under the mapping by $\fhom$ is typical in a wider context of the Ringel duality~\cite{MWood}.
Using our notations, see~\eqref{diag:cWeyl} and note that $\dim(\modWeyls_{s,r})=p(r+1)-s$, we thus have the images
\begin{equation}\label{corr-Weyls-StTL}
\begin{split}
\fhom(\XX_{s,1}) &= \StTL{\frac{s-1}{2}},\qquad\qquad\;\; 1\leq s\leq p-1,\\
\fhom(\modWeyls_{s,r})&=\StTL{\frac{p(r+1)-s-1}{2}},\qquad 1\leq s\leq p,  \quad r\geq1.
\end{split}
\end{equation}
The first line tells us that the direct summands $\XX_{s,1}$ appear in $\Hilb_N$ with a multiplicity equal to the dimension of the head of $\StTL{\frac{s-1}{2}}$, {\it i.e.}, the dimension of $\IrrTL{\frac{s-1}{2}}$, while the second line means that the direct summands $\PP_{s,r}$ which cover $\modWeyls_{s,r}$  have as multiplicity the dimension of $\IrrTL{\frac{p(r+1)-s-1}{2}}$.
Therefore,   the decomposition of
$\Hilb_{N}$ over the quantum group can be written as
\begin{multline}\label{decomp-LQG}
\Hilb_{N}|_{\rule{0pt}{7.5pt}%
\LQG} \cong   \bigoplus_{\substack{s=N\modd2+1,\\s+N=1\modd 2}}^{p-1} \dim\bigl(\IrrTL{\frac{s-1}{2}}\bigr) \XX_{s,1}
\oplus \bigoplus_{r=1}^{r_m-1}\bigoplus_{\substack{s=0,\\rp+s+N=1\modd 2}}^{p-1}\dim\bigl(\IrrTL{\frac{rp+s-1}{2}}\bigr) \PP_{p-s,r}\\
\oplus \bigoplus_{\substack{s=0,\\s+s_m=1\modd 2}}^{s_m+1}\dim\bigl(\IrrTL{\frac{r_mp+s-1}{2}}\bigr) \PP_{p-s,r_m}\,,
\end{multline}
where we defined $N=r_m p+s_m$, for $r_m\in\oN$ and $-1\leq s_m\leq p-2$. The second line in the decomposition corresponds to `boundary' terms which will be important to write down the decomposition over the centralizer (see below).
In the following, we shall denote by $\dd^0_{j} = \dim \IrrTL{j}$ the dimensions of
the irreducible modules over $\TLq{N}$.
These numbers are computed below in~\eqref{dimIrr}. 

\subsection{XXZ decomposition over TL}

We are now ready to give a decomposition of $\Hilb_N$ over $\TLq{N}$
for any root of unity and state then a correspondence for a wider family of modules than the family of standard 
modules we considered so far.

\begin{figure} 
{\footnotesize
\centering
\begin{multline*}
 \xymatrix@R=18pt@C=1.2pt@W=2pt@M=2pt{
    &&&\XX_{s,1}\ar[d]_{}\myar[dlll]\myar[drrr]  &&&&\XX_{p-s,2}\myar[dllll]\myar[drrrr] \ar[dl]\ar[dr] &&&&&\XX_{s,3}\myar[dllll]\myar[drrr] \ar[dl]\ar[dr]
    &&&&\dots\myar[dlll]\\    
    \IrrTL{\frac{s-1}{2}}\boxtimes\XX_{s,1}\myar[drrr]&&\IrrTL{\frac{2p-s-1}{2}}\;\boxtimes\!\!&\XX_{p - s, 2}\ar[d]^{}\myar[drrrr]
    &&\IrrTL{\frac{2p+s-1}{2}}\;\boxtimes\!\!&\XX_{s,1}\ar[dr]\myar[dlll]
    &
    &\XX_{s,3}\ar[dl]\myar[drrrr]&&\IrrTL{\frac{4p-s-1}{2}}\;\boxtimes\!\!&\XX_{p-s,2}\ar[dr]\myar[dllll]
    &
    &\XX_{p-s,4}\ar[dl]\myar[drrr]&&\dots\myar[dlll]&
    \\
    &&&\XX_{s,1}&
    &&&\XX_{p-s,2}&&
    &&&\XX_{s,3}&&&&\dots
  }\\\mbox{}\\
 \xymatrix@R=18pt@C=1pt@W=2pt@M=2pt{
    &\dots&\XX_{p-s,r-1}\ar[dr]_{}\myar[drrrr] & &&&&\XX_{s,r}\myar[dllll]\myar[drrrr] \ar[dl]\ar[dr] &&&&&\XX_{p-s,r+1}\myar[dllll]\myar[drrr] \ar[dl]\ar[dr]
    &&&&\dots\myar[dlll]\\    
    &\dots&\IrrTL{\frac{(r-1)p+s-1}{2}}\;\boxtimes\!\!&\XX_{s,r}\ar[dl]^{}\myar[drrrr]
    &&\IrrTL{\frac{(r+1)p-s-1}{2}}\;\boxtimes\!\!&\XX_{p-s,r-1}\ar[dr]\myar[dllll]
    &
    &\XX_{p-s,r+1}\ar[dl]\myar[drrrr]&&\IrrTL{\frac{(r+1)p+s-1}{2}}\;\boxtimes\!\!&\XX_{s,r}\ar[dr]\myar[dllll]
    &
    &\XX_{s,r+2}\ar[dl]\myar[drrr]&&\dots\myar[dlll]&
    \\
    &\dots&\XX_{p-s,r-1}&&
    &&&\XX_{s,r}&&
    &&&\XX_{p-s,r+1}&&&&\dots
  }
\end{multline*}
}
\caption{The bimodules $\bmodP_s$, with $1\leq s\leq p-1$, appear as
  direct summands in a decomposition of the spin-chain $\chVv$ over
  the pair of mutual centralizers $\TLq{N}\boxtimes\LQG$. Solid arrows
  denote the action of $\LQG$ while the dotted ones represent the
  action of $\TLq{N}$. We suppose
  that the tensor product $\boxtimes$ is applied to the full
  indecomposable direct summand (over $\LQG$) on the right -- in
  particular, we have on the top the subquotients
  $\IrrTL{\frac{2p-s-1}{2}}\boxtimes\XX_{s,1}$,
  $\IrrTL{\frac{2p+s-1}{2}}\boxtimes\XX_{p-s,2}$, and so on. It is
  assumed that the segment of the diagram in the second row has an odd value of $r$. We also
  set $\IrrTL{j>L}\equiv 0$, where we recall that $L=N/2$.} 
\label{fig:XXZ-bimod-TL}
\end{figure}
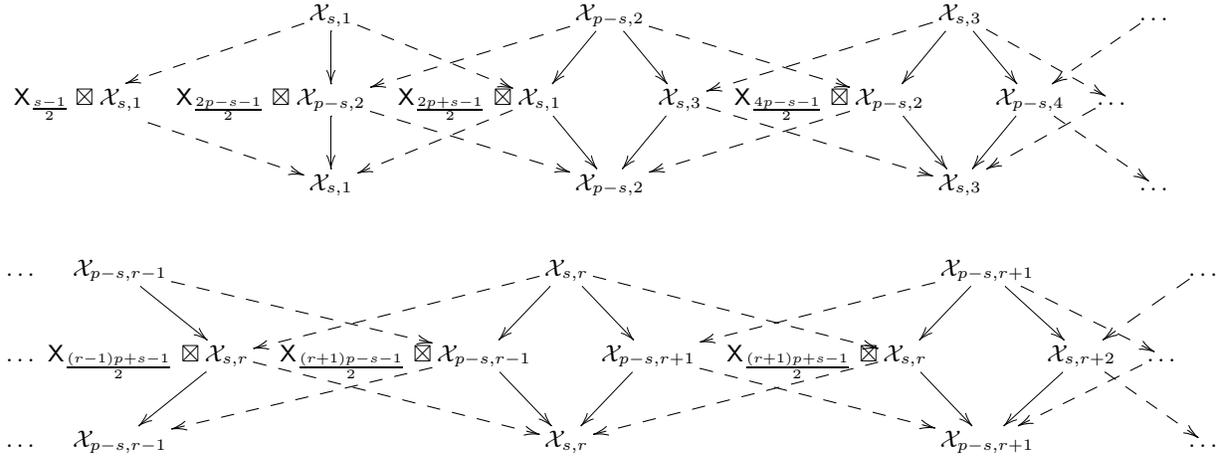

The decomposition over the centralizer of $\LQG$ -- the algebra $\TLq{N}$ -- can be obtained by studying the algebra  
 $\Endq(\Hilb_N)$ of intertwiners which is isomorphic to $\TLq{N}$. We
have an isomorphism for vector spaces of intertwining operators
\begin{equation}\label{homs-LQG}
V_{s,r;s',r'}\equiv\Hom_{\rule{0pt}{7.5pt}%
\LQG}(\PP_{s,r},\PP_{s',r'}) \cong
\begin{cases}
\oC, \qquad s'=p-s,\; r'=r\pm1,\\
\oC^2, \qquad s'=s,\; r'=r,\\
0,\qquad \text{otherwise},
\end{cases}
\end{equation}
which easily follows from the subquotient structure of the
projective modules $\PP_{s,r}$ given in~\eqref{schem-proj}. The
homomorphism spaces introduced in~\eqref{homs-LQG} have the following
basis. A basis element in $V_{s,r;s,r}$ is given by a
homomorphism $\phi$ with an image $\im\phi\cong\PP_{s,r}$ or
$\im\phi\cong\XX_{s,r}$ while a basis element in $V_{s,r;p-s,r\pm1}$ is given by a
homomorphism $\phi_{\pm}$ with an image being the indecomposable
submodule $\im\phi_{\pm}\cong\XX_{s,r}\to\XX_{p-s,r\pm1}$.
These homomorphisms describe intertwining operators mapping between {\it non}-isomorphic indecomposable direct summands in $\Hilb_N$, and they are shown by dotted arrows in the diagram in Fig.~\ref{fig:XXZ-bimod-TL}. There are of course homomorphisms in $\Endq(\Hilb_N)$ that correspond to maps between copies of the direct summands and these maps are not shown explicitly in the diagram. We also note that any homomorphism mapping a module from the top to the bottom of the diagram can be expressed as a composition of  $\phi_{\pm}$, and they are thus also not shown explicitly.

Using the homomorphisms between all direct summands in $\chVv$ just described and taking care of the `boundary' terms in~\eqref{decomp-LQG}, we
obtain finally the decomposition over $\TLq{N}$ as
\begin{multline}\label{decomp-TL}
\Hilb_{N}|_{\rule{0pt}{7.5pt}%
\TLq{N}} \cong \bigoplus_{r=1}^{r_m-1} \bigoplus_{\substack{s=0,\\rp+s+N=1\modd 2}}^{p-1} \dim\bigl(\XX_{p-s,r}\bigr)\PrTL{\frac{rp+s-1}{2}}
\oplus\bigoplus_{\substack{s=0,\\s+s_m=1\modd 2}}^{s_m+1} \dim\bigl(\XX_{p-s,r_m}\bigr)\PrTL{\frac{r_m p+s-1}{2}}\\
\oplus\bigoplus_{\substack{s=1,\\s+s_m=1\modd 2}}^{s_m+1} \dim\bigl(\XX_{s,r_m+1}\bigr)\IrrTL{\frac{r_m p+s-1}{2}}
\oplus\bigoplus_{\substack{s=s_m+2,\\s+s_m=1\modd 2}}^{p-1} \dim\bigl(\XX_{p-s,r_m}\bigr)\IrrTL{\frac{r_m p-s-1}{2}},
\end{multline}
where we recall that $N=r_m p+s_m$, for $r_m\in\oN$ and $-1\leq s_m\leq p-2$.

We note that the direct summands $\PrTL{\frac{rp+s-1}{2}}$ in~\eqref{decomp-TL} should be projective covers of the irreducible $\TLq{N}$-modules $\IrrTL{\frac{rp+s-1}{2}}$, and we show below that it is so, and their multiplicities are $\dim\bigl(\XX_{p-s,r}\bigr)=r(p-s)$.

\subsubsection{The spin-chain as a bimodule}
Using both decompositions~\eqref{decomp-LQG} and~\eqref{decomp-TL} together with the intertwiners described after~\eqref{homs-LQG} above, we obtain that
the spin-chain $\Hilb_N$  is decomposed  as a bimodule over the pair
$\TLq{N}\boxtimes\LQG$ of  mutual centralizers as
\begin{equation}\label{decomp-TL-LQG}
\Hilb_{N}|_{\rule{0pt}{7.5pt}%
\TLq{N}\boxtimes\LQG} \cong  \bigoplus_{\substack{s=N\modd2+1,\\s+N=1\modd
    2}}^{p-1}\!\!\bmodP_s\;\; \oplus
\! \bigoplus_{\substack{r\geq1,\\rp+N=1\modd2}}\IrrTL{\frac{rp-1}{2}}\boxtimes \XX_{p,r},
\end{equation}
where the indecomposable bimodules $\bmodP_s$ are given by the diagram in Fig.~\ref{fig:XXZ-bimod-TL}. The solid lines denote the $\LQG$ action while the dotted lines correspond to all possible homomorphisms between the direct summands over $\LQG$ and thus correspond to the $\TLq{N}$ action because of the double-centralizing property.  Subquotient structure of $\TLq{N}$-modules $ \PrTL{\frac{rp+s-1}{2}}$ can then be deduced from the study of $\Endq(\Hilb_N)$. We will give explicit diagrams in Sec.~\ref{sec:TL-repnongeneric} below.

\subsection{TL representation theory at root of unity}\label{sec:TL-repnongeneric}
We give here some general results about the representation theory
of $\TLq{N}$ at the root of unity case $\q=\rme^{i\pi/p}$,
with integer $p>2$. Most of these results could be deduced directly from~\cite{M1,Westbury,MWood}, 
but we feel that it is both straightforward and instructive
to show how they can be obtained directly from our previous results. 
Our approach relies again on the  $\q$-Schur--Weyl duality with the quantum group $\LQG$ and we also use the language of quasi-hereditary algebras~\cite{DlRin}.  Note that the number $N$ of
sites $N\geq2$ can be either even or odd in the following.

We use 
the representation theory of 
$\LQG$ described above in order to describe subquotient structure for all TL standard
modules and their projective covers. 

\subsubsection{Standard modules}\label{sec:stand-mod}
The bimodule $\chVv$ explicitly described in Fig.~\ref{fig:XXZ-bimod-TL} 
gives `a computational tool' for studying the TL representation theory.
 We begin with the description of the subquotient structure for the standard
modules $\StTL{j}[N]$, with integer $0\leq j\leq N/2$ for even $N$ and
half-integer $1/2\leq j\leq N/2$ when $N$ is odd. Their subquotient structure can be obtained by studying submodules in
the images~\eqref{corr-Weyls-StTL}.
We use the decomposition~\eqref{decomp-TL-LQG} of the bimodule $\chVv$
on direct summands which are given in
Fig.~\ref{fig:XXZ-bimod-TL} to find the subquotient structure of the
$\Hom$-space as a $\TLq{N}$-module. First, we see that there are two
types of homomorphisms from $\chVv$, considered as a left
$\LQG$-module, to $\XX_{s,1}$. A homomorphism of the first type maps the top
of a direct summand $\PP_{s,1}$ onto $\XX_{s,1}$ while the second type of
homomorphism maps a  direct summand $\XX_{s,1}$ onto $\XX_{s,1}$. The
multiplicity of $\PP_{s,1}$ in $\chVv$ is then given by the simple
$\TLq{N}$-subquotient $\IrrTL{\frac{2p-s-1}{2}}$, while the
multiplicity of $\XX_{s,1}$ is $\IrrTL{\frac{s-1}{2}}$. Note also that
a preimage of a first-type homomorphism is mapped by the
$\TLq{N}$-action to a preimage of a second-type
homomorphism, see Fig.~\ref{fig:XXZ-bimod-TL}. Therefore, the arrow
representing the action of $\TLq{N}$ between the two subquotients in
the $\Hom$-space is reversed. We thus obtain that $\fhom(\XX_{s,1})$
is isomorphic to $\StTL{\frac{s-1}{2}}$. All other cases are studied
in a similar way.
We conclude that the standard modules have the
subquotient structure, with $\IrrTL{j}$ denoting an irreducible
subquotient,
\begin{equation}\label{TL-St-def}
\StTL{j}:\quad\IrrTL{j}\longrightarrow\tIrrTL{j+p-1-2(j\modd p)}
\qquad \text{where}\quad\tIrrTL{j'}=
\begin{cases}
\IrrTL{j'},\quad&\text{if}\;j'>j,\\
0,\quad&\text{if}\;j'=j,\\
\IrrTL{j'+p},\quad&\text{if}\;j'<j,
\end{cases}
\end{equation}
and we additionally set $\StTL{j}=0$ for all $j>N/2$ which is crucial
when the number of through lines $2j$ is close to its maximum value
$2j=N$. We note also that the standard modules are irreducible whenever
$j\modd p = \frac{kp-1}{2}$ with $k=0,1$. In particular, all the
standard modules are irreducible for $p=2$ and odd $N$.

The subquotient structure~\eqref{TL-St-def} allows then to compute
the dimension $\dd^0_j$ of the irreducible modules taking standard
alternating sums:
\begin{multline}\label{dimIrr}
\dim(\IrrTL{j}) \equiv \dd^0_j = \sum_{n\geq0}\dd_{j+np} - 
\sum_{n\geq t(j)+1}\dd_{j+np-1-2(j\modd p)}\\
=\sum_{\substack{j'\geq j,\, (j'-j)\modd p =0,\\-2(j\modd
    p)-1}}  (-1)^{(j'-j)\modd p}\,\dd_{j'},\qquad j\modd p \ne
\ffrac{kp-1}{2},\quad k=0,1,
\end{multline}
where we introduce the step function $\stf(j)\equiv\stf$ as
\begin{equation}
\stf=
\begin{cases}\label{stf-def}
1,&\text{for}\quad  j\modd p> \ffrac{p-1}{2},\\
0,&\text{for}\quad  j\modd p< \ffrac{p-1}{2}.
\end{cases}
\end{equation}
Recall that $\dd_{j}$ is given by~\eqref{eqDimStdTL}.

\subsubsection{Projective covers} It was shown first in~\cite{Westbury} that the algebra $\TLq{N}$ is
quasi-hereditary~\cite{[Donk]} for any $\q$ unless $\q=i$ ($p=2$).
Quasi-hereditarity means  in
particular that there is a one-to-one correspondence between (isomorphism classes of) irreducible
modules and the standard modules, and that all projective modules should be filtered by (or composed of) the standard ones. What is important for us is that the structure of projective modules can be easily deduced due to a reciprocity result~\cite{D93} for any quasi-hereditary algebra. Let $[\StTL{j}:\IrrTL{j'}]$ and  $[\PrTL{j}:\StTL{j'}]$ denote the number of appearance of $\IrrTL{j'}$ in a diagram for $\StTL{j}$ and the number of appearance of $\StTL{j'}$ in a diagram for the projective cover $\PrTL{j}$, respectively. Then, the reciprocity result reads
\begin{equation}
[\PrTL{j}:\StTL{j'}] = [\StTL{j'}:\IrrTL{j}].
\end{equation}
In simple words, the projective
modules $\PrTL{j}$ over $\TLq{N}$ 
are composed of the standard modules that have the irreducible module
$\IrrTL{j}$ as a subquotient. 
Using the diagram~\eqref{TL-St-def} and recalling that $\PrTL{j}$  should cover $\StTL{j}$, we thus have,
for any integer or half integer $j$, the structure
\begin{align}
&\PrTL{j}=\StTL{j},&  0\leq j < \ffrac{p}{2},&\label{TL-proj-st}\\
&\PrTL{j}=\StTL{j}\longrightarrow\tStTL{j-1-2(j\modd p)},&  \ffrac{p}{2}\leq j\leq L,&
\end{align}
where we set 
\begin{equation}\label{TL-tSt-def}
\tStTL{j'}=
\begin{cases}
\StTL{j'},\quad&\text{if}\;j'+p>j,\\
0,\quad&\text{if}\;j'+p=j,\\
\StTL{j'+p},\quad&\text{if}\;j'+p<j.
\end{cases}
\end{equation}
So, using~\eqref{TL-St-def} and~\eqref{TL-tSt-def} together with the identity $(j-1-2(j\modd p))\modd p = p-1-j\modd p$, we get, for
 $j \geq p/2$ and $j\modd p\ne \frac{kp-1}{2}$ with $k=0,1$,
\begin{align}\label{prTL-pic-gen-dense-even}     
   \xymatrix@C=5pt@R=15pt@M=2pt{%
    &&\\
    &\PrTL{j}: &\\
    &&
 }      
&  \xymatrix@C=2pt@R=15pt@M=2pt@W=2pt{%
    &&{\IrrTL{j}}\ar[dl]\ar[dr]&\\
    &\IrrTL{j+\stf p-1-2(j\modd p)}\ar[dr]&&\IrrTL{j+(1+\stf)p-1-2(j\modd p)}\ar[dl]\\
    &&\IrrTL{j}&
 } \quad
&   \xymatrix@C=5pt@R=15pt@M=2pt{%
    &&\\
    &&\\
    &&
 }&      
\end{align}
where the right subquotient is absent whenever its subscript $j$ is greater than $L$. Note also that the TL modules in~\eqref{prTL-pic-gen-dense-even}   are self-contragredient, {\it i.e.}, $\PrTL{j}^*$ is isomorphic to $\PrTL{j}$, where a contragredient module was introduced just before~\eqref{corr-Weyls-StTL}.

We note finally that non-self-contragredient projectives $\PrTL{j}$ with
 $j<p/2$ are embedded into projectives $\PrTL{j'}$ with appropriate
 $j'>j$. It turns out that all the self-contragredient projective covers
 introduced above appear as direct summands in the decomposition of the
 spin-chain~\eqref{decomp-TL-LQG}, and
of course their subquotient structure extracted from the bimodule diagram in Fig.~\ref{fig:XXZ-bimod-TL} -- one should follow the dotted lines -- agrees with the one in~\eqref{prTL-pic-gen-dense-even} which uses a more convenient parametrization.

\subsubsection{Zig-zag modules over $\TLq{N}$}\label{sec:zig-zag}
We note that a wider family of indecomposable  TL modules can be also studied.
In particular, we can introduce a one-parameter family of
$\TLq{N}$-modules $\FFmodTL{j}{k}$ of  ``zig-zag'' shape defined (up to an isomorphism) by the following diagrams
\begin{equation}\label{K-mod}
   \xymatrix@C=0pt@R=4pt{
&\\
\boldsymbol{\FFmodTL{j}{k}:}&\\
&
   }
   \xymatrix@C=10pt@R=12pt{
     \IrrTL{\frac{s-1}{2}}\ar@/^/[dr]&
     &\IrrTL{\frac{2p+s-1}{2}}\ar@/_/[dl]
     \ar@/^/[dr]&&\dots\ar@/_/[dl]\ar@/^/[dr]&
     &\IrrTL{\frac{2kp+s-1}{2}}
     \ar@/_/[dl]\\
     &\quad\IrrTL{\frac{2p-s-1}{2}}\quad&&
     \quad\IrrTL{\frac{4p-s-1}{2}}\quad\,&\dots&
     \quad\IrrTL{\frac{2kp-s-1}{2}}\quad&
   }
\end{equation}
where we set $j=(s-1)/2$ and $1\leq s\leq p-1$.
Each of these modules can be obtained as an appropriate quotient of the direct sum $\PrTL{\frac{s-1}{2}}\oplus\PrTL{\frac{2p+s-1}{2}}\oplus\dots\oplus\PrTL{\frac{2kp+s-1}{2}}$
of projective covers. Indeed using the diagram~\eqref{prTL-pic-gen-dense-even}, it is quite easy to deduce by which submodule we should take the quotient. We only note that each pair of neighbor terms in the direct sum of projective covers shares one (isomorphic) irreducible subquotient in the middle level of their diagrams. A non-trivial linear combination of these subquotients should be quotiented out, among other obvious subquotients, in order to get the $\FFmodTL{j}{k}$ module.

We conclude this section by giving the precise correspondence 
between $\LQG$ and $\TLq{N}$ modules provided by the functor $\fhom$~\eqref{fhom-def}. Note that the definition of $\fhom$ involves an explicit decomposition of the spin-chain $\chVv$ as a bimodule. Then using the decomposition~\eqref{decomp-TL-LQG} of the bimodule $\chVv$
on direct summands, which are given in
Fig.~\ref{fig:XXZ-bimod-TL}, we  find the subquotient structure of different
$\Hom$-spaces as $\TLq{N}$-modules. An example of such a computation for $\fhom(\repX_{s,1})$ was already given in Sec.~\ref{sec:stand-mod}.
Recall finally that the TL module $\StTL{j}^*$  contragredient to the standard module $\StTL{j}$  (also called costandard) was defined just before~\eqref{corr-Weyls-StTL}. 

\begin{Prop}\label{prop:funh-img}
The functor $\fhom$ from the category $\catUq$ of left
finite-dimensional modules over (a finite-dimensional image of) $\LQG$ to the category $\catTL$ of
left $\TLq{N}$-modules has the following images
\begin{align}
\fhom(\XX_{s,1}) &= \StTL{\frac{s-1}{2}}, &\fhom(\XX_{s,r}) =
\IrrTL{\frac{(r+1)p-s-1}{2}}, \qquad &1\leq s\leq p, \quad r> 1,\\
\fhom(\modWeyls_{s,r}) &= \StTL{\frac{(r+1)p-s-1}{2}}, &
&1\leq s\leq p-1, \quad r\geq 1,\\
\fhom(\modWeyl_{s,1}) &= \FFmodTL{\frac{s-1}{2}}{1}\,, &\fhom(\modWeyl_{s,r}) = \StTL{\frac{(r+1)p-s-1}{2}}^*,\qquad
&1\leq s\leq p-1, \quad r > 1,\\
\fhom(\PP_{s,r}) &= \PrTL{\frac{(r+1)p-s-1}{2}},&  &1\leq s\leq
p,\quad\quad r\geq 1,
\end{align}
where we imply that values of the $s$-index satisfy $(r+1)p + s +
N=1\modd 2$; and the module $\FFmodTL{j}{1}$ was introduced in~\eqref{K-mod}.
\end{Prop}
We note that this correspondence at the level of projective modules was stated in~\cite{ReadSaleur07-2}  for $p=2$ and $p=3$,
while we have extended the result to a wider family of modules for any root of unity.

\section{Lattice fusion rules}
\label{Sec::fusion}

This section is devoted to a more consistent and conceptual way of computing TL fusion rules than what was done above in the preliminary Sec.~\ref{sec:TL-fusion}. 
This approach uses our results concerning the bimodule structure in the previous section
and recent results~\cite{[BGT]} about tensor-products decompositions for any pair of finite-dimensional $\LQG$-modules. We shall not give a precise
proof of TL fusion formulas  here, but rather give an idea of the proof and illustrate the formulas for TL fusion by direct calculations of the induced modules.  All necessary proofs will be presented in a separate paper~\cite{toapp}.

\subsection{From $\LQG$ tensor products to TL fusion rules}\label{sec:TL-fusion-gen}
We recall first the concept of the fusion functor of Temperley-Lieb modules discussed in  Sec.~\ref{sec:TL-fusion}. The fusion on the TL side is given by the computation of induced modules while, by duality, the fusion on the quantum group side is given by the restriction from the $\q$-Schur algebra on $(N_1+N_2)$ sites to the product of two $\q$-Schur algebras on $N_1$ and $N_2$ sites, respectively. This restriction gives an embedding of the image of $\LQG$ into the product of two images on shorter spin-chains and it corresponds exactly to the comultiplication of the Hopf algebra $\LQG$.

 The application of the induction
and restriction functors on both sides, as it is explained
in~\cite{ReadSaleur07-1}, allows us to state that fusion rules for a pair
of projective/tilting objects coincide for both algebras. In order to compute fusion functors for a wider family of TL modules, not only projectives, {\it e.g.},
for a pair of standard or irreducible modules, we need to establish an equivalence between the category   $\catUq$ of modules over the  $\q$-Schur algebra and a (sub)category of TL modules. It can be shown~\cite{toapp} that such an equivalence indeed exists between  $\catUq$  and $\fhom(\catUq)$ for any $\q$. We remind the reader that the $\Hom$-functor $\fhom$ was introduced in~\eqref{fhom-def}.  Note that the category $\fhom(\catUq)$ coincides with the category $\catTL$  of all left $\TLq{N}$ modules  for $\q$ generic and only  for $p=2$ if $\q$ is a root of unity.  For integer $p>2$, the subcategory $\fhom(\catUq)$ in particular does not contain simple TL modules such as $\IrrTL{j}$, for $0\leq j\leq \frac{p-1}{2}$.

The equivalence between the category $\catUq$ and the subcategory $\fhom(\catUq)$ of $\catTL$ means that there exists a functor from $\fhom(\catUq)$ to
$\catUq$ which composed with $\fhom$ is isomorphic to the
identity functor.  
 This functor
 is the adjoint of the functor $\fhom$. We denote this functor as $\funt$ and it is given by
\begin{equation}
\funt: \catTL \supset \fhom(\catUq) \to \catUq,\qquad M\mapsto \chVv\tensor_{\TLq{N}} M
\end{equation}
(of course, the functor $\funt$ is well defined on $\catTL$ and not only on $\fhom(\catUq)$).
The image of $\funt$ is obviously in the category of $\LQG$-modules because the quantum
group acts on the left side of the bimodule $\chVv$, and we take the
balanced product  over $\TLq{N}$ of the right module $\chVv$ and the left module $M$ (see  the definitions of the right TL action on $\chVv$ just after~\eqref{fhom-def} and of the balanced product in Sec.~\ref{sec:TL-fusion}). 

We note that the functors $\fhom$ and $\funt$ as well as
the categories $\catUq$ and $\catTL$ actually depend on the value of $\q$ and on the number of sites $N$,
but we do not specify these numbers in the notations of functors and
categories for brevity. For a fixed value of $N$ and for all $\q$, except $\q=i$, the number of simple objects in $\catTL$ is the same (at $\q=i$ the number  is less by one, see {\it e.g.}~\cite{ReadSaleur07-2}). The crucial difference comes from morphisms. The category $\catTL$ (and $\catUq$ as well) has different sets of morphisms  for different roots  of unity.

The functor $\funt$ restricted to the subcategory $\fhom(\catUq)$
has an essential advantage that it turns out to be a \textit{tensor} functor\footnote{Note that we do not actually have a tensor-category structure on $\catTL$ in the usual sense, see {\it e.g.}~\cite{Kassel},  and hence we also do not have rigorously a usual tensor functor that respects the tensor-category structures but rather generalizations of these notions to a situation where the ``tensor-product'' $\fus$ maps to a category bigger than the two initial ones.}
which respects fusions on both sides. Indeed, let $M_1\tensor M_2$ be a module over the product $\TLq{N_1}\tensor\TLq{N_2}$ of the two TL algebras. 
 We can then demonstrate the connection between TL fusion and its quantum-group counterpart, which is the restriction operation, by the diagram
\begin{equation}\label{fusion-diag}
   \xymatrix@C=50pt@R=35pt@M=6pt@W=6pt{%
   M_1\tensor M_2\ar[d]^{\funt\tensor\funt} \ar[r]^{\Ind}& M_1\fus M_2\ar[d]_{\funt} \\
   \funt(M_1)\tensor \funt(M_2)\ar@/^2ex/@{-->}[u]^{\fhom\tensor\fhom}\ar[r]^{\Res}
   &\funt(M_1\fus M_2)\ar@/_2ex/@{-->}[u]_{\fhom}
 }      
\end{equation}
where the top horizontal arrow corresponds to the TL fusion functor denoted by $\Ind$ which sends $M_1\tensor M_2$ to its fusion module $\TLq{N_1+N_2}\tensor_{\TLq{N_1}\tensor\TLq{N_2}} M_1\tensor M_2$ following the definition~\eqref{fusfunc-def} while the bottom horizontal arrow corresponds to the tensor-product decomposition of $\LQG$-modules which is the restriction for corresponding finite-dimensional quotients of $\LQG$, which are faithfully represented on the spin-chains with $N_1$, $N_2$, and $N_1+N_2$ sites. We recall that as the product of two TL algebras is naturally a proper subalgebra in $\TLq{N_1+N_2}$, then by duality the centralizer of the latter is a subalgebra in the product of the centralizers of the two TL algebras. Then, the composition $\funt\circ\Ind$ gives  a $\LQG$-module 
\begin{multline}\label{fusion-corr}
\funt\circ\Ind(M_1\tensor M_2) = \chVv\tensor_{\TL{N}}\Bigl(\TL{N}\tensor_{\TL{N_1}\tensor\TL{N_2}} M_1\tensor M_2\Bigr)\\
\cong\Bigl(\chVv\tensor_{\TL{N}}\TL{N}\Bigr)\tensor_{\TL{N_1}\tensor\TL{N_2}} M_1\tensor M_2 
\cong \Bigl(\Hilb_{N_1}\tensor\Hilb_{N_2}\Bigr)\tensor_{\TL{N_1}\tensor\TL{N_2}} M_1\tensor M_2  
\end{multline}
where we simplified our notations and set $N=N_1+N_2$ and $\TL{N}=\TLq{N}$. The first equality holds just by definition while we use associativity of the (balanced) tensor product in establishing the isomorphism in the second line, the first `$\cong$'. For the second `$\cong$', we use the usual isomorphism of bimodules $\chVv\tensor_{\TL{N}}\TL{N}\cong\chVv: v\tensor a\mapsto v \rightact a=v'$ (any element of TL can be passed through the balanced tensor product over TL itself). We remind the reader that $\TL{N}$ is considered here as a left $\TL{N}$-module and right $\TL{N_1}\tensor\TL{N_2}$-module, so $\chVv$ on the right-hand side is also a right $\TL{N_1}\tensor\TL{N_2}$-module and is isomorphic  to $\Hilb_{N_1}\tensor\Hilb_{N_2}$. This finally gives~\eqref{fusion-corr}. It is then easy to see that the final result~\eqref{fusion-corr} is \textit{isomorphic} to the image of another composition $\Res\circ\bigl(\funt\tensor\funt\bigr)$ which uses `fusion' or comultiplication 
 on the quantum-group side. This shows that the diagram~\eqref{fusion-corr} composed of solid arrows is commutative, up to an isomorphism.

For all $\q$, the composition $\fhom\circ\funt$ restricted to $\fhom(\catUq)\subset\catTL$ is the identity map, up to an isomorphism of course, and it gives an equivalence -- a one-to-one invertible consistent map (a functor) between the categories
 -- which is shown by a second pair of (dotted) arrows in~\eqref{fusion-diag}. The equivalence in the generic case can be also seen by a direct calculation of $\funt$ (the computation of images of $\fhom$ is trivial). Indeed, one can write $\funt(\StTL{j}) = \oplus_{j'}\modWeyl_{j'}\boxtimes\StTL{j'}\otimes_{\TL{\q}}\StTL{j}=\modWeyl_{j}$ where we used the generic decomposition~\eqref{Hilb-decomp-gen-bimod} and the simple equalities $\StTL{j'}\otimes_{\TL{\q}}\StTL{j} = \StTL{j'}\otimes_{\TL{\q}}\mathrm{id}_j\StTL{j} = \StTL{j'}\mathrm{id}_j\otimes_{\TL{\q}}\StTL{j} = \delta_{j,j'}\oC$. We used $\mathrm{id}_j$ which is the central idempotent corresponding to the (matrix) subalgebra $\StTL{j}\boxtimes\StTL{j}^*$ in the TL algebra. For the root of unity cases, the computation of $\funt$ is much more involved and all the necessary details will be given in~\cite{toapp}.
 
Having the ``fusion-correspondence'' diagram~\eqref{fusion-corr} at hand, we can first of all easily recover the generic TL fusion~\eqref{eqFusionGen} because in this case $\fhom(\catUq)=\catTL$. Therefore, {\sl any} fusion on the TL side can be computed by quantum-group comultiplication:
\begin{equation*}
\StTL{j_1}\tensor\StTL{j_2} \xrightarrow{\;\funt\otimes\funt\;} \modWeyl_{j_1}\otimes\modWeyl_{j_2}=\oplus_j \modWeyl_{j}\xrightarrow{\;\fhom\;} \oplus_j \StTL{j}
 \end{equation*}
where the right hand side is isomorphic to the fusion module $\StTL{j_1}\fus\StTL{j_2}$ following our commutative  diagram~\eqref{fusion-corr}.
In non-generic cases, this picture still holds but not for the full TL category $\catTL$ but for its subcategory $\fhom(\catUq)\subset\catTL$. 
Recall that the image of $\fhom$ on objects in $\catUq$ is given in~\eqref{corr-Weyls-StTL} and in Prop.~\ref{prop:funh-img}. We also note that $\funt(\StTL{\frac{s-1}{2}})\cong\repX_{s,1}$, for $1\leq s\leq p$, which is a direct analogue of the generic results.
Then using the decomposition~\eqref{fusion-XX} of the tensor products of the (irreducible) Weyl modules at $r_1=r_2=1$, the TL fusion rules for the ``first'' standard modules $\StTL{j}$, with $0\leq j\leq (p-1)/2$, are
 \begin{equation}
 \StTL{\frac{s_1-1}{2}} \fus\StTL{\frac{s_2-1}{2}}=
\bigoplus_{\substack{s=|s_1-s_2|+1\\\step=2}}^{p - |p-s_1 - s_2| - 1}\!\!\!\StTL{\frac{s-1}{2}}\oplus
\!\!\! \bigoplus_{\substack{s=\gamma_2\\\step=2}}^{s_1 + s_2 - p - 1}\!\!\!\!\!\!\PrTL{\frac{p+s-1}{2}}.
 \end{equation}
We will simplify our notations setting $M_1\fus M_2$ for $M_1[N_1]\fus M_2[N_2]$ and $M$ for $M[N_1+N_2]$, for any $M$ that appears in the fusion of $M_1$ and $M_2$. Such a simplification is motivated by the connection with the quantum group where fusion rules do not depend on the choice of $N_1$ and $N_2$. In particular, we see from the explicit decompositions that the fusion module $M_1[N_1]\fus M_2[N_2]$ itself depends only on $N=N_1+N_2$, up to an isomorphism. This is true at least for any pair of TL modules from $\fhom(\catUq)$ and can be easily proven with the use of the commutative diagram~\eqref{fusion-diag} and the fact that $\funt$ is the ``inverse'' to the $\fhom$ functor on $\fhom(\catUq)$.

Using the results about the tensor products of all
finite-dimensional $\LQG$-modules~\cite{[BGT]} reviewed in Thm.~\ref{thm:tens-prod-intro} 
and the  bimodule structure for $\Hilb_N$ together with $\fhom$-images computed in Prop.~\ref{prop:funh-img}, it is actually possible to prove~\cite{toapp} rigorously fusion
rules for most of the TL-modules following the lines explained just above.
General formulas for the TL fusion functor  are a bit cumbersome so
we introduce the following notations
\begin{equation}\label{gamma-def}
\gamma_1=(s_1+s_2+1)\!\!\!\!\mod2,\qquad \gamma_2=(s_1+s_2+p+1)\!\!\!\!\mod2,
\end{equation}
\begin{gather}
\mathop{\bigoplus{\kern-3pt}'}\limits_{r=a}^{b}f(r) = 
\bigoplus_{r=a}^{b}(1 - \ffrac{1}{2}\delta_{r,a} -
\ffrac{1}{2}\delta_{r,b})f(r),\label{not-1}\\
\mathop{\bigoplus{\kern-3pt}''}\limits_{r=a}^{b}f(r) =
\bigoplus_{r=a}^{b} \bigl(1 - \ffrac{3}{4}\delta_{r,a} -
\ffrac{1}{4}\delta_{r,a+2}(1+\delta_{a,-1}) - \ffrac{1}{4}\delta_{r,b-2} -
\ffrac{3}{4}\delta_{r,b}\bigr)f(r),\label{not-2}
\end{gather}
along with the convention that $\oplus_a^b f \equiv 0$ if $b<a$, and with the step-function
\begin{equation}\label{sg-def}
\sg(r)=\begin{cases}\phantom{-}1,\quad r>0,\\ 
      \phantom{-}0,\quad r=0,\\-1,\quad r<0.\end{cases}
\end{equation}

In this paper, we shall concentrate on the fusion of standard and projective TL modules, that we collect in the 
following proposition. 
 \begin{prop}\label{prop:fus-StTL-root}
\mbox{}\\
 $\bullet$\;For $1\leq s_1,s_2\leq p$
and $r_1,r_2\geq0$, the fusion functor on two standard modules over $\TLq{N}$ is
\begin{multline}\label{fusion-StTL}
\StTL{\frac{r_1 p + s_1-1}{2}} \fus\StTL{\frac{r_2 p+s_2-1}{2}}=
\bigoplus_{\substack{s=|s_1-s_2|+1\\\step=2}}^{p - |p-s_1 - s_2| - 1}\!\!\!\StTL{\frac{(r_1+r_2)p+s-1}{2}}\oplus
\bigoplus_{\substack{r=|r_1-r_2|+1\\\step=2}}^{r_1+r_2-1} \bigoplus_{\substack{s=\gamma_2\\\step=2}}^{p-|s_1-s_2|-1}\!\!\!\PrTL{\frac{r p+s-1}{2}}\\
\oplus\!\!\! \bigoplus_{\substack{s=\gamma_2\\\step=2}}^{s_1 + s_2 - p - 1}\!\!\!\!\!\!\PrTL{\frac{(r_1+r_2+1) p+s-1}{2}}
\oplus\bigoplus_{\substack{r=|r_1-r_2+\sg(s_1-s_2)|+1\\\step=2}}^{r_1+r_2}
\bigoplus_{\substack{s=\gamma_1\\\step=2}}^{|s_1-s_2|-1}\!\!\!\PrTL{\frac{r p+s-1}{2}},
\end{multline}
where we set $\StTL{j}$ and $\PrTL{j}$ to zero whenever $j>N/2$.\\
$\bullet$\; For $1\leq s_1,s_2\leq p-1$,  $r_1,r_2\geq1$, the fusion of two projective modules is
\begin{multline}\label{fusion-PrTL-PrTL}
\PrTL{\frac{r_1p+s_1-1}{2}}\fus \PrTL{\frac{r_2p+s_2-1}{2}}
=4\!\!\!\mathop{\bigoplus{\kern-3pt}''}\limits_{\substack{r=|r_1-r_2|-1\\\step=2}}^{r_1+r_2+1}
\bigoplus_{\substack{s=2p-s_1 - s_2 + 1\\\step=2}}^{p-\gamma_2}
\!\!\!\!\!\!\PrTL{\frac{(r+1)p-s-1}{2}}\\
\oplus 2 \bigoplus_{\substack{r=|r_1-r_2|+1\\\step=2}}^{r_1+r_2-1} 
\Biggl(\bigoplus_{\substack{s=|s_1-s_2|+1\\\step=2}}^{\substack{
\min(s_1 + s_2 - 1,\\ 2p - s_1 - s_2 - 1)}}\PrTL{\frac{(r+1)p-s-1}{2}}
\oplus2\bigoplus_{\substack{s=s_1 + s_2 + 1\\\step=2}}^{p-\gamma_2}
\PrTL{\frac{(r+1)p-s-1}{2}}\Biggr)\\
\oplus2\mathop{\bigoplus{\kern-3pt}'}\limits_{\substack{r=|r_1-r_2|\\
\step=2}}^{r_1+r_2}\Biggl(
\bigoplus_{\substack{s=|p-s_1-s_2|+1\\\step=2}}^{\substack{
\min(p-s_1 + s_2 - 1,\\ p + s_1 - s_2 - 1)}}\!\!\!\!\!\!\PrTL{\frac{(r+1)p-s-1}{2}}
\oplus2\!\!\!\!\!\!\bigoplus_{\substack{s=\min(p-s_1 + s_2 + 1,\\ p + s_1 - s_2 + 1)}}^
{p-\gamma_1}\!\!\!\!\!\!\PrTL{\frac{(r+1)p-s-1}{2}}\Biggl),
\end{multline}
where we set  $\PrTL{\frac{s-1}{2}}=0$, for $1\leq s\leq
p-1$, and $\PrTL{j<0}=0$.\\
$\bullet$\; For $1\leq s_1\leq p$, $1\leq s_2\leq p-1$ and $r\geq1$, the fusion of
a ``Kac-table'' standard module and  a projective module is
\begin{multline}\label{fusion-StTL-PrTL}
\StTL{\frac{s_1-1}{2}}\fus\PrTL{\frac{(r+1)p-s_2-1}{2}} = 
\bigoplus_{\substack{s=|s_1-s_2|+1\\\step=2}}^{\substack{
\min(s_1 + s_2 - 1,\\ 2p - s_1 - s_2 - 1)}}\!\PrTL{\frac{(r+1)p-s-1}{2}}
\;\oplus\;2\!\bigoplus_{\substack{s=2p-s_1-s_2+1\\\step=2}}^{p-\gamma_2}
\!\!\!\PrTL{\frac{(r+1)p-s-1}{2}}\\
\oplus\bigoplus_{\substack{s=p-s_1+s_2+1\\\step=2}}^{p-\gamma_1}
\!\!\bigl(\PrTL{\frac{rp-s-1}{2}}\oplus\PrTL{\frac{(r+2)p-s-1}{2}}\bigr).
\end{multline}
Finally, the fusion of a standard module $\StTL{j}$, for
$j>\frac{p-1}{2}$, with a projective module is
\begin{multline*}
\StTL{\frac{r_1p+s_1-1}{2}}\fus\PrTL{\frac{r_2p+s_2-1}{2}} = 
\bigoplus_{\substack{r=|r_1-r_2|+1\\\step=2}}^{r_1+r_2-1}\Biggl(
\bigoplus_{\substack{s=|s_1-s_2|+1\\\step=2}}^{\substack{
\min(s_1 + s_2 - 1,\\ 2p - s_1 - s_2 - 1)}}\!\!\!\!\!\!\PrTL{\frac{(r+1)p-s-1}{2}}
\oplus2\!\!\!\!\!\!\bigoplus_{\substack{s=s_1+s_2+1\\\step=2}}^{p-\gamma_2}
\!\!\!\!\!\!\PrTL{\frac{(r+1)p-s-1}{2}}\Biggr)\\
\oplus\bigoplus_{\substack{r=|r_1-r_2+1|+1\\\step=2}}^{r_1+r_2}\Biggl(
\bigoplus_{\substack{s=|p-s_1-s_2|+1\\\step=2}}^{\substack{
\min(p-s_1 + s_2 - 1,\\ p + s_1 - s_2 - 1)}}\!\!\!\!\!\!\PrTL{\frac{(r+1)p-s-1}{2}}
\oplus2\!\!\!\!\!\!\bigoplus_{\substack{s=p-s_1+s_2+1\\\step=2}}^{p-\gamma_2}
\!\!\!\!\!\!\PrTL{\frac{(r+1)p-s-1}{2}}\Biggr)\\
\oplus2\mathop{\bigoplus{\kern-3pt}'}\limits_{\substack{r=|r_1-r_2|\\\step=2}}^{r_1+r_2}
\bigoplus_{\substack{s=p+s_1-s_2+1\\\step=2}}^{p-\gamma_1}
\!\!\!\!\!\!\PrTL{\frac{(r+1)p-s-1}{2}}
\oplus2\mathop{\bigoplus{\kern-3pt}'}\limits_{\substack{r=|r_1-r_2+1|\\\step=2}}^{r_1+r_2+1}
\bigoplus_{\substack{s=2p-s_1-s_2+1\\\step=2}}^{p-\gamma_1}
\!\!\!\!\!\!\PrTL{\frac{(r+1)p-s-1}{2}},
\end{multline*}
where  $1\leq s_1,s_2\leq p-1$
and $r_1,r_2\geq1$.
 \end{prop}
It is also possible to derive similar fusion rules for (some of the) irreducible modules, we refer the interested reader to  
App.~\ref{appendix-IrrFusion} for more details.

We also note that the fusion of the ``zig-zag'' TL modules $\FFmodTL{j}{k}$  (introduced above in Sec.~\ref{sec:zig-zag}) with themselves or with standard and irreducible modules can also be computed applying the Hom functor $\fhom$. This is quite straightforward having the explicit bimodules at hand, and using the quantum-group results~\cite{[BGT]} for its images.

\subsection{Direct computation of TL fusion rules}
We next illustrate our results for TL fusion rules by several direct calculations of the fusion functor, {\it i.e.}, of the induced TL modules. We recall that the standard module with $2j$ through-lines on $N$ sites is denoted by $\StTL{j}[N]$ and  has a basis given by all possible nested configurations of $(N-j)$ arcs, like \
$\psset{xunit=2mm,yunit=2mm}
\begin{pspicture}(0,0)(7,1)
 \psellipticarc[linecolor=black,linewidth=1.0pt]{-}(0.5,1.0)(0.5,1.42){180}{360}
 \psellipticarc[linecolor=black,linewidth=1.0pt]{-}(2.5,1.0)(0.5,1.42){180}{360}
 \psellipticarc[linecolor=black,linewidth=1.0pt]{-}(5.5,1.0)(0.5,0.71){180}{360}
 \psellipticarc[linecolor=black,linewidth=1.0pt]{-}(5.5,1.0)(1.5,1.42){180}{360}
 \\
\end{pspicture}$\;.
 Meanwhile, through-lines are denoted by a vertical line $\thl$ and are not allowed to intersect any arc. We also denote the generators of the algebra $\TLq{N_1}\tensor\TLq{N_2}$ by $e_j$, with $1\leq j\leq N_1-1$ and $N_1+1\leq j\leq N_1+N_2-1$, in accordance with the natural embedding of this product into $\TLq{N_1+N_2}$. 

Examples of induced modules considered in Sec.~\ref{sec:TL-fusion} suggest the following general strategy to compute TL fusion rules. First, we assume that the fusion does not depend on $N_1$ and $N_2$, and this is indeed true due to the connection with the $\LQG$ tensor product. Second, to compute $\StTL{j_1}\fus\StTL{j_2}$ we consider $\StTL{j_1}[2j_1]\fus\StTL{j_2}[2j_2]$ which is a module induced from the one-dimensional module $\StTL{j_1}[2j_1]\tensor\StTL{j_2}[2j_2]$. An advantage of this choice is that the basis element in $\StTL{j_1}[2j_1]\tensor\StTL{j_2}[2j_2]$ is not in the image of  $\TLq{2j_1}\tensor\TLq{2j_2}$-action because $$\TLq{2j_1}\tensor\TLq{2j_2}\bigl(\StTL{j_1}[2j_1]\tensor\StTL{j_2}[2j_2]\bigr)=0.$$ Therefore, the induced module has a basis generated by the following tree
\begin{equation}\label{TL-tree}
{\footnotesize
   \xymatrix@C=0pt@R=15pt@M=1pt@W=1pt{%
    &&&e_j\ar@{-}[d]&&&\\
    &&&e_{j\pm1}e_j\ar@{-}[dr]\ar@{-}[dl]&&&\\
    &&e_{j-1}e_{j+1}e_j\ar@{-}[dr]\ar@{-}[dl]&&e_{j\pm2}e_{j\pm1}e_j\ar@{-}[dr]&&\\
    &e_{j\pm2}e_{j-1}e_{j+1}e_j\ar@{-}[dr]\ar@{-}[drrr]\ar@{-}[dl]&&e_je_{j-1}e_{j+1}e_j\ar@{-}[dr]&& e_{j\pm3}e_{j\pm2}e_{j\pm1}e_j\ar@{-}[dr]&\\
    e_{j\pm3}e_{j\pm2}e_{j-1}e_{j+1}e_j&&e_{j-2}e_{j+2}e_{j-1}e_{j+1}e_j&&e_{j\pm2}e_je_{j-1}e_{j+1}e_j&
    &\!\!\!\!\!\!\!\!\!e_{j\pm4}e_{j\pm3}e_{j\pm2}e_{j\pm1}e_j\\
    &\vdots&&\vdots&&\vdots&
 }      
 }
\end{equation}
where we set $j=2 j_1$. We will see below that it is more 
convenient to represent the states in the induced module graphically
using two different colors of through lines as we did in Sec.~\ref{paragraphOPElattice}. 

The crucial step is then to
find a Jordan form of the Hamiltonian
\begin{equation*}
H=-\sum_{j=1}^{2(j_1+j_2)-1}e_j
\end{equation*}
 on the induced module -- eigenvalues allow one to identify which standard modules are present in the fusion while 
 Jordan cells should give us the content of projective modules in the fusion. Let us see how it works with a few examples
 in order to compare with Prop.~\ref{prop:fus-StTL-root}.


\subsubsection{Example 1}\label{subsec:example-one}

We come back (see Sec.~\ref{paragraphOPElattice}) to the fusion $\StTLn{1}{2}\fus\StTLn{1}{2}$ with basis
\begin{equation*}
\langle\, l,\, e_2 l,\, e_1e_2 l,\,  e_3e_2 l,\, e_1e_3e_2 l,\, e_2e_1e_3e_2 l\,\rangle.
\end{equation*}
The Hamiltonian $H=-e_1-e_2-e_3$ in this basis reads
\begin{equation}
H = - \left( \begin{array}{cccccc}  0 & 0 & 0 & 0 & 0 & 0  \\ 1 & \fug & 1 & 1 & 0 & 0  \\ 0 & 1 & \fug & 0 & 0 & 0  \\
0 & 1 & 0 & \fug & 0 & 0  \\ 0 & 0 & 1 & 1 & 2 \fug & 2  \\ 0 & 0 & 0 & 0 &  1 & \fug  \\  \end{array} \right).
\end{equation}
Recall that the action of the Hamiltonian can be readily obtained using the graphical representation
\begin{equation*}
\langle\, \psset{xunit=2mm,yunit=2mm}
\begin{pspicture}(0,0)(3,1)
\psline[linecolor=red,linewidth=1.0pt](0,-0.5)(0,1)
\psline[linecolor=red,linewidth=1.0pt](1,-0.5)(1,1)
\psline[linecolor=blue,linewidth=1.0pt](2,-0.5)(2,1)
\psline[linecolor=blue,linewidth=1.0pt](3,-0.5)(3,1)
\end{pspicture}
\ ,\,\psset{xunit=2mm,yunit=2mm}
\begin{pspicture}(0,0)(3,1)
\psline[linecolor=red,linewidth=1.0pt](0,-0.5)(0,1)
\psellipticarc[linecolor=black,linewidth=1.0pt]{-}(1.5,1.0)(0.5,1.42){180}{360}
\psline[linecolor=blue,linewidth=1.0pt](3,-0.5)(3,1)
\end{pspicture}
\ ,\, \psset{xunit=2mm,yunit=2mm}
\begin{pspicture}(0,0)(3,1)
 \psellipticarc[linecolor=black,linewidth=1.0pt]{-}(0.5,1.0)(0.5,1.42){180}{360}
\psline[linecolor=red,linewidth=1.0pt](2,-0.5)(2,1)
\psline[linecolor=blue,linewidth=1.0pt](3,-0.5)(3,1)
\end{pspicture}
\ , \, \psset{xunit=2mm,yunit=2mm}
\begin{pspicture}(0,0)(3,1)
\psline[linecolor=red,linewidth=1.0pt](0,-0.5)(0,1)
\psline[linecolor=blue,linewidth=1.0pt](1,-0.5)(1,1)
\psellipticarc[linecolor=black,linewidth=1.0pt]{-}(2.5,1.0)(0.5,1.42){180}{360}
\end{pspicture}
,\, 
\psset{xunit=2mm,yunit=2mm}
\begin{pspicture}(0,0)(3,1)
 \psellipticarc[linecolor=black,linewidth=1.0pt]{-}(0.5,1.0)(0.5,1.42){180}{360}
 \psellipticarc[linecolor=black,linewidth=1.0pt]{-}(2.5,1.0)(0.5,1.42){180}{360}
\end{pspicture}
,\, \psset{xunit=2mm,yunit=2mm}
\begin{pspicture}(0,0)(3,1)
 \psellipticarc[linecolor=black,linewidth=1.0pt]{-}(1.5,1.0)(1.5,1.42){180}{360}
 \psellipticarc[linecolor=black,linewidth=1.0pt]{-}(1.5,1.0)(0.5,0.71){180}{360}
\end{pspicture}
\,\rangle,
\end{equation*}
where we recall that only through lines with different colors can be contracted by TL generators.
It is of course a simple matter to find the eigenvalues of $H$ in the standard modules $\StTL{0}[4]$, $\StTL{1}[4]$ and $\StTL{2}[4]$ on $N=4$ sites.
The Hamiltonian $H$ applied to the one-dimensionnal module $\StTL{2}[4]$ always yields $0$. When $p=3$, we find that $\StTL{0}[4]$ corresponds 
to the eigenvalues $\{0,-3\}$ and $\StTL{1}[4]$ to $\{-1,-1-\sqrt{2},-1+\sqrt{2}\}$. Moreover, we find that $H$ has the following Jordan form in the 
fusion basis
\begin{equation}
H = \left( \begin{array}{cccccc}  -3 & 0 & 0 & 0 & 0 & 0  \\ 0 & 0 & 1 & 0 & 0 & 0  \\ 0 & 0 & 0 & 0 & 0 & 0  \\
0 & 0 & 0 & -1 & 0 & 0  \\ 0 & 0 & 0 & 0 & -1-\sqrt{2} & 0  \\ 0 & 0 & 0 & 0 &  0 & -1+\sqrt{2}  \\  \end{array} \right).
\end{equation}
Using the subquotient structure $\PrTL{2}[4]= \StTL{2}[4] \longrightarrow \StTL{0}[4]$, we conclude that 
\begin{equation}
\StTLn{1}{2}\fus\StTLn{1}{2} =  \StTL{1}[4] \oplus \PrTL{2}[4],\qquad \text{for}\; p=3,
\end{equation}
in agreement with the approach used in the Sec.~\ref{paragraphOPElattice}.

We can repeat this easy calculation for the extended Ising model ($p=4$), in which case we find that  $\StTL{0}[4]$ corresponds to the 
Hamiltonian eigenvalues $\bigl\{ - \ffrac{3+\sqrt{5}}{\sqrt{2}}, \ffrac{1}{2}(\sqrt{10} - 3 \sqrt{2}) \bigr\}$ and $\StTL{1}[4]$ to
$\{ 0, -\sqrt{2}, -2 \sqrt{2} \}$. We can infer the fusion rules from the Jordan form of $H$ in the fusion basis
\begin{equation}
H = \left( \begin{array}{cccccc}  -2 \sqrt{2} & 0 & 0 & 0 & 0 & 0  \\ 0 & -\sqrt{2} & 0 & 0 & 0 & 0  \\ 0 & 0 & 0 & 1 & 0 & 0  \\
0 & 0 & 0 & 0 & 0 & 0  \\ 0 & 0 & 0 & 0 & - \ffrac{3+\sqrt{5}}{\sqrt{2}} & 0  \\ 0 & 0 & 0 & 0 &  0 & \ffrac{1}{2}(\sqrt{10} - 3 \sqrt{2}) \}  \\  \end{array} \right).
\end{equation}
This is consistent with the fusion rule
\begin{equation}
\StTLn{1}{2}\fus\StTLn{1}{2} =  \StTL{0}[4] \oplus \PrTL{2}[4],\qquad \text{for}\; p=4,
\end{equation}
where $\PrTL{2}[4]= \StTL{2}[4] \longrightarrow \StTL{1}[4]$ in this case. This result
is in perfect agreement with Prop.~\ref{prop:fus-StTL-root}.
In all the other cases, we recover the generic fusion rule~\eqref{fus:StTL-3-gen}.

\subsubsection{Example 2}

Let us turn to a slightly more complicated example with the fusion $\StTLn{1}{2}\fus\StTLn{2}{4}$ on 6 sites.
In the generic case, we have the following fusion
\begin{equation}
\StTLn{1}{2}\fus\StTLn{2}{4} = \StTL{1}[6] \oplus \StTL{2}[6] \oplus \StTL{3}[6].
\end{equation}
The space $\StTLn{1}{2}\fus\StTLn{2}{4}$ has dimension $15$ and we choose the basis corresponding to the tree~\eqref{TL-tree}:
\begin{multline*}
\StTLn{1}{2}\fus\StTLn{2}{4} = \langle\, l,\, e_2 l,\, e_1e_2 l,\,  e_3e_2 l,\,   e_4e_3e_2 l,\, e_5e_4e_3e_2 l, \, e_1e_3e_2 l,\, e_4e_1e_3e_2 l,\, e_5e_4e_1e_3e_2 l,\, \\
e_2e_1e_3e_2 l,\, e_4e_2e_1e_3e_2 l,\,  e_5e_4e_2e_1e_3e_2 l,\, e_3e_4e_2e_1e_3e_2 l,\,  e_5e_3e_4e_2e_1e_3e_2 l,\, e_4e_5e_3e_4e_2e_1e_3e_2 l \rangle.
\end{multline*}
Once again, it is useful to use a graphical representation of these states
\begin{multline*}
\StTLn{1}{2}\fus\StTLn{2}{4} = 
\langle\, \psset{xunit=2mm,yunit=2mm}
\begin{pspicture}(0,0)(5,1)
\psline[linecolor=red,linewidth=1.0pt](0,-0.5)(0,1)
\psline[linecolor=red,linewidth=1.0pt](1,-0.5)(1,1)
\psline[linecolor=blue,linewidth=1.0pt](2,-0.5)(2,1)
\psline[linecolor=blue,linewidth=1.0pt](3,-0.5)(3,1)
\psline[linecolor=blue,linewidth=1.0pt](4,-0.5)(4,1)
\psline[linecolor=blue,linewidth=1.0pt](5,-0.5)(5,1)
\end{pspicture}
\ ,  \, \psset{xunit=2mm,yunit=2mm}
\begin{pspicture}(0,0)(5,1)
\psline[linecolor=red,linewidth=1.0pt](0,-0.5)(0,1)
\psellipticarc[linecolor=black,linewidth=1.0pt]{-}(1.5,1.0)(0.5,1.42){180}{360}
\psline[linecolor=blue,linewidth=1.0pt](3,-0.5)(3,1)
\psline[linecolor=blue,linewidth=1.0pt](4,-0.5)(4,1)
\psline[linecolor=blue,linewidth=1.0pt](5,-0.5)(5,1)
\end{pspicture}
\ ,\, \psset{xunit=2mm,yunit=2mm}
\begin{pspicture}(0,0)(5,1)
\psellipticarc[linecolor=black,linewidth=1.0pt]{-}(0.5,1.0)(0.5,1.42){180}{360}
\psline[linecolor=red,linewidth=1.0pt](2,-0.5)(2,1)
\psline[linecolor=blue,linewidth=1.0pt](3,-0.5)(3,1)
\psline[linecolor=blue,linewidth=1.0pt](4,-0.5)(4,1)
\psline[linecolor=blue,linewidth=1.0pt](5,-0.5)(5,1)
\end{pspicture}
\ ,\, \psset{xunit=2mm,yunit=2mm}
\begin{pspicture}(0,0)(5,1)
\psline[linecolor=red,linewidth=1.0pt](0,-0.5)(0,1)
\psline[linecolor=blue,linewidth=1.0pt](1,-0.5)(1,1)
\psellipticarc[linecolor=black,linewidth=1.0pt]{-}(2.5,1.0)(0.5,1.42){180}{360}
\psline[linecolor=blue,linewidth=1.0pt](4,-0.5)(4,1)
\psline[linecolor=blue,linewidth=1.0pt](5,-0.5)(5,1)
\end{pspicture}
\ ,\, \psset{xunit=2mm,yunit=2mm}
\begin{pspicture}(0,0)(5,1)
\psline[linecolor=red,linewidth=1.0pt](0,-0.5)(0,1)
\psline[linecolor=blue,linewidth=1.0pt](1,-0.5)(1,1)
\psline[linecolor=blue,linewidth=1.0pt](2,-0.5)(2,1)
\psellipticarc[linecolor=black,linewidth=1.0pt]{-}(3.5,1.0)(0.5,1.42){180}{360}
\psline[linecolor=blue,linewidth=1.0pt](5,-0.5)(5,1)
\end{pspicture}
\ ,\, \psset{xunit=2mm,yunit=2mm}
\begin{pspicture}(0,0)(5,1)
\psline[linecolor=red,linewidth=1.0pt](0,-0.5)(0,1)
\psline[linecolor=blue,linewidth=1.0pt](1,-0.5)(1,1)
\psline[linecolor=blue,linewidth=1.0pt](2,-0.5)(2,1)
\psline[linecolor=blue,linewidth=1.0pt](3,-0.5)(3,1)
\psellipticarc[linecolor=black,linewidth=1.0pt]{-}(4.5,1.0)(0.5,1.42){180}{360}
\end{pspicture}
\ ,\, \psset{xunit=2mm,yunit=2mm}
\begin{pspicture}(0,0)(5,1)
\psellipticarc[linecolor=black,linewidth=1.0pt]{-}(0.5,1.0)(0.5,1.42){180}{360}
\psellipticarc[linecolor=black,linewidth=1.0pt]{-}(2.5,1.0)(0.5,1.42){180}{360}
\psline[linecolor=blue,linewidth=1.0pt](4,-0.5)(4,1)
\psline[linecolor=blue,linewidth=1.0pt](5,-0.5)(5,1)
\end{pspicture}
\ ,\, \psset{xunit=2mm,yunit=2mm}
\begin{pspicture}(0,0)(5,1)
\psellipticarc[linecolor=black,linewidth=1.0pt]{-}(0.5,1.0)(0.5,1.42){180}{360}
\psline[linecolor=blue,linewidth=1.0pt](2,-0.5)(2,1)
\psellipticarc[linecolor=black,linewidth=1.0pt]{-}(3.5,1.0)(0.5,1.42){180}{360}
\psline[linecolor=blue,linewidth=1.0pt](5,-0.5)(5,1)
\end{pspicture}
\ ,
 \\
 \, \psset{xunit=2mm,yunit=2mm}
\begin{pspicture}(0,0)(5,1)
\psellipticarc[linecolor=black,linewidth=1.0pt]{-}(0.5,1.0)(0.5,1.42){180}{360}
\psline[linecolor=blue,linewidth=1.0pt](2,-0.5)(2,1)
\psline[linecolor=blue,linewidth=1.0pt](3,-0.5)(3,1)
\psellipticarc[linecolor=black,linewidth=1.0pt]{-}(4.5,1.0)(0.5,1.42){180}{360}
\end{pspicture}
\ ,\, \psset{xunit=2mm,yunit=2mm}
\begin{pspicture}(0,0)(5,1)
 \psellipticarc[linecolor=black,linewidth=1.0pt]{-}(1.5,1.0)(1.5,1.42){180}{360}
 \psellipticarc[linecolor=black,linewidth=1.0pt]{-}(1.5,1.0)(0.5,0.71){180}{360}
\psline[linecolor=blue,linewidth=1.0pt](4,-0.5)(4,1)
\psline[linecolor=blue,linewidth=1.0pt](5,-0.5)(5,1)
\end{pspicture}
\ ,\, \psset{xunit=2mm,yunit=2mm}
\begin{pspicture}(0,0)(5,1)
\psline[linecolor=blue,linewidth=1.0pt](0,-0.5)(0,1)
\psellipticarc[linecolor=black,linewidth=1.0pt]{-}(1.5,1.0)(0.5,1.42){180}{360}
\psellipticarc[linecolor=black,linewidth=1.0pt]{-}(3.5,1.0)(0.5,1.42){180}{360}
\psline[linecolor=blue,linewidth=1.0pt](5,-0.5)(5,1)
\end{pspicture}
\ ,\, \psset{xunit=2mm,yunit=2mm}
\begin{pspicture}(0,0)(5,1)
\psline[linecolor=blue,linewidth=1.0pt](0,-0.5)(0,1)
\psellipticarc[linecolor=black,linewidth=1.0pt]{-}(1.5,1.0)(0.5,1.42){180}{360}
\psline[linecolor=blue,linewidth=1.0pt](3,-0.5)(3,1)
\psellipticarc[linecolor=black,linewidth=1.0pt]{-}(4.5,1.0)(0.5,1.42){180}{360}
\end{pspicture}
\ ,\, \psset{xunit=2mm,yunit=2mm}
\begin{pspicture}(0,0)(5,1)
\psline[linecolor=blue,linewidth=1.0pt](0,-0.5)(0,1)
 \psellipticarc[linecolor=black,linewidth=1.0pt]{-}(2.5,1.0)(1.5,1.42){180}{360}
 \psellipticarc[linecolor=black,linewidth=1.0pt]{-}(2.5,1.0)(0.5,0.71){180}{360}
\psline[linecolor=blue,linewidth=1.0pt](5,-0.5)(5,1)
\end{pspicture}
\ ,\, \psset{xunit=2mm,yunit=2mm}
\begin{pspicture}(0,0)(5,1)
\psline[linecolor=blue,linewidth=1.0pt](0,-0.5)(0,1)
\psline[linecolor=blue,linewidth=1.0pt](1,-0.5)(1,1)
\psellipticarc[linecolor=black,linewidth=1.0pt]{-}(2.5,1.0)(0.5,1.42){180}{360}
\psellipticarc[linecolor=black,linewidth=1.0pt]{-}(4.5,1.0)(0.5,1.42){180}{360}
\end{pspicture}
\ ,\, \psset{xunit=2mm,yunit=2mm}
\begin{pspicture}(0,0)(5,1)
\psline[linecolor=blue,linewidth=1.0pt](0,-0.5)(0,1)
\psline[linecolor=blue,linewidth=1.0pt](1,-0.5)(1,1)
 \psellipticarc[linecolor=black,linewidth=1.0pt]{-}(3.5,1.0)(1.5,1.42){180}{360}
 \psellipticarc[linecolor=black,linewidth=1.0pt]{-}(3.5,1.0)(0.5,0.71){180}{360}
\end{pspicture}
\,\rangle. 
\end{multline*}

We can readily express $H$ in this basis and find its Jordan-form basis. We then identify the eigenvalues 
with those obtained within the standard modules $\StTL{1}[6]$, $\StTL{2}[6]$, and $ \StTL{3}[6]$ to 
deduce the structure of the fusion product. For example, in the tricritical Ising case, we find
\begin{equation}
\StTLn{1}{2}\fus\StTLn{2}{4} =  \StTL{2}[6] \oplus \PrTL{3}[6],\qquad \text{for}\; p=5,
\end{equation}
with the subquotient structure $\PrTL{3}[6]= \StTL{3}[6] \longrightarrow \StTL{1}[6]$. 
This is again in agreement with Prop.~\ref{prop:fus-StTL-root} because $\PrTL{2}[6]= \StTL{2}[6]$.
One can also check that 
\begin{equation}
\StTLn{1}{2}\fus\StTLn{2}{4} =  \StTL{3}[6] \oplus \PrTL{2}[6],\qquad \text{for}\; p=2 \ {\rm or} \ p=4,
\end{equation}
and
\begin{equation}
\StTLn{1}{2}\fus\StTLn{2}{4} =  \StTL{1}[6] \oplus \PrTL{3}[6],\qquad \text{for}\; p=3.
\end{equation}
Both equations are again in agreement with Prop.~\ref{prop:fus-StTL-root}.

\subsubsection{Example 3}
Finally, we give here some examples of fusion for which more than just one projective module is generated.
Let us consider  $\StTLn{\frac{3}{2}}{3}\fus\StTLn{\frac{3}{2}}{3}$ on 6 sites. In the generic case, the fusion reads 
\begin{equation}
\StTLn{\frac{3}{2}}{3}\fus\StTLn{\frac{3}{2}}{3} = \StTL{0}[3] \oplus \StTL{1}[3] \oplus \StTL{2}[3] \oplus \StTL{3}[3].
\end{equation}
The fusion basis contains 20 states 
\begin{multline}
\StTLn{\frac{3}{2}}{3}\fus\StTLn{\frac{3}{2}}{3} = \langle\, l,\, e_3 l,\, e_2e_3 l,\,  e_1e_2e_3 l,\,   e_4e_3 l,\, e_5e_4e_3 l, \, e_4e_2e_3 l,\, e_3e_4e_2e_3 l,\, e_1e_3e_4e_2e_3 l,\, \\
e_2e_1e_3e_4e_2e_3 l,\, e_4e_1e_3e_4e_2e_3 l,\,  e_5e_4e_1e_3e_4e_2e_3 l,\, e_5e_3e_4e_2e_1e_3 l,\,  e_4e_5e_3e_4e_2e_3 l,\, e_2e_5e_4e_3 l \\
e_5e_1e_3e_4e_2 l,\, e_2e_5e_1e_3e_4e_2e_3 l,\, e_4e_5e_1e_3e_4e_2e_3 l,\, e_4e_2e_5e_1e_3e_4e_2e_3 l,\, e_3e_4e_2e_5e_1e_3e_4e_2e_3 l  \rangle.
\end{multline}
The jordanization of the Hamiltonian in this basis is straightforward using Mathematica$^\copyright$ and the comparison with the eigenvalues obtained in the standard modules
  $\StTL{0}[6]$, $\StTL{1}[6]$, $\StTL{2}[6]$, and $ \StTL{3}[6]$ allows us to deduce the structure of the fusion product.
For percolation theory and the tricritical Ising model, we find 
\begin{equation}
\StTLn{\frac{3}{2}}{2}\fus\StTLn{\frac{3}{2}}{4} = \StTL{1}[6] \oplus  \StTL{3}[6] \oplus \PrTL{2}[6],\qquad \text{for}\; p=3.
\end{equation}
\begin{equation}
\StTLn{\frac{3}{2}}{3}\fus\StTLn{\frac{3}{2}}{3} =  \StTL{0}[6] \oplus  \StTL{2}[6] \oplus \PrTL{3}[6],\qquad \text{for}\; p=5.
\end{equation}
The situation is more interesting for the logarithmic Ising model as there are two projective modules created in the fusion process
\begin{equation}
\StTLn{\frac{3}{2}}{3}\fus\StTLn{\frac{3}{2}}{3} =  \PrTL{2}[6] \oplus \PrTL{3}[6],\qquad \text{for}\; p=4,
\end{equation}
with $\PrTL{3}[6]= \StTL{3}[6] \longrightarrow \StTL{0}[6]$ and  $\PrTL{2}[6]= \StTL{2}[6] \longrightarrow \StTL{1}[6]$. 
All these results are consistent with Prop.~\ref{prop:fus-StTL-root}.

\bigskip


\section{Scaling limit and indecomposable Operator Product Expansions}
\label{Sec::ScalingLimit}

In this section, we will discuss how to interpret our results from a LCFT point of view
by considering the scaling limit of the XXZ spin chain that we have studied so far. As the size of the 
system becomes large, the different Temperley-Lieb modules tend to Virasoro modules, and 
the fusion rules for the local operators of the underlying conformal field theory can be inferred from 
the results of the previous section, as the TL fusion rules do not depend on the length of the chains. 
In particular, it is now well-admitted~\cite{ReadSaleur07-2} that the (self-contragredient) projective modules over the 
Temperley-Lieb algebra correspond in the scaling limit to the so-called staggered modules over the 
Virasoro algebra~\cite{KytolaRidout}. Our aim is therefore to discuss how the fusion rules computed in the previous
section apply to the various primary fields in the theory, with a special emphasize on specific physically-relevant examples.
To conclude this paper, we shall also show that the resolution of the ``catastrophes''
in OPEs closely parallels the lattice construction of Sec.~\ref{paragraphOPElattice}.

\subsection{Scaling limit of the XXZ spin chain, Virasoro staggered modules and logarithmic couplings}
\label{SubsecStaggeredBeta}

As discussed in Sec.~\ref{sec:scal-lim-intro}, the continuum limit of the XXZ spin chain at $\q=\mathrm{e}^{i\pi/p}$ is
described by a CFT with central charge $c_{p-1,p}$ given by~\eqref{eqCentralCharge}.
In particular, the standard modules $\StTL{j}$
of the Temperley-Lieb algebra at $\q=\rme^{i\pi/p}$ converge 
(in a way discussed in Sec.~\ref{sec:scal-lim-intro})
 to Kac modules $\VK_{1,2j+1}$ over the Virasoro 
algebra $\Vir(p-1,p)$ with central charge $c_{p-1,p}$.
These Kac modules have a highest-weight vector with conformal weight
$h_{1,2j+1}$ and are defined
as quotients of  Verma modules: $\VK_{1,2j+1} \equiv \Verma_{h_{1,2j+1}}/\Verma_{h_{1,-2j-1}}$\footnote{Recall that for generic central 
charge, the Verma module $\Verma_{h_{1,2j+1}}$ with $2j$ integer is reducible with a proper submodule isomorphic to $\Verma_{h_{1,-2j-1}}$.}. 
When $\q$ is generic, these Kac modules are irreducible.
When $\q$ is a root of unity, the Kac modules $\VK_{1,2j+1}$ may have one additional singular vector, while the second singular vector
at the level $2j+1$ was set to zero in the
corresponding Verma module. Scaling limit keeping $N$ to be
even yields Virasoro modules with integer values of $j\geq0$,
whereas the spectrum with half-integer values of
$j$ can be obtained through the odd-$N$ limit. 
Then, the finite alternating sum~\eqref{dimIrr} for the dimension of the
irreducible TL module $\IrrTL{j}$ corresponds in the limit to an infinite alternating sum
of Kac characters, giving rise to the Rocha--Caridi formula for the character of the irreducible Virasoro
representation $\VX_{1,2j+1}$ with conformal weight $h_{1,2j+1}$.


It is straightforward to see that at least at the level of generating functions (characters) and subquotient structures,
the projective modules $\PrTL{j}$ over $\TLq{N}$, 
for  $j \geq p/2$,
converge to
the so-called staggered modules $\VP_{1,2j+1}$ over the Virasoro algebra 
which are composed of two Kac modules, {\it i.e.}, they have a subquotient
structure with a diamond shape.
We note that $\PrTL{j}$, for $0\leq j < p/2$, are isomorphic to the standard modules $\StTL{j}$, see~\eqref{TL-proj-st},  and this case was already discussed above. Using the subquotient structure~\eqref{prTL-pic-gen-dense-even} for TL modules and having the correspondence $\IrrTL{j}\leftrightarrow\VX_{1,2j+1}$ between TL and Virasoro irreducible modules,
we thus have the diagram, for $j\modd p\ne \frac{kp-1}{2}$ with
$k=0,1$,
\begin{align}\label{stagg-pic-gen-dense-even}     
   \xymatrix@C=5pt@R=15pt@M=2pt{%
    &&\\
    &\VP_{1,2j+1}: &\\
    &&
 }      
&  \xymatrix@C=2pt@R=15pt@M=2pt@W=2pt{%
    &&{\stackrel{h_{1,2j+1}}{\bullet}}\ar[dl]\ar[dr]&\\
    &{\stackrel{h_{1,2(j+\stf p)-4(j\modd p)-1}}{\bullet}}\ar[dr]&&\stackrel{h_{1,2(j+p+\stf p)-4(j\modd p)-1}}{\bullet}\ar[dl]\\
    &&{\stackrel{h_{1,2j+1}}{\bullet}}&
 } \quad
&   \xymatrix@C=5pt@R=15pt@M=2pt{%
    &&\\
    &\text{for}\quad  j \geq \ffrac{p}{2},&\\
    &&
 }&      
\end{align}
where the function $\stf(j)$ was defined in~\eqref{stf-def}, and nodes `$\bullet$' together with conformal weights $h_{1,j}$ denote irreducible Virasoro subquotients $\VX_{1,j}$. We omit for brevity
the explicit dependence of $\stf$ on $j$ in the diagrams. A complete theory of staggered modules was developed by Kyt\"{o}l\"{a} and Ridout~\cite{KytolaRidout},
following the pioneering work of Rohsiepe~\cite{Rohsiepe}.
We give a brief review of the theory of Virasoro staggered modules in App.~\ref{app:stag-mod}.

We denote the vectors in $\VP_{1,2j+1}$ generating its four irreducible subquotients in terms of the fields $\primt_{j}(z)$, $\primr_{j}(z)$, $\priml_{j}(z)$, and $\primb_{j}(z)$ as
\begin{equation}\label{eq_Pbasis}
   \xymatrix@C=5pt@R=15pt@M=2pt{%
    &&\\
    &\VP_{1,2j+1}  : &\\
    &&
 }      
  \xymatrix@C=15pt@R=15pt@M=2pt{%
    &&{\primt_{j}}\ar[dl]\ar[dr]&\\
    &{\primr_{j}}\ar[dr]&&{\priml_{j}}\ar[dl]\\
    &&{{\primb_{j}}}&
 } 
\end{equation}
where we set $\primt_{j}=\lim_{z\to0}\primt_{j}(z)\vacr$, {\it etc}.
 The state $\primr_{j}$ has the lowest conformal weight in $\VP_{1,2j+1}$.
The whole module can be generated by acting with Virasoro generators on $\primt_{j}$, in particular, the dilatation
operator $L_0$ mixes the fields $\primt_{j}(z)$ and $\primb_{j}(z)$ into a rank-$2$ Jordan cell. In our basis $(\primt_{j},\primb_{j})$, 
it reads
\begin{equation}
L_0 = 
\left( \begin{array}{cc}
h_{1,1+2j} & 1 \\
0 & h_{1,1+2j} \end{array} \right).
\end{equation}
We say that $\primt_{j}$ is the
logarithmic partner of the state (or field) $\primb_{j}$.
The state $\primb_{j}$ is a singular
descendant of $\primr_{j}$ so that we write
\begin{equation}
\displaystyle \primb_{j} = A_j \primr_{j},
\end{equation}
where the operator $A_j$ belongs to the universal enveloping algebra of Virasoro. The singular-vector 
condition $L_1 \primb_{j}= L_2 \primb_{j}=0$ fixes uniquely $A_j$ once a normalization has been properly chosen.
The module $\VP_{1,2j+1}$ is uniquely characterized by a number called logarithmic coupling $\beta_{1,1+2j}$~\cite{MathieuRidout1, KytolaRidout}, see a review in App.~\ref{app:stag-mod}, also  called 
$\beta$-invariant or indecomposability parameter. It is defined through the equation
\begin{equation}
\label{eqDefBeta}
\displaystyle A_j^\dag \primt_{j} = \beta_{1,1+2j} \primr_{j}.
\end{equation}
Note that it can also be simply expressed using Virasoro bilinear form $\beta_{1,1+2j} = \langle \primb_{j} \left. \right| \primt_{j} \rangle$,
with the normalization $\langle \primr_{j} \left. \right| \primr_{j} \rangle = 1$.
These logarithmic couplings are closely related to the coefficients that appear in front of logarithmic terms in correlation functions and OPEs. In fact, using global conformal invariance,
one can show that the logarithmic pair $( \primt_{j}, \primb_{j})$ satisfies
\begin{subequations}\label{eqCorrFunctions}
\begin{eqnarray}
\left\langle \primb_{j}(z) \primb_{j}(0)\right\rangle &=& 0 \\
\left\langle \primb_{j}(z) \primt_{j}(0)\right\rangle &=& \frac{\tilde{\beta}_{1,1+2j}}{z^{2 h_{1,1+2j}}} \\
\left\langle \primt_{j}(z) \primt_{j}(0)\right\rangle &=& \frac{\theta-2 \tilde{\beta}_{1,1+2j} \log z} {z^{2 h_{1,1+2j}}}.
\end{eqnarray}
\end{subequations}
While the constant $\theta$ can be absorbed through a transformation $ \primt_{j} \rightarrow \primt_{j} + \gamma \primb_{j}$, the 
coefficient $\tilde{\beta}_{1,1+2j}$  is a fundamental characteristic of the logarithmic pair $(\primt_{j},\primb_{j})$.
In general and despite usual expectations, we have $\tilde{\beta}_{1,1+2j} \neq \beta_{1,1+2j} = \langle \primb_{j} \left. \right| \primt_{j} \rangle $; this corrects some expressions in~\cite{VJS}. Nevertheless, these two numbers are related
by a simple coefficient that depends only on the operator $A_j$. This distinction was discussed
in~\cite{MathieuRidout1} -- it is due to the fact that once the adjoint operator is defined for the state $\primr_{j}$ then
the adjoint for its descendant $\primb_{j}$ is fixed, and generally it will not coincide with the naive definition $\lim_{z\to\infty}z^{2h_{\primb_{j}}}\langle0|\primb_{j}(z)$. 
We refer the interested reader to~\cite{MathieuRidout1} for more details, or to  App.~\ref{AppBetaCorr} of this paper for a discussion of the difference between $\beta_{1,1+2j}$ and $\tilde{\beta}_{1,1+2j}$ from the perspective of the OPE approach to be developed in Sec.~\ref{IndecompOPE}. From the algebraic point of view, the fundamental coefficient is $\beta_{1,1+2j}$,
not the parameter that appears in correlation functions.
In principle, the parameters $\beta_{1,1+2j}$ can be directly obtained from lattice models; this was actually
done in~\cite{VJS, DJS}. 

\subsection{Virasoro fusion rules}
Following heuristic arguments described in Sec.~\ref{sec:scal-lim-intro},
we can conjecture that the fusion rules computed in the previous sections for the Temperley-Lieb algebra modules also apply to their corresponding Virasoro modules, and thus describe the usual fusion operation in CFT.
Let us describe as an example the fusion rules of Kac modules over Virasoro that correspond to OPEs of the corresponding primary fields.
Fusion rules of all  Kac modules  $\VK_{1,2j+1}$ with lowest conformal weight $h_{1,2j+1}$ for any 
half-integer $j$ are given by the substitutions $\StTL{j}\to \VK_{1,2j+1}$ and  $\PrTL{j}\to \VP_{1,2j+1}$ in Prop.~\ref{prop:fus-StTL-root}. 

 \begin{conj}\label{conj:fus-Kac}
 For $1\leq s_1,s_2\leq p$ 
and $r_1,r_2\geq0$, the fusion of two Kac modules over the Virasoro algebra $\Vir(p-1,p)$ with central charge $c_{p-1,p}$ is
\begin{multline}\label{fusion-StTLKac}
\VK_{1,r_1 p + s_1} \fus\VK_{1,r_2 p+s_2}=
\bigoplus_{\substack{s=|s_1-s_2|+1\\\step=2}}^{p - |p-s_1 - s_2| - 1}\!\!\!\VK_{1,(r_1+r_2)p+s}\oplus
\bigoplus_{\substack{r=|r_1-r_2|+1\\\step=2}}^{r_1+r_2-1} \bigoplus_{\substack{s=\gamma_2\\\step=2}}^{p-|s_1-s_2|-1}\!\!\!\VP_{1,r p+s}\\
\oplus\!\!\! \bigoplus_{\substack{s=\gamma_2\\\step=2}}^{s_1 + s_2 - p - 1}\!\!\!\!\!\!\VP_{1,(r_1+r_2+1) p+s}
\oplus\bigoplus_{\substack{r=|r_1-r_2+\sg(s_1-s_2)|+1\\\step=2}}^{r_1+r_2}
\bigoplus_{\substack{s=\gamma_1\\\step=2}}^{|s_1-s_2|-1}\!\!\!\VP_{1,r p+s},
\end{multline}
where we use the notations~\eqref{gamma-def}-\eqref{sg-def} of Sec.~\ref{Sec::fusion}. 
 \end{conj}
This formula is consistent with results obtained in the percolation case~\cite{MathieuRidout}, 
where a straightforward approach to Virasoro fusion rules, the Nahm--Gaberdiel--Kausch algorithm~\cite{KauschGaberdiel}, was used. 
Our conjectured fusion rules agree also with the results in~\cite{RP1,RP2}.
Similar formulas such as the fusion of Virasoro staggered and/or simple modules 
can also be readily obtained from the Temperley-Lieb fusion results of Sec.~\ref{Sec::fusion} and Prop.~\ref{prop:TLfusion-irrep} 
using the substitutions $\IrrTL{j}\to \VX_{1,2j+1}$, $\StTL{j}\to \VK_{1,2j+1}$ and  $\PrTL{j}\to \VP_{1,2j+1}$. For example, one can readily obtain
the fusion of staggered modules from our lattice results. 

 \begin{conj}
 For $1\leq s_1,s_2\leq p-1$ 
and $r_1,r_2\geq1$, the fusion of two staggered modules over the Virasoro algebra $\Vir(p-1,p)$ with central charge $c_{p-1,p}$ is
\begin{multline*}
\VP_{1,r_1p+s_1}\fus \VP_{1,r_2p+s_2}
=4\!\!\!\mathop{\bigoplus{\kern-3pt}''}\limits_{\substack{r=|r_1-r_2|-1\\\step=2}}^{r_1+r_2+1}
\bigoplus_{\substack{s=2p-s_1 - s_2 + 1\\\step=2}}^{p-\gamma_2}
\!\!\!\!\!\!\VP_{1,(r+1)p-s}\\
\oplus 2 \bigoplus_{\substack{r=|r_1-r_2|+1\\\step=2}}^{r_1+r_2-1} 
\Biggl(\bigoplus_{\substack{s=|s_1-s_2|+1\\\step=2}}^{\substack{
\min(s_1 + s_2 - 1,\\ 2p - s_1 - s_2 - 1)}}\VP_{1,(r+1)p-s}
\oplus2\bigoplus_{\substack{s=s_1 + s_2 + 1\\\step=2}}^{p-\gamma_2}
\VP_{1,(r+1)p-s}\Biggr)\\
\oplus2\mathop{\bigoplus{\kern-3pt}'}\limits_{\substack{r=|r_1-r_2|\\
\step=2}}^{r_1+r_2}\Biggl(
\bigoplus_{\substack{s=|p-s_1-s_2|+1\\\step=2}}^{\substack{
\min(p-s_1 + s_2 - 1,\\ p + s_1 - s_2 - 1)}}\!\!\!\!\!\!\VP_{1,(r+1)p-s}
\oplus2\!\!\!\!\!\!\bigoplus_{\substack{s=\min(p-s_1 + s_2 + 1,\\ p + s_1 - s_2 + 1)}}^
{p-\gamma_1}\!\!\!\!\!\!\VP_{1,(r+1)p-s}\Biggl).
\end{multline*}
 \end{conj}

\subsection{Logarithmic operator product expansions}
\label{IndecompOPE}
To illustrate these Virasoro fusion rules in more concrete terms, we discuss 
how they 
can be \textit{independently} deduced from the OPEs in the logarithmic theory.
We revisit here the argument of~\cite{VJS}, initially designed to calculate
logarithmic couplings, to illustrate the indecomposability arising in OPEs at logarithmic points. 
We shall see that 
this method nicely mimics the lattice fusion calculations of Sec.~\ref{paragraphOPElattice}.

The main idea of this approach is to consider LCFTs as a limit of ordinary CFTs, this goes back
to the work of Gurarie and Ludwig on $c=0$ theories~\cite{Gurarie, Gurarie2, GurarieLudwig1,GurarieLudwig2}. This was also
discussed by Cardy in the context of disordered systems~\cite{Cardylog}. 
Let us consider as an example the logarithmic Ising model ($p=4$, $c=1/2$), with the following fusion rule
\begin{equation}\label{eqIsingFus}
\VK_{1,3} \fus\VK_{1,3}=  \VK_{1,1} \oplus \VP_{1,5}, 
\end{equation}
given by equation~\eqref{fusion-StTLKac}.
We deform our logarithmic theory into a generic ordinary CFT ($p=4+\epsilon$) and find~\cite{BPZ,FZ}
\begin{equation}
\VK_{1,3} \fus\VK_{1,3}= \VK_{1,1} \oplus \VK_{1,3} \oplus \VK_{1,5}. 
\end{equation}
This fusion rule means that when two primary fields $\Phi_{1,3}(z)$ are brought close to each other in a correlation function,
their product can be expanded onto the fields $\Phi_{1,1}(z)$, which is the identity field $\one$, $\Phi_{1,3}(z)$, $\Phi_{1,5}(z)$ and their descendants. At the critical (``indecomposable'') point $p=4$, the fields
$\Phi_{1,3}(z)$ and $\Phi_{1,5}(z)$ (and their descendants) are mixed into the staggered module $\VP_{1,5}$ and we thus expect logarithms
to appear.

More explicitly, the OPE of $\Phi_{1,3}(z)$ with itself in the generic case reads
\begin{multline}\label{eqIsingOPEgeneric}
\displaystyle \Phi_{1,3}(z) \Phi_{1,3}(0) \sim \frac{C^{\Phi_{1,3}}_{\Phi_{1,3},\Phi_{1,3}}}{z^{h_{1,3}}} \left[\Phi_{1,3}(0) +\half z \partial \Phi_{1,3}(0) + \alpha^{(-2)} z^2 L_{-2} \Phi_{1,3}(0) \right. \\  \left. + \alpha^{(-1,-1)} z^2 L_{-1}^2 \Phi_{1,3}(0) +\dots \right] + \frac{C^{\Phi_{1,5}}_{\Phi_{1,3},\Phi_{1,3}}}{z^{2 h_{1,3}-h_{1,5}}} \left[\Phi_{1,5}(0) +\dots \right]+ \frac{1}{z^{2 h_{1,3}}} \left[\one +\dots \right],
\end{multline}
where the coefficients $\alpha^{(-2)}=\frac{6}{5 \epsilon} + \frac{16}{25} + \mathcal{O}(\epsilon)$ and $\alpha^{(-1,-1)}=-\frac{9}{10 \epsilon} - \frac{9}{50} + \mathcal{O}(\epsilon)$ are fixed by conformal invariance and are 
diverging as $\epsilon \rightarrow 0$\,\footnote{Note that structure constants like $C^{\Phi_{1,3}}_{\Phi_{1,3},\Phi_{1,3}}$ also depend on $\epsilon$ as well as on the boundary conditions, we will not make this dependence explicit as it will play no role in the following.}. However, if we introduce the field $\primb_{2} = (L_{-2} - \frac{3}{4} L_{-1}^2) \Phi_{1,3}$,
we can get rid of one of these divergences as $\frac{3}{4} \alpha^{(-2)}+\alpha^{(-1,-1)}=\frac{3}{10}+ \mathcal{O}(\epsilon)$ is well-defined:
\begin{multline}
\displaystyle \Phi_{1,3}(z) \Phi_{1,3}(0) \sim \frac{C^{\Phi_{1,3}}_{\Phi_{1,3},\Phi_{1,3}}}{z^{h_{1,3}}} \left[\Phi_{1,3}(0) +\half z \partial \Phi_{1,3}(0) + \alpha^{(-2)} z^2 \primb_{2}(0) \right. \\  \left. +  z^2 (\alpha^{(-1,-1)}+\ffrac{3}{4} \alpha^{(-2)})\partial^2 \Phi_{1,3}(0) +\dots \right] + \frac{C^{\Phi_{1,5}}_{\Phi_{1,3},\Phi_{1,3}}}{z^{2 h_{1,3}-h_{1,5}}} \left[\Phi_{1,5}(0) +\dots \right] +\dots.
\end{multline}
We dropped the identity $\one$ channel as it will play no role in the following.
We then define a new field $\primt_{2}(z)$ as\footnote{This expression might appear a bit complicated at first sight. However, 
the coefficients in front of $\primb_{2} (z)$ and $\Phi_{1,5} (z)$ can be readily obtained using the following requirements: (1) When $\Phi_{1,5}$ is expressed in terms of $\primt_{2}$ and $\primb_{2}$, the coefficient in front of $\primb_{2}$ must be chosen such that
it will combine with the divergent OPE amplitude, (2) The other coefficient can be fixed by requiring that the combination $\primt_{2} (0) +  \primb_{2} (0) \log z$ should appear in the OPE when $\epsilon \rightarrow 0$, (3) The $\tilde{\beta}$ parameter is then identified using correlation functions (see~\eqref{CorrIsingLog}). }
\begin{equation}\label{eq_defPsiIsing}
\displaystyle  \primt_{2} (z)  = \ffrac{C^{\Phi_{1,5}}_{\Phi_{1,3},\Phi_{1,3}}}{C^{\Phi_{1,3}}_{\Phi_{1,3},\Phi_{1,3}}} \ffrac{\tilde{\beta}(\epsilon)}{\alpha^{(-2)} N_{\primb_{2}}(\epsilon)}  \Phi_{1,5} (z) +   \ffrac{\tilde{\beta}(\epsilon)}{N_{\primb_{2}}(\epsilon)} \primb_{2}(z),
\end{equation}
where $\tilde{\beta} = - \frac{ N_{\primb_{2}}(\epsilon)}{h_{1,5}-h_{1,3}-2}$. 
The factor $ N_{\primb_{2}}(\epsilon)$ is defined as the coefficient that appears in the two-point function $\left\langle \primb_{2}(z) \primb_{2}(0) \right\rangle = N_{\primb_{2}}(\epsilon) z^{-2(h_{1,3}+2) }$. This number is computed in App.~\ref{AppBetaCorr},
we find $ N_{\primb_{2}}(\epsilon) = \frac{35}{96} \epsilon + \mathcal{O}(\epsilon)$.
Since $\primb_{2}$ becomes a singular state at $\epsilon=0$, we obviously have $N_{\primb_{2}}(0)=0$.
As $\tilde{\beta}(\epsilon)$ has a finite limit when $\epsilon \to 0$, we can now safely take the limit $\epsilon \rightarrow 0$ so that the logarithmic OPE reads
\begin{multline}
\displaystyle \Phi_{1,3}(z) \Phi_{1,3}(0) \sim \frac{C^{\Phi_{1,3}}_{\Phi_{1,3},\Phi_{1,3}}}{z^{1/2}} \left[ \Phi_{1,3}(0) +\ffrac{1}{2}z \partial \Phi_{1,3}(0) + \ffrac{3}{10} z^{2} \partial^2 \Phi_{1,3}(0) + \right. \\
\left. \ffrac{7}{16 \tilde{\beta}_{1,5}} z^2 (\primt_{2} (0) +  \primb_{2} (0) \log z) +\dots \right] +\dots
\end{multline}
where we have used that $\lim_{\epsilon \to 0} N_{\primb_{2}}(\epsilon) \alpha^{(-2)}=\frac{7}{16}$.
One can check that the operators $\psi_{2}$ and $\phi_{2}$ defined this way satisfy the 
usual OPEs for logarithmic operators with ``modified'' indecomposability parameter $\tilde{\beta}_{1,5} = \lim_{\epsilon \rightarrow 0} \tilde{\beta} (\epsilon) = - \frac{35}{24}$. 
In particular, it is possible to check that
\begin{subequations}
\label{CorrIsingLog}
\begin{eqnarray}
\left\langle \primb_{2}(z) \primb_{2}(0)\right\rangle &=& 0 \\
\left\langle \primb_{2}(z) \primt_{2}(0)\right\rangle &=& \frac{\tilde{\beta}_{1,5}}{z^5} \\
\left\langle \primt_{2}(z) \primt_{2}(0)\right\rangle &=& \frac{-2 \tilde{\beta}_{1,5} \log z} {z^5} ,
\end{eqnarray}
\end{subequations}
along with $L_0 \primt_{2} = \frac{5}{2} \primt_{2} + \primb_{2}$. 
Thanks to the explicit expression~\eqref{eq_defPsiIsing} of the logarithmic field $\primt_{2}(z)$,
we were thus able to identify the coefficient $\tilde{\beta}_{1,5}$ that appears in correlation functions
as the limit $\epsilon \to 0$ of the function $\tilde{\beta}(\epsilon) = - \frac{ N_{\primb_{2}}(\epsilon)}{h_{1,5}-h_{1,3}-2}$.
Note that it is also possible to compute the ``algebraic'' indecomposability parameter $\beta_{1,5}$ using
the definition~\eqref{eqDefBeta} along with~\eqref{eq_defPsiIsing}. 
For more detail concerning this calculation, with a particular emphasize on the difference between $\tilde{\beta}$
and the algebraic indecomposability parameter $\beta$, we refer the reader to App.~\ref{AppBetaCorr}.
Meanwhile, there are several important points here:
\begin{itemize} 

\item We were able to reproduce the logarithmic fusion rule~\eqref{eqIsingFus} using
a simple limit argument working directly at the level of OPEs.  In this 
language, logarithms and indecomposability arise naturally without complicated algebraic consideration.
This computation thus provides a nice check of Conj.~\ref{conj:fus-Kac}. From a rigorous algebraic point of view~\cite{[HLZ]}, the logarithmic OPE or logarithmic intertwining operator satisfying the associativity condition is unique~\cite{[HLZ2]}. As this is also true  for each ordinary CFT at generic central charge,  the OPE we obtained in the limit is consistent -- it satisfies the associativity condition, {\it etc.}

\item Although the field $\primt_{2}(z)$ is built out of two divergent quantities, it is perfectly well-defined 
in correlation functions, as long as the limit process is properly respected. It is interesting 
to see that this whole construction parallels the fusion calculations on the lattice of 
Sec.~\ref{paragraphOPElattice}. As $\epsilon \rightarrow 0$, the eigenvectors $\primb_{2}$ 
and $\Phi_{1,5}$ of $L_0$ become degenerate with the same conformal weight $h_{1,5}=\frac{5}{2}$.
The generically primary field $\Phi_{1,5}$ is ill-defined in the limit, but if one considers the appropriate combination 
with $\primb_{2}$, one can construct a new well-defined field $\primt_{2}$ that will be mixed into 
a Jordan cell with $\primb_{2}$ at $c=\frac{1}{2}$. This calculation bears a strong resemblance with the
analogous lattice construction in Sec.~\ref{paragraphOPElattice}.

\item Although the definition of $\primt_{2}(z)$ in~\eqref{eq_defPsiIsing}
seems to depend on the particular fusion that one considers in the first place (here $\Phi_{1,3}$),
all the physical properties of this field, including the OPEs and the correlation functions, do not
depend on this feature. In particular, the coefficient $C^{\Phi_{1,3}}_{\Phi_{1,3},
\Phi_{1,3}}/C^{\Phi_{1,5}}_{\Phi_{1,3},\Phi_{1,3}}$ is not very important for our matters, 
and if we were to define $\primt_{2}(z)$ to cancel the divergence in another fusion (say 
$\Phi_{1,5} \fus \Phi_{1,5}$), we would find a field $\primt_{2}(z)$ with the same
properties. We conjecture that in the limit $\epsilon \rightarrow 0$, these different definitions 
have the same limit, up to a rescaling.

\end{itemize} 

Finally, we note that this OPE construction that we presented on a specific example is actually quite general.
Indeed, the divergence of the OPE coefficients $\alpha^{(-2)}$ and $\alpha^{(-1,-1)}$ for the descendants 
of $\Phi_{1,3}$ is directly related to the vanishing of the Kac determinant $K^{(2)}(c,h)=2 h(16 h^2 + 2 h (c - 5) + c)$ at level~$2$ for $c=\frac{1}{2}$ and $h=h_{1,3}=\frac{1}{2}$.
Indeed, it can be shown in general that the OPE coefficients for the descendants are inversely proportional to the Kac
determinant at this level (see {\it e.g.}~\cite{YellowBook}). We thus expect these divergences to be general, and they must be 
canceled by the (generally unique) primary operators present in the OPE with same conformal weights at $\epsilon\to0$ as
the ill-defined descendant terms. However, there might be some additional cancellations between the Kac determinant (of the Kac modules)
and the numerator in the descendant OPE coefficients, which are hard to control in general\footnote{As a simple example, the Kac determinant $K^{(1)}(c,h)=2h$ at level $1$ is canceled in the OPE coefficient $\alpha^{(-1)}=\frac{1}{2}$ in~\eqref{eqIsingOPEgeneric}.}. In particular, using the Kac-determinant argument, we could
also expect a divergence at level $6$ in the identity channel in~\eqref{eqIsingOPEgeneric}. However, the identity module corresponds to the direct summand in~\eqref{eqIsingFus} and we do not expect fields from  it to be combined with any other Kac-module contributions. Therefore, we believe that this divergence should vanish although it is unfortunately almost impossible to check it explicitly in this case. We note that this peculiarity should happen only for Kac modules $\VK_j$, with $0 \leq j < p/2 $ (inside the finite Kac table). It would also be very interesting to understand how the field $\priml(z)$ in~\eqref{eq_Pbasis} arises from this construction\footnote{We finally note that there is a 
more rigorous way~\cite{GurarieLudwig1} to compute OPEs directly in the limit $\epsilon=0$ if  
the fusion rules are known a priori. Recovering our OPEs using a similar method should probably
shed some light on this matter.}. 

The general pattern is quite clear and can be fully understood from the single
example worked out in this section. 
To obtain the fusion~\eqref{fusion-StTLKac}, we should start from the generic result given
by~\eqref{eqFusionKacGeneric} and write down the corresponding OPE in terms of quantum fields. 
In the generic decomposition, the terms should be composed to get rid of the divergences as $\epsilon \to 0$, or in other words a proper basis among fields should be chosen before taking the limit in $\epsilon$. To be more explicit, each Kac module with conformal weight $h_{1,2(j+\stf p)-4(j\modd p)-1}$ 
in the generic fusion decomposition corresponds to a field $\primr_j$. This field has a descendant $\primb_j = A_j \primr_j$ with conformal weight $h_{1,2j+1}$ at $\epsilon=0$ that becomes singular in the limit, and it should therefore have a diverging amplitude in the OPE as $\epsilon \to 0$. 
This descendant should be mixed with the primary field $\Phi_{h_{1,1+2j}}$ with the same conformal weight which also appears in the generic fusion. A linear combination of the primary $\Phi_{h_{1,1+2j}}$ and the descendant $\primb_j = A_j \primr_j$ should then be used to define a new field that will become $\primt_j$ in the limit $\epsilon \to 0$, see~\eqref{eq_defPsiIsing}. Introducing this new field should then
cancel the ill-defined OPE amplitude in front of the field $\primb_j = A_j \primr_j$.
Using this mechanism, we finally get the staggered module~\eqref{stagg-pic-gen-dense-even} contribution in the limit of the OPE.

\subsection{OPEs and fusion rules}

The calculations of OPEs in the continuum theory are thus consistent 
with the fusion rules of Kac (standard) modules. One can understand in this way the
fusion by considering the limit of generic results, where we see that the Kac modules 
get glued together at logarithmic points, forming a maximal number of staggered modules in order
to cancel divergent terms in OPEs. 

Another fusion rule that has interesting consequences on the underlying field theory
is given by the fusion of projective $\TLq{N}$-modules.
It seems rather clear that the field associated with the  
Virasoro staggered module $\VP_{1,2j+1}$ really is $\primt_j$ in~\eqref{eq_Pbasis}.
When $\VP_{1,2j+1}$ appears in a given fusion of Kac modules 
for example, this means that the logarithmic field $\primt_j(z)$ is 
generated, along with its descendants (generated by the action 
of the $L_n$'s on $\primt_j(z)$, where $n$ can be positive) like 
$\primb_j$, $\primr_j$ or $\priml_j$. The coefficients as well as the logarithmic
terms are then fixed using conformal invariance.
Following this interpretation, it means that the fusion  
$\primt_{j_1} \fus \primt_{j_2}$ is given by $\VP_{1,2j_1+1} \fus \VP_{1,2j_2+1}$, 
which is in turn given by the fusion of projective modules of 
the Temperley-Lieb algebra on the lattice. We know exactly this fusion, 
and we know that it will generate only projective modules. 
Note that in principle, once the OPEs for $\primt_{j}$ are known, the
OPEs for all the other fields in the corresponding staggered
module could be deduced from it. 
 
Let us illustrate this  with the well-known percolation problem ($c=0$), for which we have the following fusion rule
\begin{equation}
\displaystyle \VK_{1,3} \fus\VK_{1,3}= \VK_{1,3} \oplus \VP_{1,5}.
\end{equation}
The staggered module $\VP_{1,5}$ is very important because it contains the stress energy tensor and
its logarithmic partner. The resolution of the so-called $c\rightarrow 0$ catastrophe is now well-known (see recent formulations in {\it e.g.}~\cite{KoganNichols, DJS,VJS}) so we will not repeat this calculation in details here. With a similar reasoning as for the Ising model,
we introduce a new field $t(z)$ as
\begin{equation}\label{eqDeftOPE}
\displaystyle  t (z)  = C^{\Phi_{1,5}}_{\Phi_{1,3},\Phi_{1,3}} \ffrac{\beta(\epsilon)}{h_{1,5}} \Phi_{1,5} (z)+ \ffrac{\beta(\epsilon)}{\Braket{T | T}} T(z),
\end{equation}
where $\beta(\epsilon) = \tilde{\beta}(\epsilon) = - \frac{ \Braket{T | T}}{h_{1,5}-2}$, $\Braket{T | T}=\frac{c}{2}$ and $p=3+\epsilon$. 
The OPE of two $\Phi_{1,3}$ fields as $c\rightarrow 0$ then reads
 \begin{equation}
\displaystyle \Phi_{1,3}(z) \Phi_{1,3}(0) \sim \frac{1}{z^{2/3}} \left[\one + \ffrac{ 1}{3\beta_{1,5}} z^2( T(0) \log z + t(0)) + \dots \right] + \dots,
\end{equation}
with $\beta_{1,5}=\lim_{\epsilon \rightarrow 0} \beta (\epsilon)=-\frac{5}{8}$. 

Let us consider the OPE $t(z)t(0)$
 which should correspond to the fusion of two staggered modules
 $\VP_{1,5}$. These are given as the scaling limit of the fusion rules
 for the projective module
 $\PrTL{2}$ over the Temperley--Lieb algebra. The TL fusion
\begin{equation}
\PrTL{2}\fus\PrTL{2} = \PrTL{1} \oplus 2\PrTL{2} \oplus \PrTL{3}
\oplus \PrTL{4}
\end{equation}
 obtained from~\eqref{fusion-PrTL-PrTL} allows us to state the Virasoro fusion of
 staggered modules
\begin{equation}\label{eqFusiontt}
\VP_{1,5}\fus\VP_{1,5} = \VP_{1,3} \oplus 2\VP_{1,5} \oplus \VP_{1,7}
\oplus \VP_{1,9},
\end{equation}
where $\VP_{1,3}=\VK_{1,3}$ and $\VP_{1,9}=\VK_{1,9}$ are irreducible
Virasoro modules. This fusion rule already appeared in~\cite{MathieuRidout}
where the Nahm--Gaberdiel--Kausch algorithm was used and also in~\cite{RP2}. It can be also computed
using the OPE approach with the definition~\eqref{eqDeftOPE}. Indeed, using 
OPEs in the non-logarithmic CFT with $\epsilon \neq 0$ and properly\footnote{This step involves computing the behavior of the different
structure constants in the OPE $\Phi_{1,5}(z)\Phi_{1,5}(0)$ as $\epsilon \rightarrow 0$ in order to cancel all the divergences. 
A similar analysis at the order of the correlation functions was made in~\cite{KoganNichols}.} taking the limit
$\epsilon \rightarrow 0$ yields\footnote{We transformed the fields so that $\theta=0$ in~\eqref{eqCorrFunctions}.}
\begin{multline}
\displaystyle t(z)t(0) \sim -\frac{2 \beta_{1,5} \log z}{z^4} + \frac{t(0)  \left(1-4 \log z \right)}{z^2} -  \frac{ T(0)\left(\log z + 2 \log^2 z + \alpha \right)}{z^2}
+ \frac{\partial t(0)  \left(1-4 \log z \right)}{2z} \\ - \frac{\partial T(0)\left(\log z + 2 \log^2 z + \alpha \right)}{2z} + \dots + \frac{C_{tt}^{\Phi_{1,3}}}{z^{5/3}} \left[ \Phi_{1,3}(0) + \half z \partial \Phi_{1,3}(0) + \ffrac{55}{14} z^2 \partial^2 \Phi_{1,3}(0) + \dots \right] \\+ C_{tt}^{\Phi_{1,9}} z^{22/3} \left[ \Phi_{1,9}(0) +  \dots \right].
\end{multline}
The first line of this OPE is well-known, the existence of the constant $\alpha$ in this OPE was already remarked in~\cite{KoganNichols,GurarieLudwig1}. Within our calculation, this constant is fixed and we find $\alpha = \frac{253}{100}$.
This is probably due to our precise identification of $\Phi_{1,5}$ as the field that will be mixed with $T$ in the OPEs\footnote{We also note in
passing that we found $\alpha = \frac{13}{25}$ for dilute polymers, another, so-called ``dual'',  theory with $c=0$ (see the next paragraph).}. What is more interesting here is that the OPE
approach completely determines the other channels in the OPE $t(z)t(0)$, in a way that is consistent with~\eqref{eqFusiontt}.
Note that we dropped the terms that would contribute to the $\VP_{1,7}$ part of the fusion for simplicity, although they 
also appear in the OPE formulation. Finally, we comment on the intriguing multiplicity $2$ in front of the staggered module $\VP_{1,5}$ in~\eqref{eqFusiontt}. This multiplicity could be interpreted as the existence of two different isomorphic copies of $\VP_{1,5}$ in the theory, on the other hand, one would like to have only one identity operator and one stress energy tensor in a reasonable CFT.
Using the OPE approach, one can see that there are  two channels that contribute to $\VP_{1,5}$ in the fusion, that is, that generate the field $t$. 
Indeed, using~\eqref{eqDeftOPE}, we see that the fields $\Phi_{1,5}$ and $\one$ in the OPE $t(z)t(0)$ are coming
from three different contributions $\Phi_{1,5} (z)\Phi_{1,5} (0)$, and $T(z) \Phi_{1,5} (0)$, and $T(z)T(0)$. In terms of generic fusion,
this is because the generic version of~\eqref{eqFusiontt} reads $(\VK_{1,1} \oplus \VK_{1,5}) \fus (\VK_{1,1} \oplus \VK_{1,5}) = \VK_{1,3} \oplus 2 (\VK_{1,1} \oplus \VK_{1,5}) \oplus (\VK_{1,5} \oplus \VK_{1,7}) \oplus \VK_{1,9} $, where one can already see the multiplicity $2$ at this point.
Therefore, we should just sum the amplitudes corresponding to each channel for $\Phi_{1,5}$ and $\one$ and we thus do not think this multiplicity should be interpreted as two different fields $t_1(z)$ and $t_2(z)$.

\subsection{A remark on other theories}
We finally note that although we mainly focused in this paper on theories with central charge $c_{p-1,p}$
and with conformal weights lying in the first row of the Kac table $h_{1,1+2j}$, several
simple extensions of our work are possible. The study of ``dilute'' models (see {\it e.g.}~\cite{Onc,ReadSaleur07-2})
would enable us to reach conformal weights $h_{1+2j,1}$ in the first column of the Kac table.
The underlying algebra would be in this case a dilute version of $\TLq{N}$. Another option
would be to consider the transformation $H \to -H$ in the Hamiltonian of the XXZ spin chain, this corresponds
to $\fug \to - \fug$ in the TL algebra. The scaling limit of such models are described by the so-called $(1,p)$
theories, with central charge $c_{1,p}$. The self-contragredient projective TL modules in this case also converge 
to staggered Virasoro modules in the limit, but their subquotient structure is slightly different with respect to the $(p-1,p)$ theories due to disappearance  of the irreducible TL subquotients $\IrrTL{j}$, for $0\leq j<p/2$, which were the ``minimal-models'' contributions in the $(p-1,p)$ theories. The structure of staggered modules in the $(1,p)$ theories appeared in many papers~\cite{KauschGaberdiel,[BFGT],VJS}, see also references therein, and we do not repeat it here. It turns out that our results about fusion of Kac and staggered  modules in these theories agrees with those conjectured in~\cite{JR,[BGT]}.

%


\section{Conclusion}

To conclude, we believe that the fusion functor studied in this paper is a very powerful tool.
Fusion in logarithmic CFTs is usually studied through the so-called Nahm--Gaberdiel--Kausch 
algorithm~\cite{KauschGaberdiel} which is considerably more involved than the simple-minded
lattice calculations on a few sites that we performed throughout the examples of this paper. Moreover, if the 
$\q$-Schur--Weyl
duality and the correspondence with $\LQG$ tensor products may appear quite technical to the reader,
this approach provides general rigorous results for the lattice fusion thus avoiding tentative
conjectures based on calculations performed for a few examples on a small number of sites. We also think that our paper
provides the correct algebraic framework to understand and interpret the results of~\cite{PRZ,RP1,RP2}
and we wish to emphasize the crucial importance of algebraic methods to understand the connection
between ``lattice'' associative algebras and the Virasoro algebra. 
 
Our result are nicely illustrated by 
OPE calculations performed in the corresponding LCFTs. These logarithmic OPEs are
strongly similar to lattice fusion calculations, where indecomposable points are approached
as {\it limit} of results obtained in semi-simple (generic) cases. This 
suggests that this concept of {\it limit} is crucial, and could become
an important tool in the near future both on the lattice side 
and in the field theory. An interesting task would be to apply the construction of intertwining operators in~\cite{[HLZ2],[HLZ3]} 
to our ``moduli space'' of CFTs and take the limit in a rigorous manner.

It would be most certainly interesting to generalize these results to other boundary conditions
in order to explore other regions of the Kac table than the first row. A careful understanding
of the connection between the boundary extensions of the TL algebra (see {\it e.g.}~\cite{JSBlob}) and Virasoro algebra
would then be required in order to proceed. 
Work along these lines is in progress, 
we hope to report on these issues soon.

\smallskip

{\bf Acknowledgments.}
We gratefully acknowledge valuable discussions and collaboration on
related matters with H.~Saleur, N.~Read, P.P.~Martin, J.L.~Jacobsen, and I.Yu.~Tipunin. 
We are indebted to D.~Ridout for pointing out to us the subtle distinction
between indecomposability parameters and the coefficients that appear
in correlation functions.
We also wish to thank K.~Kyt\"{o}l\"{a}
for his encouragement to publish these results.
The work of AMG was supported in 
part by Marie-Curie IIF fellowship, the RFBR grant 10-01-00408 and the RFBR--CNRS grant 09-01-93105.

\appendix

\section{Tensor products for $\LQG$}\label{appendix:tensor-prod-LQG}
In this appendix, we recall general formulas for tensor products of different $\LQG$-modules computed in~\cite{[BGT]}. We recall that translation between notations in this paper and the ones in~\cite{[BGT]} is the following: $\modWeyl_{s,r}\leftrightarrow\mathsf{N}_{s,r}$, $\modWeyls_{s,r}\leftrightarrow\text{\mifody\sf I}_{s,r}$, $\XX_{s,r}\leftrightarrow\mathsf{X}_{s,r}$, $\PP_{s,r}\leftrightarrow\mathsf{P}_{s,r}$, and $\modM_{s,r}\leftrightarrow\mathsf{M}_{s,r}$.
We use also the notations~\eqref{gamma-def}-\eqref{sg-def} of Sec.~\ref{Sec::fusion}. 

The base tensor products are collected in the following theorem.
\begin{Thm}\label{thm:tens-prod-intro}\cite{[BGT]}
 The tensor product \\
 -- of two irreducible modules for $s_1,s_2=1,\dots, p$
and $r_1,r_2\in\oN$ is
\begin{equation}\label{fusion-XX}
\repX_{s_1,r_1}\tensor\repX_{s_2,r_2} =
\bigoplus_{\substack{r=|r_1-r_2|+1\\\step=2}}^{r_1+r_2-1}
\Bigr(\bigoplus_{\substack{s=|s_1-s_2|+1\\\step=2}}^{\substack{
\min(s_1 + s_2 - 1,\\ 2p - s_1 - s_2 - 1)}}\!\!\!\repX_{s,r}
\;\oplus\!\!\! \bigoplus_{\substack{s=2p - s_1 - s_2 +1\\\step=2}}^{p-\gamma_2}
\!\!\!\!\!\!\PP_{s,r}\Bigl)
\end{equation}
-- of an irreducible with a projective module
for $s_1=1,\dots, p$, $s_2=1,\dots, p-1$
and $r_1,r_2\in\oN$ is
\begin{equation}\label{fusion-XP}
\repX_{s_1,r_1}\tensor\PP_{s_2,r_2} = 
\bigoplus_{\substack{r=|r_1-r_2|+1\\\step=2}}^{r_1+r_2-1}\Bigl(
\bigoplus_{\substack{s=|s_1-s_2|+1\\\step=2}}^{\substack{
\min(s_1 + s_2 - 1,\\ 2p - s_1 - s_2 - 1)}}\!\!\!\!\!\!\PP_{s,r}
\oplus2\!\!\!\!\!\!\bigoplus_{\substack{s=2p-s_1-s_2+1\\\step=2}}^{p-\gamma_2}
\!\!\!\!\!\!\PP_{s,r}\Bigr)
\oplus2\mathop{\bigoplus{\kern-3pt}'}\limits_{\substack{r=|r_1-r_2|\\\step=2}}^{r_1+r_2}
\bigoplus_{\substack{s=p-s_1+s_2+1\\\step=2}}^{p-\gamma_1}
\!\!\!\!\!\!\PP_{s,r},
\end{equation}
where we set $\PP_{s,0}=0$.\\
-- of two contragredient Weyl modules
for $s_1,s_2=1,\dots, p-1$
and $r_1,r_2\in\oN$ is
\begin{multline}\label{fusion-NN}
\modNbar_{s_1,r_1}\tensor\modNbar_{s_2,r_2} = 
\bigoplus_{\substack{r=|r_1-r_2|+1\\\step=2}}^{r_1+r_2-1} \bigoplus_{\substack{s=|s_1-s_2|+1\\\step=2}}^{p-\gamma_2}\!\!\!\PP_{s,r}\;
\oplus\!\!\! \bigoplus_{\substack{s=s_1 + s_2 +1\\\step=2}}^{p-\gamma_2}\!\!\!\!\!\!\PP_{s,r_1+r_2+1}\\
\oplus\bigoplus_{\substack{r=|r_1-r_2+\sg(s_2-s_1)|+1\\\step=2}}^{r_1+r_2}
\bigoplus_{\substack{s=p-|s_1-s_2|+1\\\step=2}}^{p-\gamma_1}\!\!\!\PP_{s,r}
\oplus\bigoplus_{\substack{s=|p-s_1-s_2|+1\\\step=2}}^{p - |s_1 - s_2| - 1}\!\!\!\modNbar_{s,r_1+r_2}.
\end{multline}
-- of a contragredient Weyl module with a projective module
for $1\leq s_1,s_2\leq p-1$
and $r_1,r_2\geq1$ is
\begin{multline*}
\modNbar_{s_1,r_1}\tensor\PP_{s_2,r_2} = 
\bigoplus_{\substack{r=|r_1-r_2|+1\\\step=2}}^{r_1+r_2-1}\Bigl(
\bigoplus_{\substack{s=|s_1-s_2|+1\\\step=2}}^{\substack{
\min(s_1 + s_2 - 1,\\ 2p - s_1 - s_2 - 1)}}\!\!\!\!\!\!\PP_{s,r}
\oplus2\!\!\!\!\!\!\bigoplus_{\substack{s=2p-s_1-s_2+1\\\step=2}}^{p-\gamma_2}
\!\!\!\!\!\!\PP_{s,r}\Bigr)
\oplus2\mathop{\bigoplus{\kern-3pt}'}\limits_{\substack{r=|r_1-r_2|\\\step=2}}^{r_1+r_2}
\bigoplus_{\substack{s=p-s_1+s_2+1\\\step=2}}^{p-\gamma_1}
\!\!\!\!\!\!\PP_{s,r}\\
\oplus\bigoplus_{\substack{r=|r_1-r_2+1|+1\\\step=2}}^{r_1+r_2}\Bigl(
\bigoplus_{\substack{s=|p-s_1-s_2|+1\\\step=2}}^{\substack{
\min(p-s_1 + s_2 - 1,\\ p + s_1 - s_2 - 1)}}\!\!\!\!\!\!\PP_{s,r}
\oplus2\!\!\!\!\!\!\bigoplus_{\substack{s=p+s_1-s_2+1\\\step=2}}^{p-\gamma_2}
\!\!\!\!\!\!\PP_{s,r}\Bigr)
\oplus2\mathop{\bigoplus{\kern-3pt}'}\limits_{\substack{r=|r_1-r_2+1|\\\step=2}}^{r_1+r_2+1}
\bigoplus_{\substack{s=s_1+s_2+1\\\step=2}}^{p-\gamma_1}
\!\!\!\!\!\!\PP_{s,r},
\end{multline*}
-- of two projective modules for $1\leq s_1,s_2\leq p-1$
and $r_1,r_2\geq1$ is
\begin{multline}\label{fusion-PP}
\PP_{s_1,r_1}\tensor \PP_{s_2,r_2}
= 2 \bigoplus_{\substack{r=|r_1-r_2|+1\\\step=2}}^{r_1+r_2-1} 
\Bigl(\bigoplus_{\substack{s=|s_1-s_2|+1\\\step=2}}^{\substack{
\min(s_1 + s_2 - 1,\\ 2p - s_1 - s_2 - 1)}}\PP_{s,r}
\oplus2\bigoplus_{\substack{s=2p-s_1 - s_2 + 1\\\step=2}}^{p-\gamma_2}
\PP_{s,r}\Bigr)\\
\oplus2\mathop{\bigoplus{\kern-3pt}'}\limits_{\substack{r=|r_1-r_2|\\
\step=2}}^{r_1+r_2}\Bigl(
\bigoplus_{\substack{s=|p-s_1-s_2|+1\\\step=2}}^{\substack{
\min(p-s_1 + s_2 - 1,\\ p + s_1 - s_2 - 1)}}\!\!\!\!\!\!\PP_{s,r}
\oplus2\!\!\!\!\!\!\bigoplus_{\substack{s=\min(p-s_1 + s_2 + 1,\\ p + s_1 - s_2 + 1)}}^
{p-\gamma_1}\!\!\!\!\!\!\PP_{s,r}\Bigl)
\oplus\,4\!\!\!\mathop{\bigoplus{\kern-3pt}''}\limits_{\substack{r=|r_1-r_2|-1\\\step=2}}^{r_1+r_2+1}
\bigoplus_{\substack{s=s_1 + s_2 + 1\\\step=2}}^{p-\gamma_2}
\!\!\!\!\!\!\PP_{s,r},
\end{multline}
where we set $\PP_{s,0}=\PP_{s,-1}=0$ and $\modWeyls_{p-s,0}\equiv\XX_{s,1}$.
\end{Thm}
In the main text of this paper,
we  use this theorem to decompose the XXZ spin-chain Hilbert space for any root of unity and to derive exactly $\TLq{N}$ fusion rules.

\section{Fusion of irreducible TL modules}
\label{appendix-IrrFusion}

 While we were able to compute fusion for all
 standard and projective $\TLq{N}$-modules, the fusion involving the irreducible modules
 $\IrrTL{j}$ with $0\leq j\leq p-1$, except $\IrrTL{\frac{p-1}{2}}=\StTL{\frac{p-1}{2}}$,  \textit{is unfortunately not covered by the quantum-group fusion} because these modules have no quantum-group counterparts.
 On the other hand, fusion rules for $\IrrTL{j}$ with $j > p-1$ can be obtained similarly using quantum-group results
 and are collected in this appendix.  We use here the notations~\eqref{gamma-def}-\eqref{sg-def} of Sec.~\ref{Sec::fusion} and set $\q=\rme^{i\pi/p}$ for any integer $p\geq2$, as usual. 
 \begin{Prop}\label{prop:TLfusion-irrep}
\mbox{}\\
 $\bullet$\; For $0\leq s_1,s_2\leq p-1$
and $r_1,r_2>1$, the fusion of two irreducible $\TL{\q}$-modules for $r_1\ne r_2$
is
\begin{equation}\label{fusion-IrrTL}
\IrrTL{\frac{r_1p+s_1-1}{2}}\fus\IrrTL{\frac{r_2p+s_2-1}{2}}=
\bigoplus_{\substack{r=|r_1-r_2|+1\\\step=2}}^{r_1+r_2-1}
\Biggl(\bigoplus_{\substack{s=|s_1-s_2|+1\\\step=2}}^{\substack{
\min(s_1 + s_2 - 1,\\ 2p - s_1 - s_2 - 1)}}\!\!\!\IrrTL{\frac{(r+1)p-s-1}{2}}
\;\oplus\! \bigoplus_{\substack{s=s_1 + s_2 +1\\\step=2}}^{p-\gamma_2}
\!\!\!\!\!\!\PrTL{\frac{(r+1)p-s-1}{2}}\Biggr)
\end{equation}
and for $r_1= r_2=k>1$ the fusion is
\begin{multline*}
\IrrTL{\frac{kp+s_1-1}{2}}\fus\IrrTL{\frac{kp+s_2-1}{2}}=
\bigoplus_{\substack{s=|s_1-s_2|+1\\\step=2}}^{\substack{
\min(s_1 + s_2 - 1,\\ 2p - s_1 - s_2 - 1)}}\!\!\!\StTL{\frac{s-1}{2}}
\oplus \bigoplus_{\substack{r=3\\\step=2}}^{2k-1}
\bigoplus_{\substack{s=|s_1-s_2|+1\\\step=2}}^{\substack{
\min(s_1 + s_2 - 1,\\ 2p - s_1 - s_2 - 1)}}\!\!\!\IrrTL{\frac{(r+1)p-s-1}{2}}\\
\oplus\bigoplus_{\substack{r=1\\\step=2}}^{2k-1}
 \bigoplus_{\substack{s=s_1 + s_2 +1\\\step=2}}^{p-\gamma_2}
\!\!\!\!\!\!\PrTL{\frac{(r+1)p-s-1}{2}}.
\end{multline*}
$\bullet$\; For $0\leq s_1\leq p-1$, $1\leq s_2\leq p-1$,
 $r_1>1$ and $r_2\geq1$, the fusion of an irreducible with a projective $\TL{\q}$-module is
\begin{multline}\label{fusion-IrrTL-PrTL}
\IrrTL{\frac{r_1p+s_1-1}{2}}\fus\PrTL{\frac{r_2p+s_2-1}{2}} = 
\bigoplus_{\substack{r=|r_1-r_2|+1\\\step=2}}^{r_1+r_2-1}\Biggl(
\,\bigoplus_{\substack{s=|s_1-s_2|+1\\\step=2}}^{\substack{
\min(s_1 + s_2 - 1,\\ 2p - s_1 - s_2 - 1)}}\!\!\!\PrTL{\frac{(r+1)p-s-1}{2}}
\oplus2\!\!\bigoplus_{\substack{s=s_1+s_2+1\\\step=2}}^{p-\gamma_2}
\!\!\PrTL{\frac{(r+1)p-s-1}{2}}\Biggr)\\
\oplus 2\mathop{\bigoplus{\kern-3pt}'}\limits_{\substack{r=|r_1-r_2|\\\step=2}}^{r_1+r_2}
\bigoplus_{\substack{s=p+s_1-s_2+1\\\step=2}}^{p-\gamma_1}
\!\!\PrTL{\frac{(r+1)p-s-1}{2}},
\end{multline}
where we additionally  set  $\PrTL{\frac{s-1}{2}}=0$, for $1\leq s\leq
p-1$, in the sum at the second row and the
primed sum is given in~\eqref{not-1}.\\
\end{Prop}

 \section{Virasoro staggered modules theory}\label{app:stag-mod}
Let $\Vir(p',p)$ denote the Virasoro algebra with central charge
$c_{p',p}$. We review here the theory~\cite{KytolaRidout} of the so-called \textit{staggered} modules over
$\Vir(p',p)$. In general, a staggered module is a gluing (an extension) of two
highest-weight modules with a non-diagonalizable action of $L_0$. Using Prop.~$4.6$
and Cor.~$4.7$ from~\cite{KytolaRidout}, the study of any
staggered module can actually be reduced to the simpler analysis of ``Verma-type'' staggered modules which are gluing of two Verma modules
\begin{equation}\label{Verma-stag-seq}
0\longrightarrow \Verma_{h_l}\longrightarrow \cVP(h_l,h_r) \longrightarrow \Verma_{h_r}\longrightarrow 0,
\end{equation}
where $\Verma_{h}$ denotes the Verma module generated from the highest-weight
vector with conformal weight~$h$. Let $\Hmod_{h}$ be a quotient of the Verma module $\Verma_{h}$, then any staggered module $\VP$
defined by a non-split  exact sequence
\begin{equation*}
0\longrightarrow \Hmod_{h_l}\longrightarrow \VP(h_l,h_r) \longrightarrow \Hmod_{h_r}\longrightarrow 0,
\end{equation*}
 \textit{is} a quotient of the ``Verma-type'' staggered module $\cVP$ defined in~\eqref{Verma-stag-seq} above.

In what follows, we consider 
the case $p'=p-1$ and describe a general strategy to fully determine the action of the Virasoro generators on the 
staggered modules $\VP_{1,2j+1}$ from~\eqref{stagg-pic-gen-dense-even}.
In order to get the staggered module $\VP_{1,2j+1}$, consider first a Verma-type staggered module
with the subquotient structure
\begin{equation}\label{Vstagg-pic-gen-dense-odd}
   \xymatrix@C=5pt@R=15pt@M=2pt@W=2pt{%
    &&\\
    &\cVP_j: &\\
    &&
 }      
  \xymatrix@C=15pt@R=15pt@M=2pt{%
    &&{\stackrel{\reldim{\lev_1}}{\bullet}}\ar[dl]\ar[dr]\ar[drr]&&&\\
    &{\stackrel{\reldim{0}}{\bullet}}\ar[dr]\ar[ddr]&
    &{\stackrel{\reldim{\lev_2}}{\bullet}}\ar[ddl]\ar[dr]\ar[drr]&{\stackrel{\reldim{\lev_1+2j+1}}{\times}}\ar[d]\ar[dr]\ar[ddll]&\\
    &&{\stackrel{\reldim{2(j+\stf p)-4(j\modl p)-1}}{\times}}\ar[ddrr]\ar[ddr]&&\ddots&\ddots\\
    &&  {\stackrel{\reldim{\lev_1}}{\bullet}}\ar[dr]\ar[drr] &&&\\
    && &\ddots&\ddots&
 } 
\end{equation}
where we set $j\modl p\equiv j\modd p$
 and  numbers in braces show the relative level with respect to the
field with lowest conformal dimension denoted by $\reldim{0}$, filled dots `$\bullet$' denote
preimages of the corresponding subquotients in $\VP_{1,2j+1}$ while the
crosses $\times$ denote the  singular vectors that we need to set to zero in order to get the module $\VP_{1,2j+1}$\footnote{To be more explicit, the crosses represent eigenvectors of $L_0$ that should be annihilated by positive Virasoro modes $L_{>0}$.}. We note that arrows in down-right direction correspond to action of negative Virasoro generators while arrows    in down-left direction correspond to action of $L_n$'s with positive $n$. 
For the case $p=1$, we  should replace the ``braid''-type Verma modules in~\eqref{Vstagg-pic-gen-dense-odd} by   the ``chain''-type Verma modules corresponding to diagrams like $\bullet\to\bullet\to\times\to\dots$.
Here, we also introduce  the relative conformal dimensions $\lev_i=\lev_i(j)$:
\begin{multline}\label{lev1-def}
\lev_1(j)\equiv h_{1,2j+1}-h_{1,2(j+\stf p)-4(j\modd p)-1}\\
=\ffrac{\bigl(1 + 2 (j\modd p) - \stf p\bigr) \bigl(2 j (p-1) + p ((p-1)\stf -1 ) - 
   2(p-1) (j\modd p)\bigr)}{p}
\end{multline}
and
\begin{multline}\label{lev2-def}
\lev_2(j)\equiv h_{1,2(j+p+\stf p)-4(j\modd p)-1}-h_{1,2(j+\stf p)-4(j\modd p)-1}\\
=1 + 2(p-1)(j-2(j\modd p)) - p(3+2\stf) + p^2(1 + 2 \stf).
\end{multline}
where we recall that the function $\stf \equiv \stf(j)$ was defined in~\eqref{stf-def}.

Using Thm. $6.14$  from~\cite{KytolaRidout}, we see that
there exists a one-parameter family of non-isomorphic staggered modules $\cVP_j$ with the
subquotient structure given by~\eqref{Vstagg-pic-gen-dense-odd}. 
This parameter is called the \textit{logarithmic coupling} or the $\beta$-invariant
$\bb_{h_r}$ of the module $\cVP(h_l,h_r)$ and was introduced in Sec.~\ref{SubsecStaggeredBeta}.
The necessary and sufficient condition for the existence of the module $\VP_{1,2j+1}$
is described in Prop.~7.2 of~\cite{KytolaRidout}: the vector $\reldim{\lev_1+2j+1}$ denoted by $\times$
should be a singular vector, {\it i.e.} the arrow representing an action of
positive modes $L_{>0}$ in the diagram
in~\eqref{Vstagg-pic-gen-dense-odd} should actually be absent. Except for the case $p=1$, this  happens only
for a particular value of the $\bb$-invariant that is a solution of
the system of linear equations produced by the condition
\begin{equation}\label{eq-sing-gen-dense}
L_1 \reldim{\lev_1+2j+1} = L_2 \reldim{\lev_1+2j+1} =0.
\end{equation}
Using an appropriate ansatz for the singular vector $\reldim{\lev_1+2j+1}$,
one can solve the system~\eqref{eq-sing-gen-dense} to obtain the value of $\beta$,
and to fix the full action of the Virasoro generators on the module $\VP_{1,2j+1}$ (and
in particular the action  of positive Virasoro modes on
the top vector $\reldim{\lev_1}$). We refer the interested reader to the very complete reference~\cite{KytolaRidout}
for more details.

\section{Formulas for the indecomposability parameters  
$\beta_{1,1+2j}$ and $\tilde{\beta}_{1,1+2j}$}
\label{AppBetaCorr}

We have shown in Sec.~\ref{IndecompOPE} that  the appearance of  
logarithmic OPEs in
the (extended) Ising model could be  understood by taking a limit  
$\epsilon \to 0$ (with $p=4+\epsilon$ and $c_{p-1,p}=1-6/p(p-1)$).
In particular, we found a field  $\primt_{2}$ with logarithmic  
two-point function,
mixed at $c=\frac{1}{2}$ with a descendant  
$\primb_{2}=(L_{-2}-\frac{3}{4}L_{-1}^2)\primr_{2}$ of the primary  
field $\primr_{2}=\Phi_{1,3}$ with conformal weight~$h_{1,3}$. The  
two-point functions of these fields are characterized by a unique  
parameter $\tilde{\beta}_{1,5}$, that can be computed through the  
formula
\begin{equation}
\label{eqAppDefBetatilde}
\tilde{\beta}_{1,5} \equiv \lim_{z\to \infty} z^5 \langle  
\primt_{2}(z) \primb_{2}(0) \rangle = - \lim_{\epsilon \to 0}  
\dfrac{N_{\primb_{2}}(\epsilon)}{h_{1,5}-h_{1,3}-2}.
\end{equation}
In this equation, $N_{\primb_{2}}(\epsilon)$ is defined as the  
coefficient that appears in the two-point function $\left\langle  
\primb_{2}(z) \primb_{2}(0) \right\rangle = N_{\primb_{2}}(\epsilon)  
z^{-2(h_{1,3}+2) }$, where  
$\primb_{2}=(L_{-2}-\frac{3}{4}L_{-1}^2)\primr_{2}$. The calculation  
$N_{\primb_{2}}(\epsilon)$ is straightforward using standard CFT  
tools, we first write
\begin{equation}
\langle \primb_{2} (z) \primb_{2} (w) \rangle = \oint_z \dfrac{{\rm d}  
\xi}{2\pi i} \frac{1}{\xi - z} \langle T(\xi) \primr_{2} (z)  
\primb_{2} (w) \rangle - \frac{3}{4} \partial_z^2 \langle \primr_{2}  
(z) \primb_{2}(w) \rangle.
\end{equation}
The correlation function $\langle \primr_{2} (z) \primb_{2}(w) \rangle  
= \langle \Phi_{1,3} (z) (L_{-2}-\frac{3}{4}L_{-1}^2)\Phi_{1,3} (w)  
\rangle$ between the primary field $\primr_{2}$ with its descendant  
$\primb_{2}$ can be easily computed using the standard differential  
operator expressions for the Virasoro generators. On the other hand,  
the contour integral can be deformed in the usual way and computed  
using the OPE
\begin{equation}
T(\xi) \primb_{2} (w) \sim \dfrac{c - h_{1,3}}{2(\xi-w)^4}  
\primr_{2}(w) + \dfrac{3/2 - 3 h_{1,3}}{(\xi-w)^3} \partial  
\primr_{2}(w)+ \dfrac{ h_{1,3}+2}{(\xi-w)^2} \primb_{2} (w)+ \dfrac{  
1}{(\xi-w)} \partial \primb_{2} (w).
\end{equation}
As expected, the correlation function $\langle \primb_{2} (z)  
\primb_{2} (w) \rangle $ can then be expressed with the use of a  
differential operator acting on $\langle \primr_{2} (z) \primr_{2} (w)  
\rangle = (z-w)^{-2h_{1,3}} $. We find that $\langle \primb_{2} (z)  
\primb_{2} (w) \rangle = N_{\primb_{2}} (\epsilon)  
(z-w)^{-2(h_{1,3}+2)} $, with
\begin{equation}
N_{\primb_{2}} (\epsilon) = \frac{1}{4} \left( 2c + h_{1,3} (7+9   
h_{1,3}(4 h_{1,3}(1+ h_{1,3})-5)) \right).
\end{equation}
This yields $N_{\primb_{2}} (\epsilon) = \frac{35}{96} \epsilon +  
\mathcal{O}(\epsilon^2)$, so that $\tilde{\beta}_{1,5}=-\frac{35}{24}$.

Note that the ``algebraic'' indecomposability parameter characterizing  
the corresponding staggered module, defined as  
$\beta_{1,5}=\Braket{\primb_{2}|\primt_{2}}$ can also be recovered
by replacing $N_{\primb_{2}}(\epsilon)$ by $\Braket{\primb_{2} |  
\primb_{2}}$ in the expression of $\tilde{\beta} (\epsilon)$~\cite{VJS}.
To see this, we use the definition~\eqref{eqDefBeta} along with~\eqref{eq_defPsiIsing} to compute
\begin{equation}
(L_{2}-\ffrac{3}{4}L_{1}^2) \primt_{2}   = \frac{\tilde{\beta}_{1,5}  
(\epsilon)}{N_{\primb_{2}} (\epsilon)} (L_{2}-\ffrac{3}{4}L_{1}^2)  
\primb_{2} = \beta_{1,5}(\epsilon) \primr_{2},
\end{equation}
where in the limit $\epsilon \to 0$
\begin{equation}
\label{eqAppFormulaBeta}
\beta_{1,5} \equiv  \Braket{\primb_{2}|\primt_{2}} = \lim_{\epsilon  
\to 0} \beta_{1,5}(\epsilon)  = - \lim_{\epsilon \to 0}  
\dfrac{\Braket{ \primb_2 | \primb_2 }}{h_{1,5}-h_{1,3}-2}.
\end{equation}
In our case, we find $\Braket{ \primb_2 | \primb_2 } = \Braket{  
\primr_2 |  
(L_{2}-\frac{3}{4}L_{1}^2)(L_{-2}-\frac{3}{4}L_{-1}^2)|\primr_2 } =  
\frac{1}{4}(2c+h_{1,3}(18 h_{1,3} -11))$. Although $\Braket{ \primb_2  
| \primb_2 } \neq N_{\primb_2}(\epsilon)$,
these two expressions coincide up to order $\mathcal{O}(\epsilon^2)$,  
so that in our case we find  
$\beta_{1,5}=\tilde{\beta}_{1,5}=-\frac{35}{24}$. However, these two  
numbers have no reason to be the same in general, it is actually quite  
easy to find examples where they differ. The simplest example where  
this happens corresponds to the case $\primb=L_{-1}\primr$, and  
$h_{\primr}=0$ at the logarithmic point.
In this case, we find $\Braket{\primb|\primb} = 2 h_{\primr}$ and  
$N_{\primb}=-2 h_{\primr} (1+h_{\primr})=-2 h_{\primr} +  
\mathcal{O}(h_{\primr}^2)$, so that $\beta=-\tilde{\beta}$ here. This  
result was obtained using a different method in~\cite{MathieuRidout1}.

Let us summarize what we have achieved here. There are two different steps in our OPE calculation:
the first one is to properly identify the field~\eqref{eq_defPsiIsing} that will become
a logarithmic field in the limit $\epsilon \to 0$, this field is found by imposing the cancellation
of divergent terms in OPEs. The second step is then to use the explicit expression~\eqref{eq_defPsiIsing}
to compute various quantities, be it $\beta_{1,5}$ or $\tilde{\beta}_{1,5}$. The coefficient $\tilde{\beta}_{1,5}$
is obtained through the computation of two-point functions, this leads to the formula~\eqref{eqAppDefBetatilde}.
Meanwhile, one can also use the definition of $\beta_{1,5}$ from~\eqref{eqDefBeta} to obtain the different expression~\eqref{eqAppFormulaBeta}.


 \end{document}